\documentclass[preprint,12pt]{elsarticle}
\usepackage{lmodern} 
\usepackage{amssymb}
\usepackage{amsmath}
\usepackage{booktabs}
\usepackage[showonlyrefs]{mathtools}
\usepackage{hyperref}

\usepackage{siunitx}
\usepackage{makecell}
\usepackage{float}
\usepackage{caption}
\newcommand{\tablesize}{\fontsize{9pt}{10.5pt}\selectfont}
\usepackage{zref-clever}

\usepackage{tikz}

\definecolor{mygreen}{HTML}{20C997}
\definecolor{stmGreen}{HTML}{00A651}
\definecolor{stmOrange}{HTML}{f5821f}
\definecolor{stmYellow}{HTML}{ffde00}
\definecolor{stmDarkYellow}{HTML}{885600}
\definecolor{stmBlue}{HTML}{0083C9}

\affiliation[inst1]{
    organization={Department of Decision Sciences, HEC Montr\'eal},
    addressline={3000, chemin de la Côte-Sainte-Catherine}, 
    city={Montréal},
    postcode={H3T 2A7}, 
    state={QC},
    country={Canada}
}

\affiliation[inst2]{
    organization={Department of Statistics and Actuarial Science, University of Waterloo},
    addressline={200, University Avenue West}, 
    city={Waterloo},
    postcode={N2L 3G1}, 
    state={ON},
    country={Canada}
}

\affiliation[inst3]{
    organization={Department of Mathematics and Statistics, McMaster University},
    addressline={1280, Main Street West}, 
    city={Hamilton},
    postcode={L8S 4L8}, 
    state={ON},
    country={Canada}
}

\affiliation[inst4]{
    organization={Department of Mathematical and Industrial Engineering, Polytechnique Montréal}, 
    city={Montréal},
    postcode={H3T 0A3}, 
    state={QC},
    country={Canada}
}

\affiliation[inst5]{
    organization={Centre Interuniversitaire de Recherche sur les r\'Eseaux d'Entreprise, la Logistique et le Transport (CIRRELT)},
    addressline={2920, chemin de la Tour}, 
    city={Montréal},
    postcode={H3T 1J4}, 
    state={QC},
    country={Canada}
}

\begin{document}
\begin{frontmatter}
\title{A Bayesian Framework for Post-disruption Travel Time Prediction in Metro Networks} 
\author[inst1]{Shayan Nazemi} 
\author[inst1]{Aurélie Labbe}
\author[inst2]{Stefan Steiner}
\author[inst3]{Pratheepa Jeganathan}
\author[inst4,inst5]{Martin Trépanier}
\author[inst1]{Léo R. Belzile}

\begin{abstract}
Disruptions are an inherent feature of transportation systems, occurring unpredictably and with varying durations. Even after an incident is reported as resolved, disruptions can induce irregular train operations that generate substantial uncertainty in passenger waiting and travel times. Accurately forecasting post-disruption travel times therefore remains a critical challenge for transit operators and passenger information systems. This paper develops a Bayesian spatiotemporal modeling framework for post-disruption train travel times that explicitly captures train interactions, headway imbalance, and non-Gaussian distributional characteristics observed during recovery periods. The proposed model decomposes travel times into delay and journey components and incorporates a moving-average error structure to represent dependence between consecutive trains. Skew-normal and skew-$t$ distributions are employed to flexibly accommodate heteroskedasticity, skewness, and heavy-tailed behavior in post-disruption travel times. The framework is evaluated using high-resolution track-occupancy and disruption log data from the Montr\'eal metro system, covering two lines in both travel directions. Empirical results indicate that post-disruption travel times exhibit pronounced distributional asymmetries that vary with traveled distance, as well as significant error dependence across trains. The proposed models consistently outperform baseline specifications in both point prediction accuracy and uncertainty quantification, with the skew-$t$ model demonstrating the most robust performance for longer journeys. These findings underscore the importance of incorporating both distributional flexibility and error dependence when forecasting post-disruption travel times in urban rail systems.
\end{abstract}


\end{frontmatter}

\section{Introduction}
Travel times exhibit substantial temporal variability driven by factors such as travel distance, passenger demand, service frequency, and train congestion \citep{lee2017}. Disruptions, which constitute an inherent and often unavoidable aspect of metro operations, typically arise without warning and vary widely in duration. These incidents considerably degrade network performance by affecting the punctuality and reliability of the incident train, as well as the trains operating immediately before and after it \citep{Dollevoet2018, Jin2016}. Accordingly, accurate travel time prediction becomes particularly critical during the recovery phase following a disruption, as reliable forecasts support operational decision-making and improve overall service quality \citep{Zhang2022}. Our study develops a probabilistic model for predicting the travel times of post-disruption trains, thereby enabling passengers to make more informed travel decisions and facilitating realistic service expectations. We employ a Bayesian modeling framework that allows us to obtain probabilistic forecasts of travel times.

In many high-frequency metro systems, overtaking is not possible within a line segment. As a result, when a disruption occurs, a common operational response is to suspend train movements, after which service controllers manage train departures to restore regular temporal spacing once the incident is cleared. During the disruption, trains accumulate along the line, increasing congestion and leading to additional delays and longer travel times at downstream stations once operations resume. Because disruptions arise unpredictably and are multi-factorial, providing reliable arrival-time estimates at the moment the incident begins is near impossible. Our research therefore focuses instead on predicting the travel time required for trains to move from their positions at the moment the disruption resolution is announced until they reach a designated downstream station.

In the transportation literature, intercity railway systems have received considerably more research attention than urban metro networks, largely because their fixed schedules facilitate the modeling and analysis of delay prediction problems. Central to modeling train operations is the characterization of process times, which consist of \emph{dwell times}, defined as the period a train remains stationary at stations, and \emph{running times}, defined as the period required to travel between stations. A complete journey can be expressed as the sum of dwell and running times across all segments between the origin and destination. However, the primary source of variability in process-time models arises from the dwell times \citep{Li2016, Cornet2019}, whereas running times exhibit comparatively low variability \citep{Kecman2015}. 

Existing research in transportation networks has explored process time modeling and travel time prediction in rail systems. \citet{Li2016} examine both parametric and non-parametric approaches to estimating dwell times during peak and off-peak periods, incorporating temporal and spatial covariates, and conclude that dwell durations are strongly influenced by the number of boarding and alighting passengers. \citet{Cornet2019} propose a data-driven method for estimating dwell-time distributions at a station for a given passenger demand level by decomposing the process into a deterministic minimum dwell component and a stochastic component representing various disruptions during passenger boarding and alighting. In contrast, \citet{lee2017} examine passenger travel time rather than dwell times, decomposing it into walking, waiting, and riding components. Their model estimates passenger travel times using ticket tap-in and tap-out records.

Passenger flow information is not always available in real time, limiting its usefulness for real-time prediction and operational management—an issue that also applies to our case study. More recent research has therefore focused on incorporating the complex network structure of railway systems into modeling frameworks to capture dependencies among operational events such as delays, control actions, and process times. Prior work on delay propagation has employed probabilistic network models to represent these interactions \citep{Bearfield2013, SUN2015116, Ulak2020}, with many approaches relying on the Markov property within these networks \citep{Corman2018, Li2021}. 

Using a Bayesian network constructed from the railway network topology, \citet{Corman2018} model downstream delays in a real-time setting. Their approach reduces the uncertainty associated with predicting future delays as new information becomes available along a train's trajectory. From the perspective of waiting passengers at downstream stations, any updated arrival or departure information reflecting delays can erode perceived service reliability and reduce trust in the system. For this reason, our study concentrates on generating a single prediction for train travel time to downstream stations (during post-disruption periods). Modeling journey times to downstream locations is equivalent to predicting station arrival times; the distinction lies only in whether the problem is viewed from the standpoint of onboard passengers or those waiting on the platform. This problem is closely tied to delay propagation, which can be defined only in networks operating under a fixed timetable. In such systems, delays are measured as the difference between expected and actual arrival times.

\citet{Li2021} address delay propagation by extending the framework of \citet{Corman2018} to incorporate prolonged dwell and running times within their conditional Bayesian model, thereby enabling the estimation of both delay distributions and process-time distributions. \citet{Ge2024} extends this methodology by relaxing the Markov property and considering the influence of more than one preceding trains in their Bayesian network.

The non-linear and complex dynamics of train operation variables have shifted research attention toward neural-based architectures, especially with the increasing capabilities of deep neural networks. Recurrent Neural Networks (RNNs) are particularly effective in modeling the sequential dynamics of train movements and station interactions \citep{HUANG2020, li2024, Luo2023}, while Graph Neural Networks (GNNs) exploit the topological structure of railway systems by embedding station connectivity within a graph-based framework \citep{li2024, wang2023}. \citet{HUANG2020} models arrival delay by considering inter-train interactions and station level dependencies with Long Short-Term Memory (LSTM) components. \citet{li2024} segments time into intervals and construct an interaction network for the trains in each time interval and models arrival and departure delays using graph convolutional networks. \citet{wang2023} proposes a deep reinforcement learning approach to dynamically schedule dwell time at stations with the objective of minimizing both the total waiting time of passengers on the platform and the in-train travel time for onboard passengers. 

Although metro systems differ operationally from road and bus networks, the core challenge of predicting travel times under congested conditions within an interconnected network is common across modes. \citet{Ma2022} model global and local spatial dependencies using a combination of multi-attention graph neural networks and LSTM layers, enabling accurate predictions even on routes with limited data. \citet{Chen2024} investigate both local and long-range correlation structures in bus route networks by developing a Bayesian Gaussian framework for travel-time forecasting. Building on this line of work, \citet{Chen2023} propose a hierarchical Bayesian probabilistic forecasting model that represents link travel times and headways with the preceding bus through a multivariate Gaussian mixture formulation. \citet{Chen2025} further extend this perspective by modeling the joint distribution of bus link travel times and passenger occupancy using a Bayesian Markov regime-switching vector autoregressive model capable of capturing skewness and multi-modality in bus travel times.

While multivariate Gaussian assumptions offer analytical tractability, they may be restrictive in real-world settings characterized by non-linear, asymmetric, and heavy-tailed dependencies. Copula-based models provide a more flexible alternative by decoupling marginal distributions from the dependence structure. Multi-modality is particularly pronounced in road segment travel times \citep{Chen2017, Qin2020}. For example, \citet{Chen2017} address multi-modal marginals using Gaussian mixture models before fitting bivariate copulas, whereas \citet{Qin2020} propose a \(K\)-component copula mixture model that more effectively captures the multi-modal nature of joint travel-time distributions. 

To the best of our knowledge, an important research gap in the transportation literature concerns the effect of disruptions on travel times during the recovery phase. To address this gap, we propose a hierarchical Bayesian framework that captures the temporal dependence in travel times among consecutive trains while accommodating the variability and skewness that arise as a function of traveled distance. Our method decomposes post-disruption travel time into a delay component and a journey component, modeling each separately. The delay component is informed by the spatial separation between a train and its predecessors in the network, whereas the journey component accounts for passenger accumulation resulting from the longer-than-usual headways that develop during disruptions. The Bayesian structure of our framework not only yields point predictions but also provides a principled quantification of forecast uncertainty, which can improve passenger information systems and inform service planning decisions.

The remainder of this paper is organized as follows. \zcref[S]{sec:casestudy} describes the Montr\'eal metro system that serves as the case study for this research and details the various data sources employed in the analysis. \zcref[S]{sec:methodo} presents the proposed methodological framework and outlines the statistical modeling approach developed to address the problem of post-disruption travel time prediction. In \zcref[S]{sec:results}, the empirical results are reported and systematically compared with those obtained from the baseline models. Finally, \zcref[S]{sec:conclusion} concludes the paper by summarizing the main findings and discussing their implications.

\section{Case Study} \label{sec:casestudy}
This study focuses on the operations of two of the most traveled metro lines in Montr\'eal: the Green and the Orange lines, which comprise respectively 27 and 31 stations \cite{STM}. The analysis includes all stations along the Green line and 26 stations from the Orange line. The excluded stations are either located outside the Island of Montr\'eal or exhibit complex network geometries that prevent reliable extraction of train trajectories without incurring a high error rate. Key operational statistics for these two lines are summarized in \zcref[S]{tab:data_metrolinestats}. \zcref[S]{fig:map} presents the map of the Montr\'eal metro system.

\begin{table}[H]
    \centering
    \tablesize
    \begin{tabular}[t]{lcccccc}
        \toprule
        \multicolumn{2}{c}{ } & \multicolumn{5}{c}{Observed operational summaries} \\
        \cmidrule(l{3pt}r{3pt}){3-7}
        Line & Direction & \# of trains & \makecell[c]{Median\\dwell time\\(seconds)} & \makecell[c]{Median\\running time\\(seconds)} & \makecell[c]{Median\\headway\\(minutes)} & \makecell[c]{Median line\\travel time\\(minutes)}\\
        \midrule
        Green & 1 & 75,134 & 40.0 & 43.0 & 5.1 & 41.1\\
         & 2 & 75,012 & 40.0 & 44.0 & 5.2 & 40.9\\
        \addlinespace
        Orange & 1 & 79,098 & 42.0 & 46.0 & 4.9 & 39.8\\
         & 2 & 79,266 & 42.0 & 45.0 & 4.9 & 39.5\\
        \bottomrule
    \end{tabular}
    \caption{Summary of the observed operational data for all train trajectories on the Green and Orange lines of the Montr\'eal metro system over the entire year of 2018. Headway is the temporal gap between two consecutive trains.}
    \label{tab:data_metrolinestats}
\end{table}

\label{sec_data}
In this research, two complementary data sources are used to construct a dataset to model and predict train travel times during post-disruption periods. Train trajectories are extracted from track occupancy records (\zcref[S]{subsec1_data}), and are chronologically aligned with the reported disruption on the corresponding metro line (\zcref[S]{subsec2_data}). For the remainder of this article, and to avoid repetition, we present results based on data from Direction~1 of the Green metro line. Results for the remaining lines and directions are provided in the Appendices.

\subsection{Track Occupancy Data}
\label{subsec1_data}
The track occupancy dataset consists of all sensor records from the Montr\'eal metro network. The network is a fixed block system, with one sensor installed at the beginning of each section and another at its end. A block occupation signal is recorded when a train enters a block, while a block release signal is recorded once the train leaves it. The dataset includes timestamps for every occupation and release event across all blocks of the network during 2018. Each station corresponds to exactly one block, whereas tunnels between stations consist of multiple blocks. Block lengths vary considerably, ranging from several tens of meters to several hundred meters in some sections of the network. By aligning consecutive block occupation and release signals and reconstructing the corresponding sequences, train trajectories can be derived. With the inclusion of station block information, exact arrival and departure timestamps for every train movement within the system can be obtained.

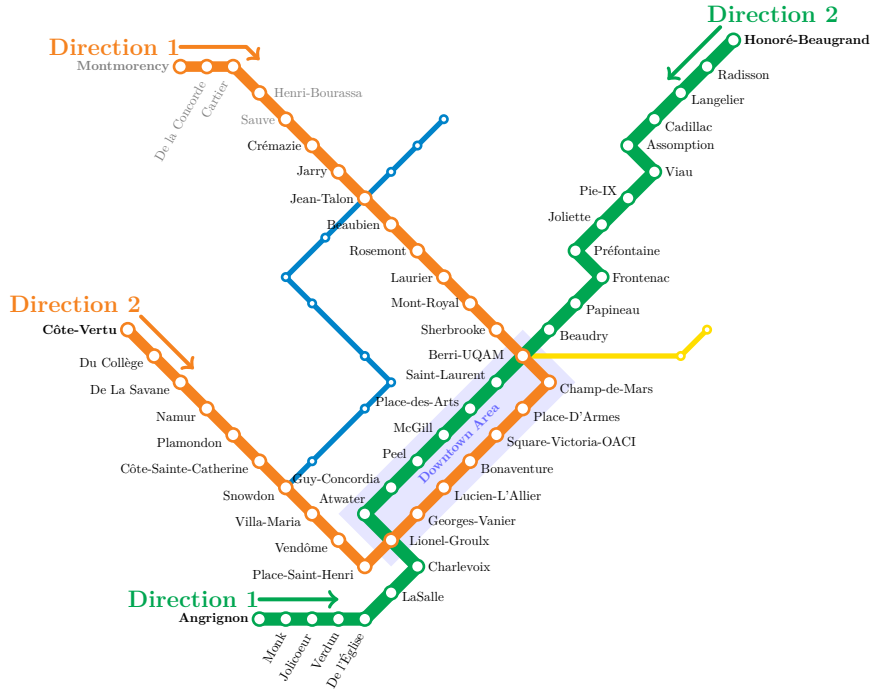
\begin{figure}
    \centering
    \begin{tikzpicture}[scale = 0.35, line cap=round, line join=round]
        % Downtown area
        \fill[blue!10] (3, 4) -- (5, 2) -- (12, 9) -- (10, 11) -- cycle;
        \node[anchor=south west, font=\footnotesize, rotate=45, xshift=-1.75mm, yshift=-0.5mm, text=blue!60] at (6.5, 5) {\scalebox{0.5}{\textbf{Downtown Area}}};

        % Yellow line
        \coordinate (y0) at (10, 10);
        \coordinate (y1) at (16, 10);
        \coordinate (y2) at (17, 11);
        \draw[stmYellow, line width=2pt]
            (y0) -- (y1) -- (y2);
        \filldraw[fill=white, draw=stmYellow, line width=1pt] (y1) circle (4pt);
        \filldraw[fill=white, draw=stmYellow, line width=1pt] (y2) circle (4pt);

        % Blue line
        \coordinate (b0) at (1, 5);
        \coordinate (b1) at (5, 9);
        \coordinate (b2) at (1, 13);
        \coordinate (b3) at (7, 19);
        
        \draw[stmBlue, line width=2pt]
            (b0) -- (b1) -- (b2) -- (b3);

        \filldraw[fill=white, draw=stmBlue, line width=1pt] (2, 6) circle (4pt);
        \filldraw[fill=white, draw=stmBlue, line width=1pt] (4, 8) circle (4pt);
        \filldraw[fill=white, draw=stmBlue, line width=1pt] (5, 9) circle (4pt);
        \filldraw[fill=white, draw=stmBlue, line width=1pt] (4, 10) circle (4pt);
        \filldraw[fill=white, draw=stmBlue, line width=1pt] (2, 12) circle (4pt);
        \filldraw[fill=white, draw=stmBlue, line width=1pt] (b2) circle (4pt);
        \filldraw[fill=white, draw=stmBlue, line width=1pt] (2.5, 14.5) circle (4pt);
        \filldraw[fill=white, draw=stmBlue, line width=1pt] (5, 17) circle (4pt);
        \filldraw[fill=white, draw=stmBlue, line width=1pt] (6, 18) circle (4pt);
        \filldraw[fill=white, draw=stmBlue, line width=1pt] (b3) circle (4pt);

        % Green line
        \coordinate (Angrignon)       at (0, 0);
        \coordinate (Monk)            at (1, 0);
        \coordinate (Jolicoeur)       at (2, 0);
        \coordinate (Verdun)          at (3, 0);
        \coordinate (De l'Eglise)     at (4, 0);
        \coordinate (LaSalle)         at (5, 1);
        \coordinate (Charlevoix)      at (6, 2);
        \coordinate (LionelGroulx)    at (5, 3);
        \coordinate (Atwater)         at (4, 4);
        \coordinate (GuyConcordia)    at (5, 5);
        \coordinate (Peel)            at (6, 6);
        \coordinate (McGill)          at (7, 7);
        \coordinate (PlaceDesArts)    at (8, 8);
        \coordinate (SaintLaurent)    at (9, 9);
        \coordinate (BerriUQAM)       at (10, 10);
        \coordinate (Beaudry)         at (11, 11);
        \coordinate (Papineau)        at (12, 12);
        \coordinate (Frontenac)       at (13, 13);
        \coordinate (Préfontaine)     at (12, 14);
        \coordinate (Joliette)        at (13, 15);
        \coordinate (PieIX)           at (14, 16);
        \coordinate (Viau)            at (15, 17);
        \coordinate (Assomption)      at (14, 18);
        \coordinate (Cadillac)        at (15, 19);
        \coordinate (Langelier)       at (16, 20);
        \coordinate (Radisson)        at (17, 21);
        \coordinate (HonoréBeaugrand) at (18, 22);

        \draw[stmGreen, line width=5pt]
            (Angrignon) -- (De l'Eglise) -- (Charlevoix) -- (Atwater) -- (Frontenac) -- (Préfontaine) -- (Viau) -- (Assomption) -- (HonoréBeaugrand);
            
        % % --- Draw Orange Line ---
        % \draw[stmYellow, line width=2pt]
        %     (BerriUQAM) -- (y1) -- (y2);
        % \node[anchor=west, text=stmDarkYellow] at (y2) {\scalebox{0.4}{Yellow Line}}; 
        
        % --- Draw stations ---
        \foreach \station/\name/\deg/\direction/\shiftX/\shiftY in {
          HonoréBeaugrand/\textbf{Honoré-Beaugrand}/0/west/0mm/0mm,
          Radisson/Radisson/0/west/0mm/-1mm,
          Langelier/Langelier/0/west/0mm/-1mm,
          Cadillac/Cadillac/0/west/0mm/-1mm,
          Assomption/Assomption/0/west/1mm/0mm,
          Viau/Viau/0/west/0mm/0mm,
          PieIX/Pie-IX/0/east/0mm/1mm,
          Joliette/Joliette/0/east/0mm/1mm,
          Préfontaine/Préfontaine/0/west/1mm/0mm,
          Frontenac/Frontenac/0/west/0mm/0mm,
          Papineau/Papineau/0/west/0mm/-1mm,
          Beaudry/Beaudry/0/west/0mm/-1mm,
          % BerriUQAM/Berri-UQAM/0/north west/0mm/0.2mm,
          SaintLaurent/Saint-Laurent/0/east/0mm/1mm,
          PlaceDesArts/Place-des-Arts/0/east/0mm/1mm,
          McGill/McGill/0/east/0mm/1mm,
          Peel/Peel/0/east/0mm/1mm,
          GuyConcordia/Guy-Concordia/0/east/0mm/1mm,
          Atwater/Atwater/0/south/-3mm/0mm,
          % LionelGroulx/Lionel-Groulx/0/west/0mm/1mm,
          Charlevoix/Charlevoix/0/west/0mm/0mm,
          LaSalle/LaSalle/0/west/0mm/-0.5mm,
          De l'Eglise/De l'Église/60/north/-6mm/2mm,
          Verdun/Verdun/60/north/-4.5mm/1.5mm,
          Jolicoeur/Jolicoeur/60/north/-5.5mm/1.5mm,
          Monk/Monk/60/north/-4mm/1.5mm,
          Angrignon/\textbf{Angrignon}/0/east/0mm/0mm
        }{
          \filldraw[fill=white, draw=stmGreen, line width=1pt] (\station) circle (7pt);
          \node[anchor=\direction, font=\footnotesize, rotate=\deg, xshift=\shiftX, yshift=\shiftY] at (\station) {\scalebox{0.5}{\name}};
        }
        \coordinate (CoteVertu)              at (-5, 11);
        \coordinate (DuCollege)              at (-4, 10);
        \coordinate (DeLaSavane)              at (-3, 9);
        \coordinate (Namur)                   at (-2, 8);
        \coordinate (Plamondon)               at (-1, 7);
        \coordinate (CoteSainteCatherine)     at (0, 6);
        \coordinate (Snowdon)                 at (1, 5);
        \coordinate (VillaMaria)              at (2, 4);
        \coordinate (Vendome)                 at (3, 3);
        \coordinate (PlaceSaintHenri)         at (4, 2);
        
        \coordinate (LionelGroulx)             at (5, 3);
        \coordinate (GeorgesVanier)            at (6, 4);
        \coordinate (LucienLallier)            at (7, 5);
        \coordinate (Bonaventure)              at (8, 6);
        \coordinate (SquareVictoriaOACI)       at (9, 7);
        \coordinate (PlaceDArmes)              at (10, 8);
        \coordinate (ChampDeMars)              at (11, 9);
        
        \coordinate (BerriUQAM)                at (10, 10);
        
        \coordinate (Sherbrooke)               at (9, 11);
        \coordinate (MontRoyal)                at (8, 12);
        \coordinate (Laurier)                  at (7, 13);
        \coordinate (Rosemont)                 at (6, 14);
        \coordinate (Beaubien)                 at (5, 15);
        \coordinate (JeanTalon)                at (4, 16);
        \coordinate (Jarry)                    at (3, 17);
        \coordinate (Cremazie)                 at (2, 18);
        \coordinate (Sauve)                    at (1, 19);
        \coordinate (HenriBourassa)             at (0, 20);
        \coordinate (Cartier)                  at (-1, 21);
        \coordinate (DeLaConcorde)             at (-2, 21);
        \coordinate (Montmorency)              at (-3, 21);

        \draw[stmOrange, line width=4pt]
            (CoteVertu) -- (PlaceSaintHenri) -- (ChampDeMars) -- (Cartier) -- (Montmorency);

        \foreach \station/\name/\deg/\direction/\shiftX/\shiftY in {
          CoteVertu/\textbf{C\^ote-Vertu}/0/east/0mm/0mm,
          DuCollege/Du Collège/0/east/0mm/-1mm,
          DeLaSavane/De~La~Savane/0/east/0mm/-1mm,
          Namur/Namur/0/east/0mm/-1mm,
          Plamondon/Plamondon/0/east/0mm/-1mm,
          CoteSainteCatherine/C\^ote-Sainte-Catherine/0/east/0mm/-1mm,
          Snowdon/Snowdon/0/east/0mm/-1mm,
          VillaMaria/Villa-Maria/0/east/0mm/-1mm,
          Vendome/Vend\^ome/0/east/0mm/-1mm,
          PlaceSaintHenri/Place-Saint-Henri/0/east/0mm/-1mm,
          LionelGroulx/Lionel-Groulx/0/west/1mm/0mm,
          GeorgesVanier/Georges-Vanier/0/west/0mm/-1mm,
          LucienLallier/Lucien-L'Allier/0/west/0mm/-1mm,
          Bonaventure/Bonaventure/0/west/0mm/-1mm,
          SquareVictoriaOACI/Square-Victoria-OACI/0/west/0mm/-1mm,
          PlaceDArmes/Place-D'Armes/0/west/0mm/-1mm,
          ChampDeMars/Champ-de-Mars/0/west/0mm/-1mm,
          BerriUQAM/Berri-UQAM/0/east/-1mm/0mm,
          Sherbrooke/Sherbrooke/0/east/0mm/0mm,
          MontRoyal/Mont-Royal/0/east/0mm/0mm,
          Laurier/Laurier/0/east/0mm/0mm,
          Rosemont/Rosemont/0/east/0mm/0mm,
          Beaubien/Beaubien/0/east/0mm/0mm,
          JeanTalon/Jean-Talon/0/east/0mm/0mm,
          Jarry/Jarry/0/east/0mm/0mm,
          Cremazie/Cr\'emazie/0/east/0mm/0mm,
          Sauve/\textcolor{gray}{Sauve}/0/east/0mm/0mm,
          HenriBourassa/\textcolor{gray}{Henri-Bourassa}/0/west/0.5mm/0mm,
          Cartier/\textcolor{gray}{Cartier}/60/north/-5mm/2mm,
          DeLaConcorde/\textcolor{gray}{De la Concorde}/60/north/-8mm/1.5mm,
          Montmorency/\textbf{\textcolor{gray}{Montmorency}}/0/east/0mm/0mm
        }{
          \filldraw[fill=white, draw=stmOrange, line width=1pt] (\station) circle (7pt);
          \node[anchor=\direction, font=\footnotesize, rotate=\deg, xshift=\shiftX, yshift=\shiftY] at (\station) {\scalebox{0.5}{\name}};
        }

        \draw[->, line width=1.5pt, draw=stmGreen] (0, 0.75) -- (3, 0.75);
        \node[text=stmGreen] at (-2.5, 0.75) {\scalebox{0.75}{\textbf{Direction 1}}};

        \draw[->, line width=1.5pt, draw=stmGreen] (17.5, 22.5) -- (15.5, 20.5);
        \node[text=stmGreen] at (19.5, 23) {\scalebox{0.75}{\textbf{Direction 2}}};

        \draw[->, line width=1.5pt, draw=stmOrange] (-3, 21.75) -- (-0.5, 21.75) -- (0, 21.25);
        \node[text=stmOrange] at (-5.5, 21.75) {\scalebox{0.75}{\textbf{Direction 1}}};

        \draw[->, line width=1.5pt, draw=stmOrange] (-4.5, 11.5) -- (-2.5, 9.5);
        \node[text=stmOrange] at (-7, 12) {\scalebox{0.75}{\textbf{Direction 2}}};
        
    \end{tikzpicture}
    \caption{Map of the Montréal metro system. The network comprises four metro lines, of which only the Green and Orange lines are analyzed in this study. The operating direction of each line is indicated. Stations excluded from the Orange line are shown in gray. The blue-shaded area represents the downtown region, and stations located within this area are referred to as downtown stations.}
    \label{fig:map}
\end{figure}

\subsection{Disruption Logs Dataset}
\label{subsec2_data}
The second dataset contains all of the disruptions reported in 2018. Each record provides the duration, location, start and end timestamps, as well as the reported cause of the disruption. However, disruptions are not systematically recorded by the staff, so some incidents may be missing. In addition, inaccuracies in the reported timestamps and durations are occasionally observed. Basic descriptive statistics for the reported disruptions are provided in \zcref[S]{tab:data_metrodisruptionstats}.

\begin{table}[H]
    \tablesize
    \centering
    \begin{tabular}[t]{lcccc}
        \toprule
        Line & Direction & \# of disruptions & \makecell[c]{Average disruption\\length (minutes)} & \makecell[c]{Average \# of\\affected trains}\\
        \midrule
        Green & 1 & 226 & 11.05 & 7.5\\
         & 2 & 224 & 10.96 & 7.3\\
        \addlinespace
        Orange & 1 & 171 & 9.86 & 7.3\\
         & 2 & 179 & 9.51 & 7.1\\
        \bottomrule
    \end{tabular}
    \caption{Summary statistics of reported disruptions on the Green and Orange metro lines.}
    \label{tab:data_metrodisruptionstats}
\end{table}

To accurately model post-disruption travel times, it is necessary to determine the precise onset of the post-disruption period. Due to the aforementioned inaccuracies in the disruption logging procedure, the reported ending timestamps cannot be relied upon exclusively. In practice, a post-disruption period begins only when train operations resume and the first train departs from its station. Accordingly, the departure of the first train following a reported disruption is used to establish the effective ending timestamp of that disruption. This approach reflects the operational procedure followed by the Soci\'et\'e de transport de Montr\'eal (STM) when managing service disruptions.

Our preliminary analysis of train trajectories around disruption periods indicates that distinct operational protocols are enforced depending on the time of day, the duration of the incident, and the underlying cause of the disruption. In some instances, trains are required to stop immediately at the nearest station, whereas in other cases, trains are permitted to continue their journey for several stations before being forced to stop due to an obstructed path ahead. In yet other situations, some trains continue operating during the disruption itself, visiting intermediate stations within the gap left by the preceding train. Such cases, where trains continue to serve gap stations during an ongoing disruption, often lead to the accumulation of multiple trains along the tracks. This phenomenon subsequently increases congestion and results in prolonged travel times in the post-disruption period.

\section{Methodology} \label{sec:methodo}
 Given the unpredictability of disruption durations and the potential for employing different operational scenarios and management protocols during a disruption, the most relevant factor in this context is the instantaneous spatial distribution of trains along the tracks and their relative spacing. When trains are positioned closer to one another, they tend to experience prolonged process times, as they are required either to dwell for longer periods at stations or to reduce their speed in tunnel sections in order to ensure that the track ahead remains clear and that a safe buffer distance is preserved relative to the preceding train. In the absence of explicit knowledge about the precise disruption management protocol, the only reliable assumption is that trains resume movement solely when the downstream path to subsequent stations is confirmed to be free of other trains.

\subsection{Model Formulation}
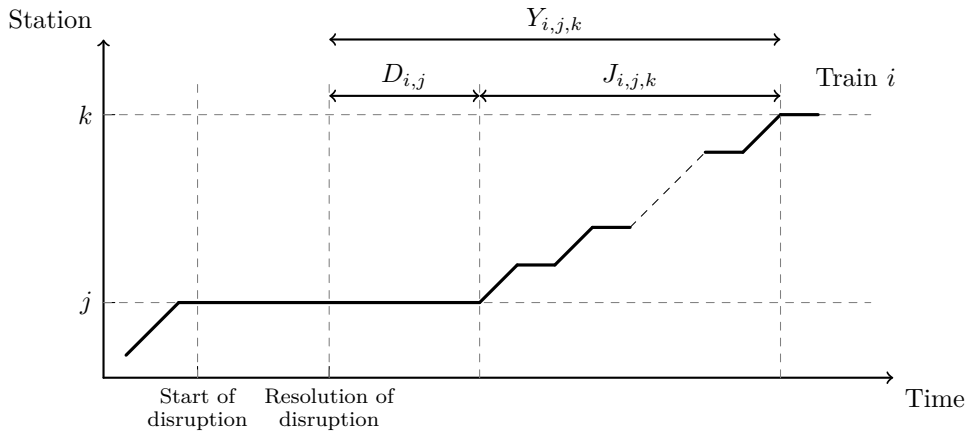
\begin{figure}
    \centering
    \begin{tikzpicture}[line cap=round, line join=round, font=\footnotesize]
      % Axes
      \draw[->, thick] (0,0) -- (10.5,0) node[below right] {Time};
      \draw[->, thick] (0,0) -- (0,4.5) node[above left] {Station};
    
      % Station markers (y-axis)
      \draw (0,1) node[left] {$j$} -- (0.15,1);
      \draw (0,3.5) node[left] {$k$} -- (0.15,3.5);

      % Optional dashed guide lines
      \draw[dashed, gray] (0,1) -- (10.2,1);
      \draw[dashed, gray] (0,3.5) -- (10.2,3.5);
    
      % Departure/arrival times on x-axis
      \draw[very thick] (0.3, 0.3) -- (1, 1);
      \draw[very thick] (1, 1) -- (5, 1);
      \draw[very thick] (5, 1) -- (5.5, 1.5);
      \draw[very thick] (5.5, 1.5) -- (6, 1.5);
      \draw[very thick] (6, 1.5) -- (6.5, 2);
      \draw[very thick] (6.5, 2) -- (7, 2);
      \draw[dashed] (7, 2) -- (8, 3);
      \draw[very thick] (8, 3) -- (8.5, 3);
      \draw[very thick] (8.5, 3) -- (9, 3.5);
      \draw[very thick] (9, 3.5)-- (9.5, 3.5);
      \node at (10, 4) {Train $i$};
      
      \draw (1.25,0) node[below] {\scriptsize\shortstack{Start of\\disruption}} -- (1.25, 0.15);
      \draw (3,0) node[below] {\scriptsize\shortstack{Resolution of\\disruption}} -- (3, 0.15);

      \draw[dashed, gray] (1.25, 0) -- (1.25, 4);
      \draw[dashed, gray] (3, 0) -- (3, 4);
      \draw[dashed, gray] (5, 0) -- (5, 4);
      \draw[dashed, gray] (9, 0) -- (9, 4);

      \draw[<->, thick] (3,3.75) -- (5,3.75);
      \draw[<->, thick] (5,3.75) -- (9,3.75);
      \draw[<->, thick] (3,4.5) -- (9,4.5);

      \node at (4, 4) {$D_{i, j}$};
      \node at (7, 4) {$J_{i, j, k}$};
      \node at (6, 4.75) {$Y_{i, j, k}$};
    
      % \node[rotate=32] at (3.3,2.35) {Train trajectory};
    \end{tikzpicture}
    \caption{Space--time diagram illustrating a representative post-disruption travel time for train~$i$ from station~$j$ to station~$k$. The travel time is decomposed into a delay component, $D_{i,j}$, defined as the interval between the time the disruption is reported as resolved and the departure of train~$i$ from station~$j$, and a journey component, $J_{i,j,k}$, corresponding to the time required for train~$i$ to travel from station~$j$ to its arrival at station~$k$.}
    \label{fig:space-time}
\end{figure}

As discussed previously, our objective is to model the travel time of trains to downstream stations, as these are displayed on screens to passengers waiting on the platform to assist them in making more informed travel decisions. The post-disruption travel time can be conceptually decomposed into two distinct components. The first component corresponds to the \textbf{delay process (D)}, which arises from the accumulation of trains along the tracks following a disruption. The second component represents the time required for a train to traverse all track segments --- including both stations and tunnel sections --- from its origin station to its destination station, which we refer to as the \textbf{journey process (J)}. We denote the post-disruption travel time by $Y_{i,j,k}$, defined as the time taken for train~$i$ to depart from station~$j$ and arrive at station~$k$. Accordingly, the expectation of $Y_{i, j, k}$, denoted by $\mu_{i, j, k}$, can be expressed as:
\begin{equation}
    \mathbb{E}[Y_{i, j, k}] = \mu_{i, j, k} = D_{i, j} + J_{i, j, k}, \label{eq:model0}
\end{equation}
where $D_{i,j}$ denotes the delay experienced by train~$i$ at station~$j$ due to train bunching, and $J_{i,j,k}$ represents the time required for train~$i$ to complete its journey from departure at station~$j$ to arrival at station~$k$. Illustrations of these parameters are provided in the space--time diagram shown in \autoref{fig:space-time}.

For each of these processes, distinct modeling strategies are employed, reflecting the specific characteristics of the underlying dynamics they represent.

\subsubsection{Modeling Journey Process (J)}
The journey process corresponds to the total time a train spends traversing both platforms and tunnel segments when traveling from an origin station to a destination station. In our formulation, the journey is defined as beginning with the departure from the origin station~($j$) and concluding upon arrival at the destination station~($k$). According to our convention, the tunnel section immediately following station~$j$ is referred to as the $j$th tunnel (i.e., the segment between stations $j$ and $j+1$).  

With these definitions, $J_{i,j,k}$ can be decomposed as the sum of the dwell times at intermediate stations $j+1, \dots, k-1$ and the running times through the corresponding tunnel sections. Running times exhibit relatively low variability (right panel of \zcref[S]{fig:process_time_eda}) because train speeds are tightly regulated by operational constraints. The primary source of variability in journey times rather arises from dwell times (left panel of \zcref[S]{fig:process_time_eda}), which are influenced by the passenger boarding and alighting process \citep{Li2016, Cornet2019}. Higher passenger volumes generally lead to longer dwell times; however, the relationship between dwell time and passenger volume is not strictly linear, as it can be affected by numerous other factors \citep{Cornet2019, Li2021}. Among these, the station layout --- specifically, the number and location of platform entrances --- is particularly influential. These structural characteristics shape the spatial distribution of passengers on the platform, which in turn impacts the duration required to complete the passenger exchange process.
\begin{figure}[!htpb]
    \centering
    \includegraphics[width=1\linewidth]{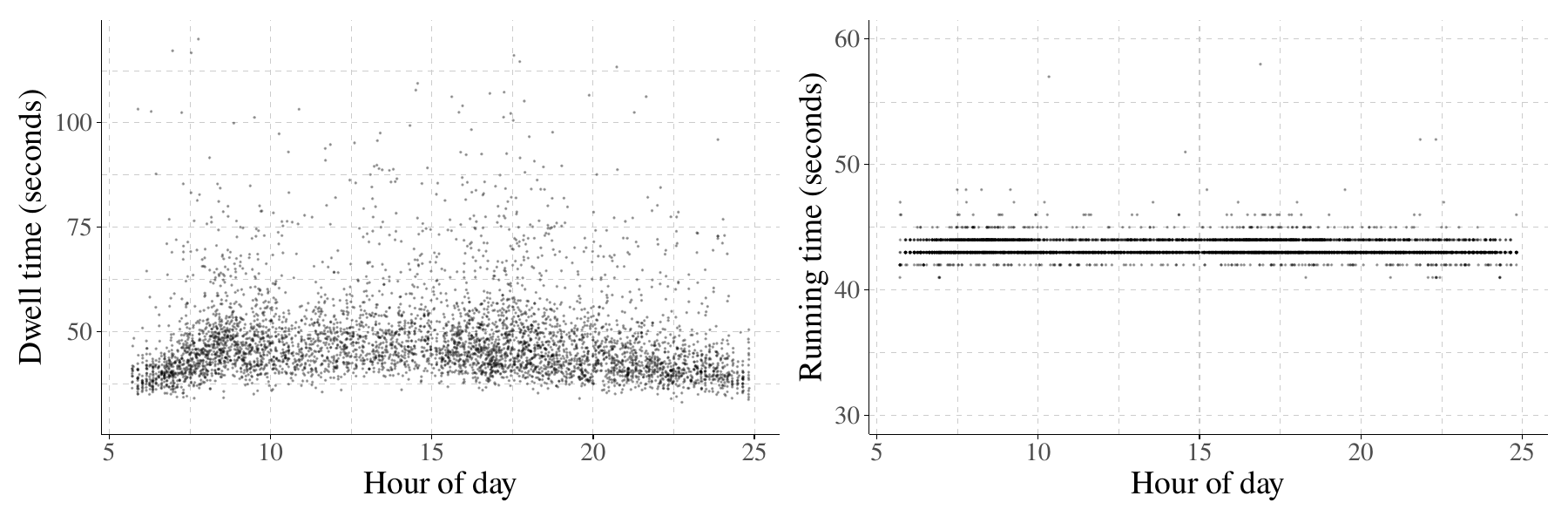}
    \caption{Process times of trains at Guy-Concordia station on the Green metro line based on the hour of the day. Dwell times (left) exhibit substantial temporal variability, with noticeably higher variance during peak periods, while their mean remains relatively stable aside from slight increases at rush hours. The running times of tunnel segment after the station platform (right) remain highly stable throughout the day, showing no dependence on time of day and maintaining nearly constant variability. The x-axis indicates the hour of day, with post-midnight operations displayed by adding 24 to the corresponding hour.
}
    \label{fig:process_time_eda}
\end{figure}

During a network disruption, trains are typically held at stations until the issue is resolved, which leads to the accumulation of passengers at other platforms. Upon the arrival of a train at a given station, the waiting crowd has been accumulating since the departure of the preceding train. This interval is called the \emph{headway}. More formally, we denote the headway $h_{i,j}$ as the time gap between the departure of train~$i-1$ from station~$j$ and the arrival of train~$i$ at the same station.

In the Montr\'eal metro network, train arrivals and departures are not governed by a fixed timetable. Instead, operations are regulated by adjusting the frequency at which trains are dispatched along a metro line. Consequently, during peak hours with high passenger demand, headways are relatively short, whereas during off-peak periods, longer time gaps are observed between consecutive trains (left panel of \zcref[S]{fig:headway_eda}). Headways are managed to prevent excessive crowding at any platform. However, during a disruption, the balance of headways is disrupted: certain stations may remain unvisited for the duration of the incident, leading to passenger accumulation. These longer-than-usual headways result in extended passenger exchange process, which in turn increase the dwell time of the train arriving at the affected station once the disruption has been resolved.  
\begin{figure}[!htpb]
    \centering
    \includegraphics[width=\linewidth]{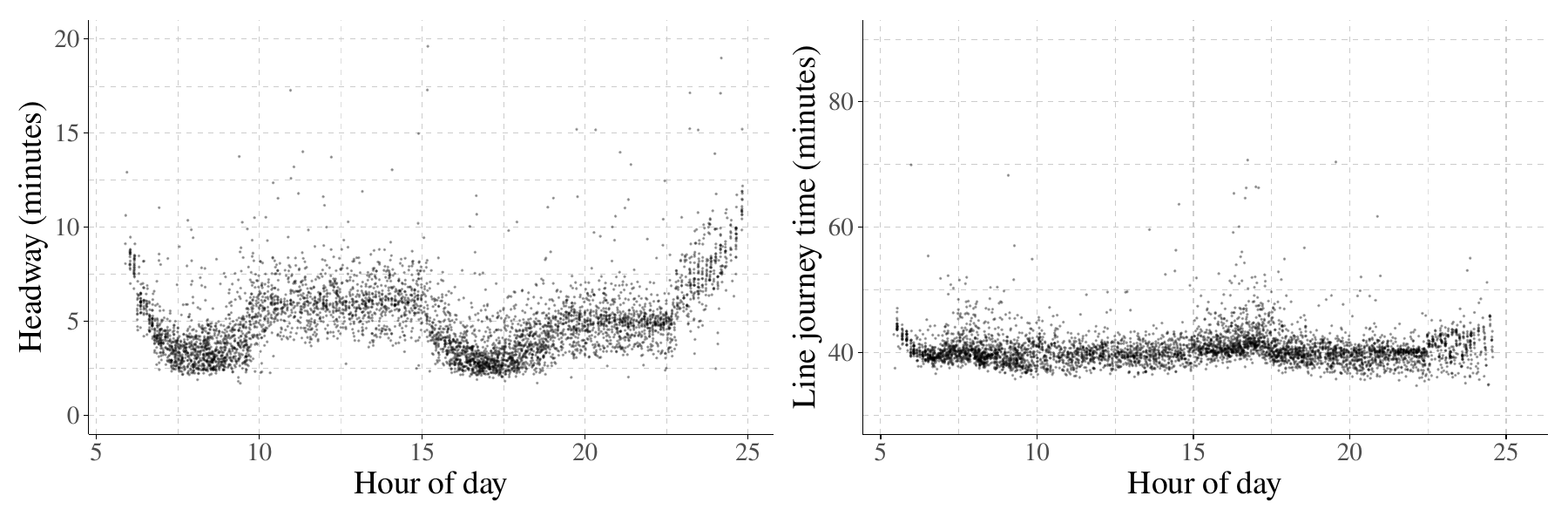}
    \caption{Headways at Guy-Concordia station (left) and full Green line journey times (right) as a function of the hour of day. Headways show a pronounced temporal pattern corresponding to operational adjustments between peak and off-peak periods. Line journey times exhibit higher mean and variance during peak hours, yet remain relatively stable over the rest of the day. The x-axis denotes the hour of day, with post-midnight operations represented by adding 24 to the corresponding hour.
}
    \label{fig:headway_eda}
\end{figure}

Based on these considerations, the following model is proposed for the journey process during post-disruption periods:
\begin{equation}
    J_{i, j, k} = t_0 + \tilde{T}_{i, j, k} + \sum_{m = j + 1}^{k - 1} \theta_m (h_{i, m} - \tilde{h}_{i, m}), \label{eq:journey}
\end{equation}
where $\tilde{T}_{i,j,k}$ denotes the median travel time from station $j$ to station $k$ under normal operating conditions, and $\tilde{h}_{i,m}$ denotes the median headway at station $m$ under normal operations evaluated at the time train $i$ arrives at $m$. Both travel time and headway are inherently time-dependent (\zcref[S]{fig:headway_eda}). To account for this temporal variability, median values are computed using a 30-minute rolling time window throughout the day. Importantly, the timestamp of train $i$’s arrival at each station determines the specific time window from which the corresponding median headway and travel time values are extracted.

An intercept term, $t_0$, is included in this formulation to account for any global offset or systematic discrepancy between post-disruption journey times and those observed under normal operating conditions.

The third term on the right-hand side of eqn.~\ref{eq:journey} represents the sum of the added journey time experienced by train~$i$ at all intermediate stations $m$ between the origin station~$j$ and the destination station~$k$. Prolonged headway due to disruption leads to more waiting passengers at the platform. The higher the headway imbalance $h_{i, m} - \tilde{h}_{i, m}$ is, the more passengers will arrive at the station platform. The underlying modeling assumption here is that there is a linear relationship between the excess duration for which passengers have been accumulating at a platform and the time required for train~$i$ to pass the platform of station~$m$ during its journey \citep{Li2021}. For every additional minute of prolonged headway at an intermediary station $m$, the journey process increases by $\theta_m$.

\subsubsection{Modeling Delay Process (D)}
When the system resumes its operations after an incident, trains begin to move again; however, due to the disruption, train bunching may occur along the tracks. Consequently, some trains experience additional delays before they can continue their journeys. These delays, in turn, increase the estimated passenger waiting times for train arrivals at the downstream stations.

Our modeling strategy assumes that train operations are influenced by the presence of other trains ahead of them in the direction of travel. We define segment $j$ as the combined track section consisting of station $j$ and its subsequent tunnel. Since the Montr\'eal metro system aims to maintain a balanced headway between consecutive trains, it introduces controlled delays to reestablish uniform spacing during the post-disruption period. For a specific train, we model the delay process by assuming that the presence of another train in several segments ahead exerts a constant effect on the delay of the current train. This effect is both distance-dependent and location-dependent, meaning that the influence of leading trains on delay varies depending on the specific station where the train was stopped during the incident.

In our model, the delay experienced by train $i$ at station $j$ is formulated as a linear combination of the influence exerted by other trains located up to $D_{\max}$ segments ahead along the line:
\begin{equation}
    D_{i, j} = \sum_{\ell = 1}^{D_{\max}} \gamma_{\ell, j} z_{i, j, \ell}, \label{eq:delay}
\end{equation}
where $z_{i, j, \ell}$ is a binary variable that takes the value $1$ if, when train $i$ is at station $j$, another train is located $\ell$ segments ahead, and $0$ otherwise.

\subsubsection{Variance Specification and Travel Time Distribution}
Travel times consist of the sum of running times and dwell times, so it is reasonable to hypothesize that its variance increases with the traveled distance. \zcref[S]{fig:eda_sd_1} presents the empirical variance of travel times for all train trajectories in 2018 as a function of the traveled distance, measured by the number of stations. The data exhibit a clear trend which is nearly linear, supporting the assumption that travel-time variance accumulates proportionally with distance and that a linear functional form in \(k - j\) for the variance specification in our model is a plausible choice. Additional empirical results supporting this pattern are reported in \ref{appx:eda_var}.

\begin{figure}[H]
  \centering
  \includegraphics[width=0.5\linewidth]{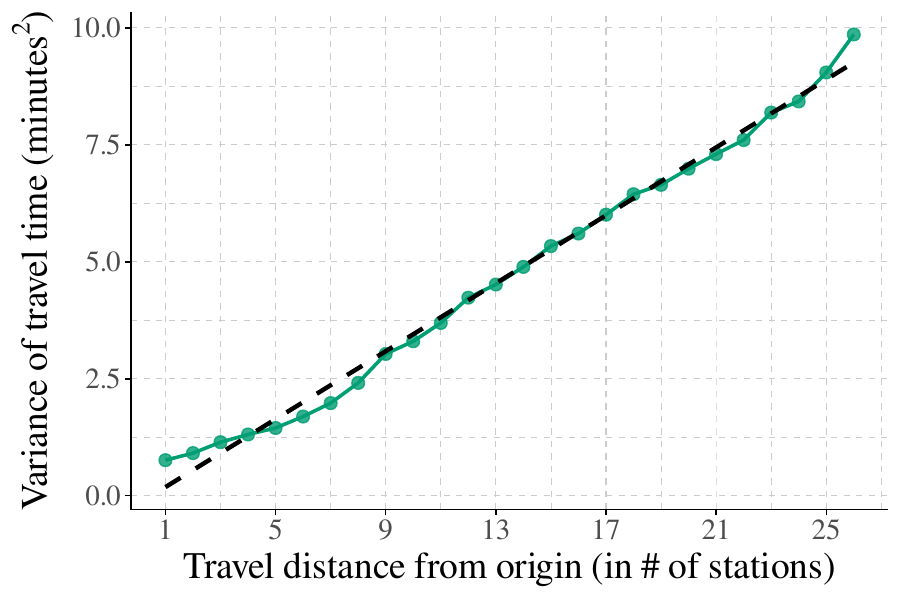}
  \caption{Empirical variance of journey times for all train trajectories in 2018 on the Green line of the Montr\'eal metro, shown for direction~1. The horizontal axis represents the traveled distance from the origin station, measured by the number of stations traveled, while the vertical axis indicates the variance of the corresponding journey times.}
  \label{fig:eda_sd_1}
\end{figure}

While travel times can, in theory, become arbitrarily long due to factors such as disruptions and congestion, trains cannot complete their journeys faster than a certain lower bound. This limitation arises from operational constraints, including tunnel speed limits and the minimum dwell time required for braking, acceleration, and door operations when entering or leaving stations. Consequently, we expect $Y_{i, j, k}$ to be asymmetric around its mean $\mu_{i, j, k}$ and to exhibit a skewed distribution.

\subsection*{Baseline model}
To gain a clearer understanding of the behavior of the model error terms, we construct a baseline model. In this baseline specification, the error terms are assumed to follow a normal distribution with a linearly increasing standard deviation, capturing the heterogeneity present in the data (\zcref[S]{fig:eda_sd_1}). By combining eqns~\ref{eq:model0}, \ref{eq:journey}, and \ref{eq:delay}, we obtain the following baseline formulation for post-disruption travel times:
\begin{align}
    Y_{i, j, k} &= \mu_{i,j, k} + \varepsilon_{i, j, k} \label{eq:baseline_model} \\  
    \mu_{i, j, k }&=  t_0 + \tilde{T}_{i, j, k} + \sum_{m = j + 1}^{k - 1} \theta_m (h_{i, m} - \tilde{h}_{i, m}) + \sum_{\ell = 2}^{D_{\max}} \gamma_{\ell, j} z_{i, j, \ell} \label{eq:mu} \\
    \varepsilon_{i, j, k} &\sim \mathcal{N}\{0, \omega_0 + \omega_1 (k - j)\}, \label{eq:err_baseline_model}
\end{align}
where $\varepsilon_{i, j, k}$ represents the baseline model error terms, assumed to follow a normal distribution $\mathcal{N}(\mu, \sigma^2)$ with mean $\mu$ and standard deviation $\sigma$. 

\zcref[S]{fig:baseline_error_1} displays the model residuals against the predicted values, along with the empirical skewness of the error terms as a function of traveled distance for Direction~1 of the Green line. The plot of residual vs predicted values reveals a slight misfit particularly for longer trips, indicating that the baseline model does not adequately capture  travel times over longer distances. This asymmetry implies that the normality assumption for the error distribution is insufficient, as the distribution becomes increasingly tilted. The residual skewness generally decreases with travel distance, although the pattern is not entirely consistent. A plausible explanation for the decreasing skewness is that delays and congestion tend to prolong travel times, while operational constraints limit the occurrence of unusually short travel times. At the same time, minor speed adjustments and shorter dwell processes can be implemented to partially mitigate the impact of delays. Consequently, over longer journeys, trains have greater opportunity to absorb initial delays, leading to a reduction in skewness as traveled distance increases. A reasonable adjustment is therefore to allow the skewness parameter to vary linearly with traveled distance. 

In \zcref[S]{fig:baseline_error_1}, skewness is shown only for traveled distances up to 20 stations. This restriction is due to operational procedures that typically prevent the insertion of new trains along the line during a disruption, resulting in very few post-disruption observations with traveled distances exceeding 20 stations. Since empirical skewness estimates are unreliable with very small samples, as they become overly sensitive to outliers and model misfit, we truncate the maximum distance considered. Additional empirical results for the remaining line and direction combinations are presented in \ref{appx:baseline_residuals}.

The presence of numerous extreme residual observations suggests that the normality assumption is inadequate, indicating the need to adopt a distribution with heavier tails. In the following sections we propose two models based on skew-normal and skew-student-\emph{t} (skew-\emph{t}) distributions.

\begin{figure}[t]
  \centering
  \includegraphics[width=1\linewidth]{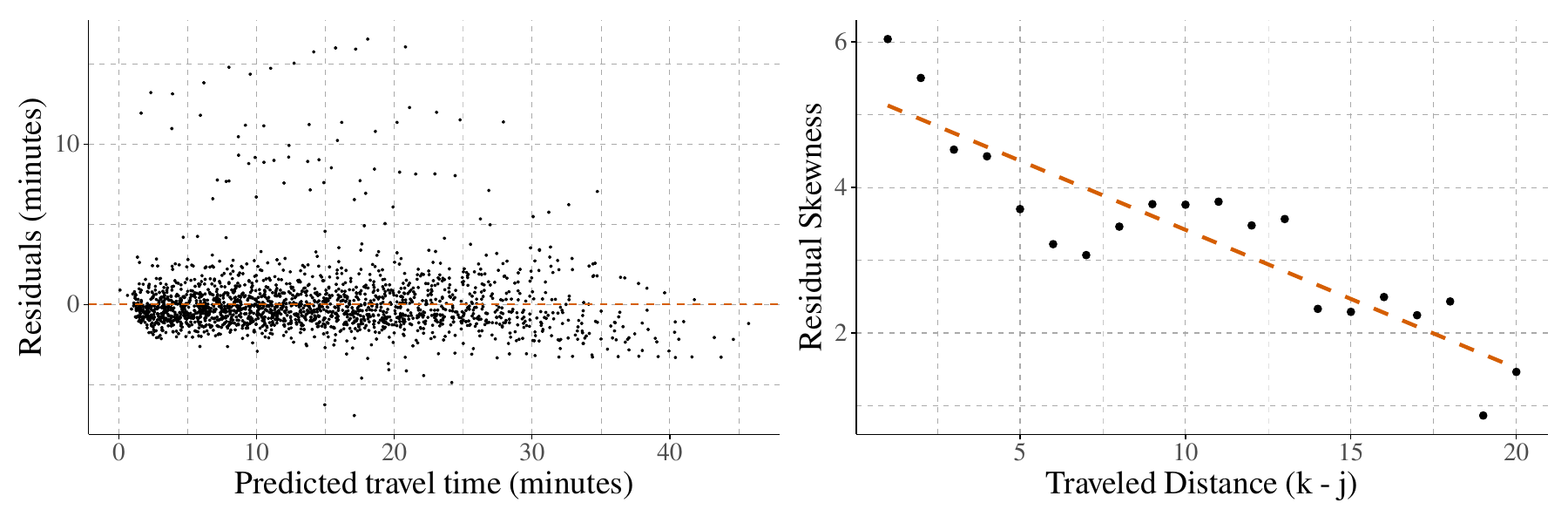}
  \caption{Baseline model residuals plotted against predicted travel times (left) and empirical skewness of the residuals (right) for Direction~1 of Line~1. The results indicate non-constant heterogeneity at larger predicted travel times and pronounced positive skewness in the residual distributions, with skewness generally decreasing as traveled distance increases.}
  \label{fig:baseline_error_1}
\end{figure}

\subsubsection{Skew-Normal Distribution}
% To more accurately characterize post-disruption travel times, we adopt a skew-normal distribution for the error term. 

The skew-normal distribution is a continuous probability distribution that extends the normal distribution by allowing for non-zero skewness. Among the various existing parametrizations, we adopt that of \citet[][Section~2.1]{Azzalini_2013}. The skew-normal distribution has density
\begin{align}
   f_{Y}(y) = \frac{2}{\omega}\phi\!\left(\frac{y - \xi}{\omega}\right) \Phi\!\left(\alpha \frac{y - \xi}{\omega}\right),
    \label{eq:sn}
\end{align}
where $\xi \in \mathbb{R}$ denotes the location parameter, $\omega \in \mathbb{R}^+$ the scale parameter, and $\alpha \in \mathbb{R}$ the skewness parameter, and where $\phi(x)$ and  $\Phi(x) = \int_{-\infty}^x \phi(t)\mathrm{d} t$ denote the standard normal density and distribution functions, respectively.

The skew-normal distribution is a location–scale family, and we may write $Y \sim \mathcal{SN}(\xi, \omega, \alpha)\equiv \xi + \omega Z$ with $Z \sim\mathcal{SN}(0, 1, \alpha)$; its expectation and variance are
\begin{align}
    \mathbb{E}(Y) &= \xi + \omega \alpha\sqrt{\frac{2}{\pi}(1 + \alpha^2)}, \label{eq:sn_mean}\\
    \mathrm{Var}(Y) &= \omega^2 \left\{1 - \frac{2\alpha^2}{\pi(1 + \alpha^2)}\right\}. \label{eq:sn_var}
\end{align}

\subsubsection{\texorpdfstring{Skew-$t$ Distribution}{Skew-t Distribution}}
The skew-normal distribution can be generalized to accommodate heavy-tailed data. Indeed, a random scale mixture of standard skew-normal with $Z=Z_0/\sqrt{V}$, where $Z_0 \sim \mathcal{SN}(0, 1, \alpha)$ and $V \sim \chi^2_\nu$, yields the skewed-$t$ distribution \citep[][Section~4.3]{Azzalini_2013}, which can be made into a location-scale family by taking $Y = \xi + \omega Z$. The skew-$t$ $\mathcal{ST}(\xi, \omega, \alpha, \nu)$ density is
\begin{align}    f(y) =& \frac{2}{\omega} t\left(z; \nu\right) T\left(\alpha z \sqrt{\frac{\nu + 1}{\nu + z^2}}; \nu + 1\right), \qquad z = \frac{y-\xi}{\omega};
\end{align}
where $\nu > 0$, $\alpha \in \mathbb{R}$, and $t(\cdot; \nu)$ and $T(\cdot; \nu)$ are the density and distribution functions of a standard Student-$t$ distribution with $\nu$ degrees of freedom, respectively. The expectation and variance of $Y \sim \mathcal{ST}(\xi, \omega, \alpha, \nu)$ are
\begin{align}
\mathbb{E}(Y)
    &= \xi 
       + \omega\, 
          \frac{\alpha\sqrt{\nu}}{\sqrt{\pi(1 + \alpha^{2})}}\frac{\Gamma\left\{(\nu - 1)/2\right\}}{\Gamma\left(\nu/2\right)},
       && \nu > 1; \label{eq:st_mean} \\[6pt]
\mathrm{Var}(Y) 
    &= \omega^{2}\!\left(
         \frac{\nu}{\nu - 2}
         - \frac{2\alpha^{2}\nu}{\pi(1 + \alpha^{2})}
           \left[
             \frac{\Gamma\left\{(\nu - 1)/2\right\}}
                  {\Gamma\left(\nu/2\right)}
           \right]^{2}
       \right),
       && \nu > 2. \label{eq:st_var}
\end{align}

\subsubsection{Travel Time Error Dependence}
\label{subsec:err_dep}
As post-disruption trains proceed toward their downstream stations, they may encounter congestion or delays at intermediate stops, resulting in longer travel times than those predicted by the model. During this recovery phase, congestion effects can propagate from one train to the next, leading subsequent trains to experience similar delays. Examination of the error terms in our baseline model (eqn.~\ref{eq:baseline_model}) for consecutive post-disruption trains reveals a positive correlation between the baseline model residuals of successive trains (\zcref[S]{fig:error_dep_1}). The additional empirical results for the error dependence across different direction and line is presented in \ref{appx:err_dep}.

\begin{figure}[t]
  \centering
  \includegraphics[width=0.5\linewidth]{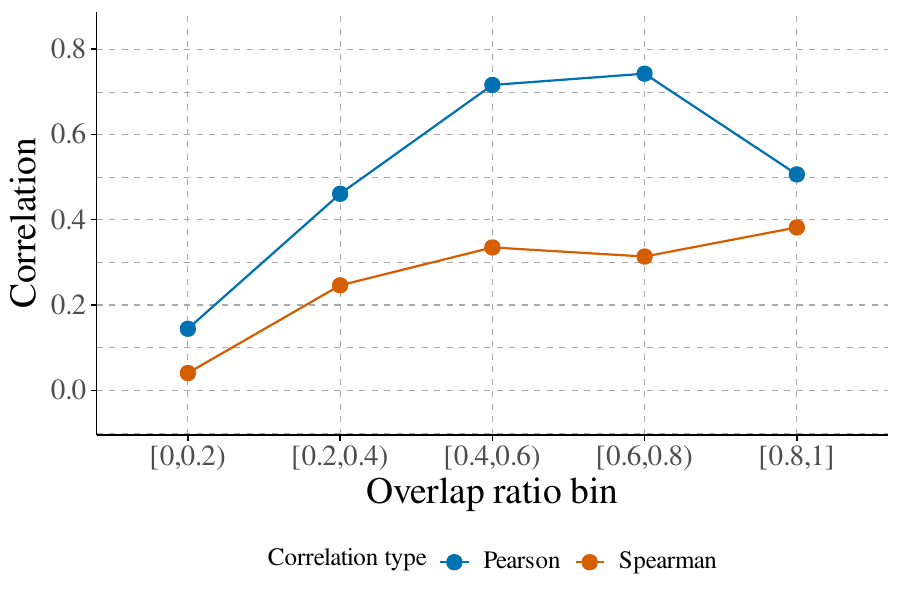}
  \caption{Pearson and Spearman correlations between the baseline model residuals and those of the preceding trains, binned by the proportion of journey overlap, for the Green line of the Montr\'eal metro system. The results reveal positive correlations, indicating that delay and congestion propagate from one train to its successor. Train journeys with a larger overlap in traveled segments with the preceding train exhibit stronger positive correlations.}
  \label{fig:error_dep_1}
\end{figure}

When a disruption is declared resolved, trains begin moving from different stations along the metro line. For a given downstream destination station $k$, all post-disruption trains that were stopped at upstream stations before $k$ during the disruption will eventually arrive at $k$. Except for the first post-disruption train reaching station $k$, all subsequent trains traverse segments that have already been traveled by another post-disruption train; thus, their journeys from origin to destination partially overlap with those of preceding trains.  

For instance, consider train $i$ stopped at station $j$ during the disruption and train $i-1$ stopped at station $j'$. If we wish to display the predicted waiting times for trains $i-1$ and $i$ on the platform screens to passengers waiting at a downstream station $k$ ($j < j' < k$), we must estimate $y_{i-1, j', k}$ and $y_{i, j, k}$. The track segments between $j'$ and $k$ are common to both trains, so both trains will pass through the same sections on their way to $k$. Since train $i-1$ precedes train $i$, any congestion encountered by train $i-1$ along these shared segments will appear in its error term, $\varepsilon_{i-1, j', k}$, and may influence the travel time of train $i$, introducing correlation between their journeys.  

\zcref[S]{fig:error_dep_1} shows the Pearson and Spearman correlation between these consecutive errors in our baseline model. Increasing pattern of both correlation coefficients with journey overlap ratio indicates that propagation of delays is both monotonic and approximately linear, and becomes more pronounced as overlap ratio increases. The magnitude of this effect depends on the proportion of shared segments between the two journeys, expressed as $(k - j') / (k - j)$  which we refer to as overlap ratio. The closer the positions of trains $i-1$ and $i$ during the disruption, the greater the overlap ratio among their paths toward station $k$, and therefore, any congestion affecting train $i-1$ is more likely to have a pronounced impact on the travel time of train $i$.

We extend our baseline model in eqn.~\ref{eq:baseline_model} to incorporate the error term of the preceding train. There exists a one-to-one correspondence between each train and its stopping location during the incident and that of its preceding train, denoted by $(i, j) \rightarrow (i - 1, j')$, where $j < j'$. For any destination station $k$ located downstream of both $j$ and $j'$, the following mean specification is proposed:
\begin{equation}
    y_{i, j, k} = \mu_{i, j, k}' + \rho_{j, j', k} \varepsilon_{i-1, j', k} + \varepsilon_{i, j, k}
    \label{eq:mean_spec_final}
\end{equation}
where $\rho_{j,j',k}$ quantifies the correlation between the residual of a given train and that of its preceding train. Based on the observed increase in correlations with the journey-overlap ratio, $\rho_{j,j',k}$ in eqn.~\ref{eq:mean_spec_final} should be an increasing function of the overlap ratio. Although various functional forms could be considered, we adopt an exponential specification for the relationship between $\rho_{j,j',k}$ and the overlap ratio:
\begin{equation}
    \rho_{j, j', k} = \rho \left\{1 - \exp\!\left(-\lambda \frac{k - j'}{k - j}\right)\right\}
    \label{eq:rho}
\end{equation}
where $\rho \ge 0$ and $\lambda >0$ are model parameters.

\subsubsection{Model Hierarchy}
\label{section:model_hier}
We integrate our baseline model in eqn.~\ref{eq:baseline_model} with skewed error terms, and incorporate the error correlation term defined in eqn.~\ref{eq:rho} to model the travel times of post-disruption trains traveling toward downstream stations. From eqns~\ref{eq:sn_mean}, \ref{eq:sn_var}, ~\ref{eq:st_mean}, and \ref{eq:st_var}, it is evident that the mean–variance relationship in the skew-normal and skew-\emph{t} distributions differs from that of the normal and Student-$t$ distributions, and to ensure that the expectation of $Y_{i, j, k}$ equals $\mu_{i, j, k}$, an adjustment term in eqn.~\ref{eq:baseline_model} must be incorporated. 

We assume a linearly changing specification for the skewness parameter, denoted by $\alpha_0 + \alpha_1 (k - j)$. For the scale parameter, our modeling strategy differs from that of the baseline model. In the baseline model of eqn.~\ref{eq:baseline_model}, the motivation for a linearly increasing variance specification stems from the additive structure of travel times: total travel time is the sum of running and dwell times between the origin and destination stations. Under the assumption that these running and dwell processes are approximately independent across segments, the overall variance of travel time can be expressed as the sum of segment-level variances, implying a linear dependence on the traveled distance $k - j$.

For the skew-normal and skew-$t$ distributions, the variance in eqns~\ref{eq:sn_var} and~\ref{eq:st_var} takes a more complex form, as it depends not only on the scale parameter $\omega$, but also on the skewness parameter and, in the case of the skew-$t$ distribution, the degrees of freedom. Nevertheless, in both distributions the variance can be written as $\omega^2$ multiplied by a function of the remaining parameters. To preserve the same intuitive interpretation of variance accumulation with distance while accounting for the distribution-specific variance structure, we therefore model the squared scale parameter as a linear function of the traveled distance, $\omega_{kj}^2 = \omega_0 + \omega_1 (k - j)$. This specification allows the overall variance to increase approximately linearly with distance, while flexibly accommodating the effects of skewness and tail heaviness inherent to the skew-normal and skew-$t$ distributions.

A distinction must also be made between the first post-disruption train that reaches a destination station $k$ and all subsequent post-disruption trains. The first post-disruption train is the initial train to arrive at station $k$ following the resolution of an incident. Because it has no preceding train with which it shares any portion of its journey, no lagged residual term is included; its travel time is therefore modeled using the mean specification in eqn.~\ref{eq:baseline_model}. In contrast, the travel times of all subsequent post-disruption trains arriving at station $k$ are modeled according to eqn.~\ref{eq:mean_spec_final}. We therefore express $y_{i,j,k}$ as follows:
\begin{align}
    Y_{i, j, k} &= 
    \begin{cases}
        \mu_{i, j, k}' + \varepsilon_{i, j, k}, & \text{if $i$ is the first post-disruption train} \\[4pt]
        \mu_{i, j, k}' + \rho_{j, j', k}\,\varepsilon_{i-1, j', k} + \varepsilon_{i, j, k}, & \text{otherwise}
    \end{cases}
    \label{eq:y}
\end{align}
where, under the proposed skew-normal specification,
\begin{align}
    \mu_{i, j, k}' &= \mu_{i, j, k} 
    -  \sqrt{\frac{2}{\pi}} \,
       \frac{\{\alpha_0 + \alpha_1(k - j)\}\sqrt{\omega_0 + \omega_1(k - j)}}
            {\sqrt{1 + (\alpha_0 + \alpha_1(k - j))^{2}}} \label{eq:sn_mu}\\
    \varepsilon_{i, j, k} 
        &\sim \mathcal{SN}\!\left\{0,\; \sqrt{\omega_0 + \omega_1(k - j)},\; \alpha_0 + \alpha_1(k - j)\right\} \label{eq:sn_innov}
\end{align}
and under the proposed skew-\(t\) specification,
\begin{align}
    \mu_{i, j, k}' &= \mu_{i, j, k} 
    - \frac{\sqrt{\nu}\,\Gamma\left\{(\nu - 1)/2\right\}}
           {\sqrt{\pi}\,\Gamma\left(\nu/2\right)}
      \frac{\{\alpha_0 + \alpha_1(k - j)\}\sqrt{\omega_0 + \omega_1(k - j)}}
           {\sqrt{1 + \{\alpha_0 + \alpha_1(k - j)\}^{2}}} \label{eq:st_mu}\\
    \varepsilon_{i, j, k} 
        &\sim \mathcal{ST}\!\left\{0,\; \sqrt{\omega_0 + \omega_1(k - j)},\; \alpha_0 + \alpha_1(k - j),\; \nu\right\} \label{eq:st_innov}
\end{align}
where, in both models, $\mu_{i, j, k}$ follows eqn.~\ref{eq:mu}, and $\rho_{j, j', k}$ follows eqn.~\ref{eq:rho}. The full set of model parameters includes $\theta_m$, $\gamma_{\ell, j}$, $t_0$, $\omega_0$, $\omega_1$, $\alpha_0$, $\alpha_1$, $\rho$, and $\lambda$ for both models, with the skew-\(t\) model containing one additional degree-of-freedom parameter $\nu$.

\subsection{Predictive distribution with dependent errors}
Recall that the post-disruption travel times are modeled using a moving-average structure across consecutive trains, conditional on spatial overlap, given by
\begin{equation}
Y_{i,j,k} = \mu_{i,j,k} + \rho_{j,k}\,\varepsilon_{i-1,j',k} + \varepsilon_{i,j,k},
\end{equation}
where $\varepsilon_{i,j,k}$ denotes an independent innovation term following either a skew-normal or a skew-$t$ distribution with scale $\omega_{kj}=\{\omega_0 + \omega_1 (k - j)\}^{1/2}$, skewness $\alpha_{kj} =\alpha_0 + \alpha_1 (k - j)$, and, in the case of skew-$t$ innovations, degrees of freedom $\nu$.

Conditioning on the innovation of the preceding train, the predictive distribution of the realization $y_{i,j,k}$ takes the form of a location-shifted distribution,
\begin{equation}
y_{i,j,k} \mid \varepsilon_{i-1,j',k} \sim \mathcal{SN}\!\left(\mu'_{i,j,k} + \rho_{j,k}\varepsilon_{i-1,j',k}, \omega_{kj}, \alpha_{kj} \right),
\end{equation}
for skew-normal innovations, where $\mu'_{i,j,k}$ is defined in eqn.~\ref{eq:sn_mu}, and
\begin{equation}
y_{i,j,k} \mid \varepsilon_{i-1,j',k} \sim \mathcal{ST}\!\left(\mu'_{i,j,k} + \rho_{j,k}\varepsilon_{i-1,j',k}, \omega_{kj}, \alpha_{kj}, \nu \right),
\end{equation}
for skew-$t$ innovations, where $\mu'_{i,j,k}$ follows from eqn.~\ref{eq:st_mu}.

The unconditional predictive distribution is obtained by integrating out the latent innovation of the preceding train,
\begin{equation}
p(y_{i,j,k})
= \int p\!\left( y_{i,j,k} \mid \varepsilon_{i-1,j',k} \right) \,
p\!\left( \varepsilon_{i-1,j',k} \right) \, \mathrm{d}\varepsilon_{i-1,j',k},
\end{equation}
which results in a location mixture of skew-$t$ distributions and generally does not admit a closed-form expression. In practice, posterior predictive inference is carried out via simulation by jointly sampling the model parameters and latent error terms, thereby fully capturing both heavy-tailed behavior and dependence across trains.

\subsection{Prior specification}
We adopt weakly informative priors for all model parameters to regularize estimation while remaining agnostic about their precise values. Location, intercept, and regression parameters are assigned zero-centered Gaussian priors, reflecting prior symmetry and moderate uncertainty. Specifically, the baseline travel-time intercept $t_0$ and regression coefficient $\theta_m$, for $m = 2, \dots, J -1$ where $J$ is the number of stations in each case study, are assigned standard normal priors, $t_0, \theta_m \sim \mathcal{N}(0, 1)$.
The covariate effect parameters $\gamma_{\ell, j}$ are assigned independent zero-centered Gaussian priors with relatively large variance, $\gamma_{\ell, j} \sim \mathcal{N}(0, 5)$,
which provides sufficient flexibility in their magnitudes while maintaining mild regularization. Here, $\ell = 1, \ldots, D_{\max}$ and $j = 1, \ldots, J$ index the distance and stations, respectively.

The scale parameters governing variance accumulation are constrained to be positive and are assigned weakly informative Gaussian priors truncated above zero, namely $\omega_0 \sim \mathcal{N}_{+}(1,1)$ and $\omega_1 \sim \mathcal{N}_{+}(1,1)$, which ensures positivity of the squared scale parameter over the observed range of traveled distances. It is important to note that although this constraint enforces a monotonic increase in the squared scale of both the skew-normal and skew-$t$ distributions, it does not necessarily imply monotonic behavior of the variance itself. As shown in eqns~\ref{eq:sn_var} and~\ref{eq:st_var}, the variance also depends on the skewness parameter, which, depending on the chosen skewness specification, may induce a variety of functional forms with respect to the distance traveled.

For the skewness specification, the intercept and slope parameters are assigned standard normal priors $\alpha_0, \alpha_1 \sim \mathcal{N}(0, 1)$, which encourages moderate skewness while allowing the data to determine both its magnitude and direction. 

The correlation parameter $\rho \in (-1,1)$ is obtained through a logistic transformation of an unconstrained parameter,
\begin{equation}
\rho = 2\,\text{logit}^{-1}(\rho_{\text{raw}}) - 1, \qquad
\rho_{\text{raw}} \sim \mathcal{N}(0, 1), \nonumber    
\end{equation}
where $\text{logit}^{-1}(x) = (1 + e^{-x})^{-1}$, implying a prior that is approximately uniform over the admissible correlation range. The additional regression coefficient $\lambda$ is assigned a standard normal prior $\lambda \sim \mathcal{N}(0, 1)$. Finally, the degrees of freedom $\nu$ of the skew-$t$ distribution is assigned a Gamma prior with a shape--rate parameterization, $\nu \sim \text{Gamma}(2, 0.1)$. This prior favors moderately heavy-tailed distributions while retaining sufficient probability mass over larger values of $\nu$, thereby allowing the model to approximate the Gaussian case when supported by the data.

\subsection{Model Estimation and Computation}
We fitted the proposed model described by eqns~\ref{eq:y}, \ref{eq:sn_mu}, and \ref{eq:st_mu} to both the Green and Orange lines of the Montr\'eal metro network data. Posterior inference for all model parameters was carried out using Hamiltonian Monte Carlo \citep[cf.][]{Neal:2011,Betancourt.Girolami:2015} as implemented in \texttt{Stan} through \textsf{R} \citep{Stan,cmdstanr}. The full set of parameters drawn from the joint posterior via Markov chain Monte Carlo includes the mean specification parameters $(t_0, \theta_m, \gamma_{\ell, j})$, the scale parameters $(\omega_0, \omega_1)$, the skewness parameters $(\alpha_0, \alpha_1)$, the correlation parameter ($\rho$, $\lambda$), and, for the skew-$t$ specification, the degrees of freedom $\nu$. 

Each model was fitted using four independent chains, with 1{,}000 warm-up iterations followed by 4{,}000 sampling iterations per chain. Convergence was assessed using visual inspection of trace plots, which are reported in \ref{appx:estimation}. The lowest reported effective sample size (ESS) \citep[cf.][]{Geyer:2011}, which accounts for autocorrelation, exceeded 10{,}500 for skew-normal model parameters and 7{,}000 for skew-$t$ model, indicating satisfactory mixing of the Markov chains.

\section{Results} \label{sec:results}

To construct the dataset, the positional encoding was defined using a maximum look-ahead distance of five segments ($D_{max} = 5$). The analysis focused exclusively on the trajectories of trains that were in operation at the time of a disruption, as identified from the disruption logs, and their travel times to downstream stations. Each travel time to a downstream station was treated as an individual observations in the dataset. As previously noted, only weekday operations were considered, with public holidays excluded according to the official Soci\'et\'e de transport de Montr\'eal (STM) holiday calendar.  

Each metro line and both travel directions were modeled separately, resulting in distinct datasets for each case. To evaluate model performance, each dataset was divided into in-sample and out-of-sample subsets based on the disruption log identifiers. To minimize correlations between training and testing samples, the data were split by disruption event rather than by individual observation. For each dataset, $90\%$ of the disruptions were used for model fitting, while the remaining $10\%$ were reserved for out-of-sample validation. \zcref[S]{tab:dataset_size} shows the number of observations in our in-sample and out-of-sample subsets in each case study. 

\begin{table}[!ht]
    \centering
    \tablesize
    % \resizebox{\ifdim\width>\linewidth\linewidth\else\width\fi}{!}{
    \begin{tabular}[t]{lcc|cc}
        \toprule
        \multicolumn{1}{c}{ } & \multicolumn{2}{c}{Green line (line 1)} & \multicolumn{2}{c}{Orange line (line 2)} \\
        \cmidrule(l{3pt}r{3pt}){2-3} \cmidrule(l{3pt}r{3pt}){4-5}
        Sample & Direction 1 & Direction 2 & Direction 1 & Direction 2\\
        \midrule
        In-sample & 18,942 & 17,362 & 12,439 & 12,618\\
        Out-of-sample & 2,043 & 2,092 & 1,637 & 1,457\\
        \bottomrule
    \end{tabular}
    % }
    \caption{Number of in-sample and out-of-sample travel time observations}
    \label{tab:dataset_size}
\end{table}

\subsection{\texorpdfstring{The effect of longer-than-usual headway, $\theta_m$}{The effect of longer-than-usual headway}}
\zcref[S]{tab:theta_11} present the estimated $\theta_m$ parameters in skew-normal and skew-\(t\) models on the Green line of the Montr\'eal metro system in Direction~1. It is important to note that these parameters represent the effect of longer-than-usual headway on the travel time of trains passing through intermediate stations. Therefore, the parameters are not defined for the first and last stations along each metro line. According to eqn.~\ref{eq:mu}, the parameters $\theta_m$ capture the effects associated with the term $h_{i,m} - \tilde{h}_{i, m}$. This quantity represents the deviation of the realized headway during the post-disruption period at station $m$ from the corresponding headway under normal operating conditions. It reflects the excess time gap between successive trains during the post-disruption period relative to typical operations and thus directly relates to the additional time during which passengers accumulate on station platforms compared to normal conditions. It is important to note, however, that despite being referred to as a ``more-than-usual'' gap, this quantity may occasionally take negative values, indicating that the realized post-disruption headway is shorter than the headway observed during normal operations. The $\theta_m$ values represent the additive effect on travel time (in seconds) for every additional minute of headway caused by a disruption when a train passes a specific station during the post-disruption period. For instance, a posterior mean of $\theta_{\text{Monk}} = 2.4$ for the skew-normal model implies that trains passing Monk station during post-disruption periods experience an additional $2.4$ seconds of travel time for every additional minute that the headway at Monk station was extended due to a disruption. Although these effects may appear small in isolation, they can accumulate over longer journeys following major disruptions, leading to a substantial increase in total travel time.  

An alternative perspective is to compare these effects relative to the normal operational conditions, particularly in terms of dwelling times at stations. As previously stated, total travel time consists of the sum of running and dwell times across all intermediate tunnels and stations, with the primary source of variability being the uncertainty in dwell times. The "\% of median dwell time" column of in \zcref[S]{tab:theta_11} expresses $\theta_m$ as a percentage of the median dwell time at the corresponding station. For instance in the skew-normal model, at Monk station, where $\theta$ corresponds to $6.39\%$ of the median dwell time, a 10-minute increase of headway (caused by a disruption) compared to normal operation, results in a $63.9\%$ increase in the dwell time for the trains passing through Monk once the disruption is resolved.

For the majority of stations, the posterior sample means of the corresponding $\theta_m$ parameter are observed to be close to zero, suggesting that longer-than-usual headways have minimal impact on the travel time of trains passing through stations during post-disruption periods. Although service interruptions may initially result in passenger accumulation on platforms, the waiting time often allows passengers to disperse toward less crowded areas while waiting for the next train, which results in a more efficient passenger exchange when the next train arrives. Furthermore, during prolonged disruptions, many passengers are likely to shift to alternative transportation modes, such as buses, leading to a gradual clearing of platforms \citep{mo2022}. Therefore, when normal operations resume after extended disruptions, platform crowding is typically limited, which explains the near-zero estimates of the $\theta_m$ parameters.
% address differences of skew-normal and skew-t models

\begin{table}[!ht]
\centering
\resizebox{\ifdim\width>\linewidth\linewidth\else\width\fi}{!}{%
\begin{tabular}[t]{rlr@{\quad(}r@{, }r@{)\quad}rr@{\quad(}r@{, }r@{)\quad}r}
\toprule
\multicolumn{2}{c}{ } & \multicolumn{4}{c}{Skew-normal} & \multicolumn{4}{c}{Skew-$t$} \\
\cmidrule(l{3pt}r{3pt}){3-6} \cmidrule(l{3pt}r{3pt}){7-10}
m & Parameter & \makecell[r]{Posterior mean\\(seconds)} & 5\% & 95\% & \makecell[r]{\% of median\\dwell time} & \makecell[r]{Posterior mean\\(seconds)} & 5\% & 95\% & \makecell[r]{\% of median\\dwell time}\\
\midrule
2 & $\theta_{\text{Monk}}$ & $2.4$$^{\ast}$ & $0.8$ & $3.9$ & $6.39\%$ & $1.3$$^{\hphantom{\ast}}$ & $-0.3$ & $2.9$ & $3.49\%$\\
3 & $\theta_{\text{Jolicoeur}}$ & $1.9$$^{\ast}$ & $0.5$ & $3.4$ & $5.19\%$ & $1.3$$^{\hphantom{\ast}}$ & $0.0$ & $2.7$ & $3.61\%$\\
4 & $\theta_{\text{Verdun}}$ & $-0.7$$^{\hphantom{\ast}}$ & $-1.7$ & $0.2$ & $-1.99\%$ & $-1.7$$^{\ast}$ & $-2.4$ & $-0.9$ & $-4.47\%$\\
5 & $\theta_{\text{De l’Église}}$ & $1.1$$^{\ast}$ & $0.2$ & $2.0$ & $2.93\%$ & $2.5$$^{\ast}$ & $1.9$ & $3.1$ & $6.55\%$\\
6 & $\theta_{\text{LaSalle}}$ & $-2.6$$^{\ast}$ & $-3.5$ & $-1.6$ & $-6.76\%$ & $-2.6$$^{\ast}$ & $-3.2$ & $-2.0$ & $-6.83\%$\\
\addlinespace
7 & $\theta_{\text{Charlevoix}}$ & $-0.4$$^{\hphantom{\ast}}$ & $-1.2$ & $0.4$ & $-1.03\%$ & $0.4$$^{\hphantom{\ast}}$ & $-0.2$ & $1.0$ & $1.00\%$\\
8 & $\boldsymbol{\theta_{\textbf{Lionel-Groulx}}}$ & $1.1$$^{\ast}$ & $0.6$ & $1.5$ & $2.25\%$ & $0.9$$^{\ast}$ & $0.4$ & $1.4$ & $1.94\%$\\
9 & $\boldsymbol{\theta_{\textbf{Atwater}}}$ & $0.2$$^{\ast}$ & $0.0$ & $0.4$ & $0.52\%$ & $-0.1$$^{\hphantom{\ast}}$ & $-0.6$ & $0.2$ & $-0.33\%$\\
10 & $\boldsymbol{\theta_{\textbf{Guy-Concordia}}}$ & $-0.2$$^{\ast}$ & $-0.5$ & $0.0$ & $-0.54\%$ & $0.1$$^{\hphantom{\ast}}$ & $-0.1$ & $0.3$ & $0.15\%$\\
11 & $\boldsymbol{\theta_{\textbf{Peel}}}$ & $0.3$$^{\hphantom{\ast}}$ & $0.0$ & $0.6$ & $0.70\%$ & $-0.3$$^{\hphantom{\ast}}$ & $-0.8$ & $0.3$ & $-0.63\%$\\
\addlinespace
12 & $\boldsymbol{\theta_{\textbf{McGill}}}$ & $0.6$$^{\ast}$ & $0.3$ & $0.9$ & $1.39\%$ & $1.2$$^{\ast}$ & $0.7$ & $1.5$ & $2.76\%$\\
13 & $\boldsymbol{\theta_{\textbf{Place-des-Arts}}}$ & $0.1$$^{\hphantom{\ast}}$ & $-0.2$ & $0.3$ & $0.17\%$ & $-0.3$$^{\ast}$ & $-0.5$ & $-0.1$ & $-0.81\%$\\
14 & $\boldsymbol{\theta_{\textbf{Saint-Laurent}}}$ & $-0.6$$^{\ast}$ & $-0.9$ & $-0.3$ & $-1.46\%$ & $-0.1$$^{\hphantom{\ast}}$ & $-0.3$ & $0.1$ & $-0.19\%$\\
15 & $\boldsymbol{\theta_{\textbf{Berri-UQAM}}}$ & $0.1$$^{\hphantom{\ast}}$ & $-0.1$ & $0.4$ & $0.27\%$ & $0.1$$^{\hphantom{\ast}}$ & $-0.2$ & $0.3$ & $0.09\%$\\
16 & $\theta_{\text{Beaudry}}$ & $-0.1$$^{\hphantom{\ast}}$ & $-0.4$ & $0.2$ & $-0.35\%$ & $0.3$$^{\ast}$ & $0.1$ & $0.6$ & $0.85\%$\\
\addlinespace
17 & $\theta_{\text{Papineau}}$ & $0.0$$^{\hphantom{\ast}}$ & $-0.3$ & $0.3$ & $-0.02\%$ & $-0.2$$^{\hphantom{\ast}}$ & $-0.5$ & $0.0$ & $-0.57\%$\\
18 & $\theta_{\text{Frontenac}}$ & $0.4$$^{\ast}$ & $0.1$ & $0.7$ & $1.14\%$ & $0.3$$^{\ast}$ & $0.0$ & $0.5$ & $0.72\%$\\
19 & $\theta_{\text{Préfontaine}}$ & $-0.4$$^{\ast}$ & $-0.6$ & $-0.2$ & $-1.04\%$ & $-0.3$$^{\ast}$ & $-0.6$ & $-0.1$ & $-0.90\%$\\
20 & $\theta_{\text{Joliette}}$ & $0.2$$^{\hphantom{\ast}}$ & $0.0$ & $0.4$ & $0.48\%$ & $0.0$$^{\hphantom{\ast}}$ & $-0.1$ & $0.2$ & $0.10\%$\\
21 & $\theta_{\text{Pie-IX}}$ & $-0.1$$^{\hphantom{\ast}}$ & $-0.3$ & $0.2$ & $-0.13\%$ & $0.1$$^{\hphantom{\ast}}$ & $0.0$ & $0.3$ & $0.34\%$\\
\addlinespace
22 & $\theta_{\text{Viau}}$ & $0.0$$^{\hphantom{\ast}}$ & $-0.2$ & $0.3$ & $0.02\%$ & $-0.2$$^{\hphantom{\ast}}$ & $-0.3$ & $0.0$ & $-0.42\%$\\
23 & $\theta_{\text{Assomption}}$ & $0.1$$^{\hphantom{\ast}}$ & $-0.1$ & $0.4$ & $0.36\%$ & $0.1$$^{\hphantom{\ast}}$ & $-0.1$ & $0.3$ & $0.22\%$\\
24 & $\theta_{\text{Cadillac}}$ & $0.4$$^{\ast}$ & $0.1$ & $0.7$ & $1.07\%$ & $0.2$$^{\ast}$ & $0.0$ & $0.4$ & $0.57\%$\\
25 & $\theta_{\text{Langelier}}$ & $0.1$$^{\hphantom{\ast}}$ & $-0.3$ & $0.5$ & $0.21\%$ & $0.0$$^{\hphantom{\ast}}$ & $-0.3$ & $0.3$ & $0.02\%$\\
26 & $\theta_{\text{Radisson}}$ & $0.1$$^{\hphantom{\ast}}$ & $-0.3$ & $0.5$ & $0.20\%$ & $0.0$$^{\hphantom{\ast}}$ & $-0.3$ & $0.3$ & $0.05\%$\\
\bottomrule
\end{tabular}
}
\caption{Posterior summaries of the $\theta_m$ parameters for the skew-normal and skew-\(t\) models on the Green line of the Montreal metro system in Direction~1. The estimates are interpreted as the change in travel time (in seconds) associated with a one-minute increase in headway. Posterior means marked with an asterisk ($\ast$) indicate strong evidence of either a positive or negative effect, as determined by the \(90\%\) credible interval. The column ``\% of median dwell time'' reports each $\theta_m$ estimate as a percentage of the median dwell time at the corresponding station. Rows shown in bold correspond to stations located in the downtown area.}
\label{tab:theta_11}
\end{table}

\subsection{\texorpdfstring{The effect of train formation parameters, $\gamma_{\ell, j}$}{The effect of train formation parameters}}
The effect of train presence in downstream segments is captured by the station-specific train formation parameters $\gamma_{\ell, j}$, which quantify the additional delay experienced by a train departing from station $j$ immediately after an incident is resolved, when another train is located $\ell$ segments ahead. The posterior means of these parameters are presented in \zcref[S]{fig:gamma_11} for Direction~1 of the Green metro line for both skew-normal and skew-$t$ models. $\gamma_{\ell, j}$ parameters capture the magnitude of delay propagation along the line, reflecting the spatial influence of upstream train formations on post-disruption recovery dynamics. To provide a clearer understanding of the interpretation of these parameters, consider the following example. For Peel station in the skew-$t$ model, the posterior mean of $\gamma_{\ell, j}$ at a distance of one segment is estimated to be $70.1$ seconds, indicating that a train remaining stationary at Peel during an incident will experience an additional $70.1$ seconds of travel time to its downstream stations once operations resume if another train is located one segment ahead. 

\begin{figure}[t]
    \centering
    \includegraphics[width=1\linewidth]{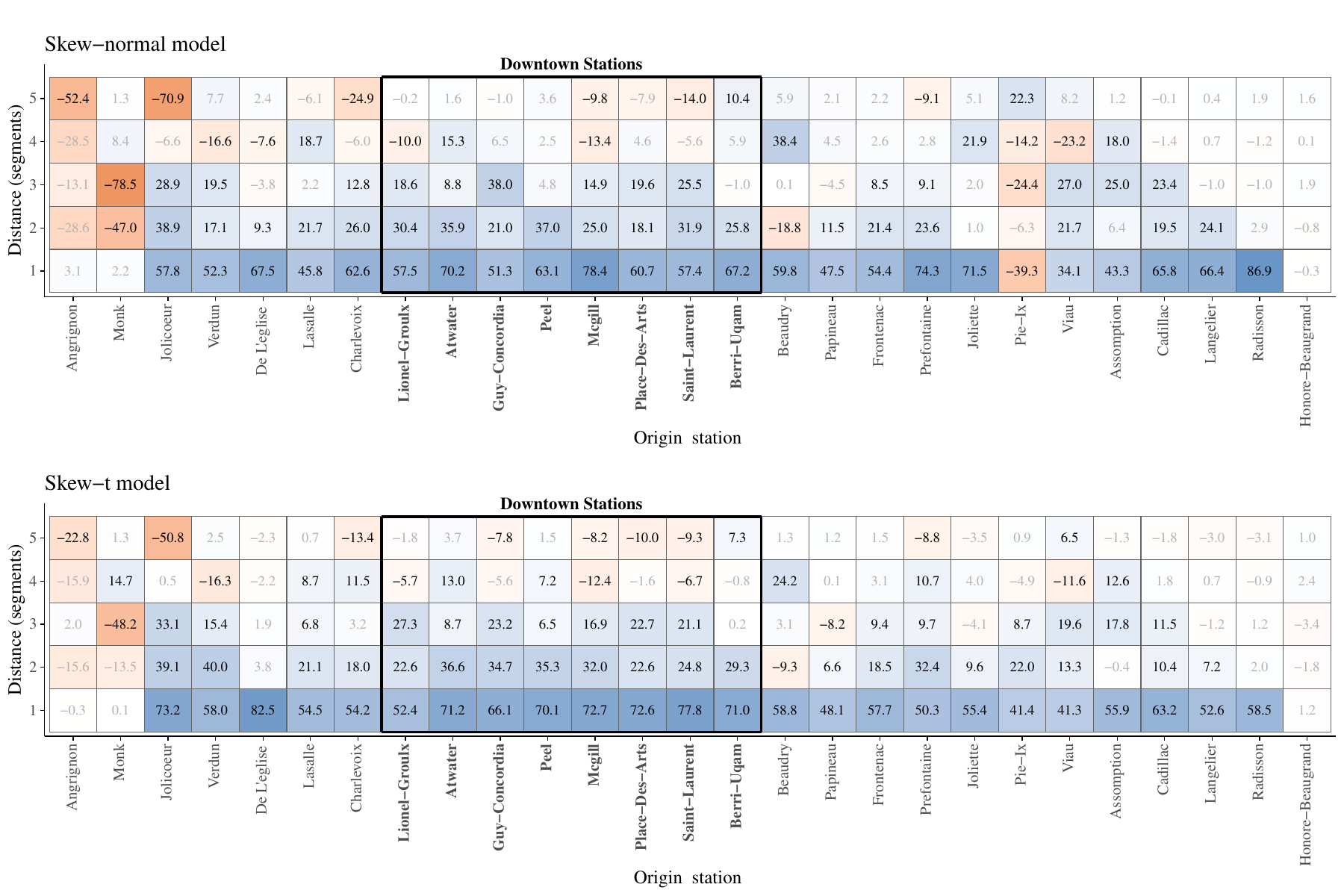}
    \caption{Posterior means of the train formation parameters $\gamma_{\ell,j}$ for post-disruption operations on Direction~1 of the Green line, shown for the skew-normal model (top) and the skew-\(t\) model (bottom). All estimates are reported in seconds and represent the average additional delay incurred by a train at a given location when another train is positioned $\ell$ segments ahead. Stations in the downtown area are highlighted in bold, and estimates displayed in black indicate strong evidence of either a positive or negative effect based on the \(90\%\) credible interval. Cell colors represent the sign and magnitude of the delay effect, with blue tones corresponding to positive values and orange tones to negative values; color intensity reflects magnitude and fades toward zero.}
    \label{fig:gamma_11}
\end{figure}

We can observe in  \zcref[S]{fig:gamma_11}, that the influence of other trains on the delay process of the current train varies substantially across stations and over different distances. Smaller delay effects are observed near the terminal sections of the lines. This pattern can be attributed to the fact that the endpoints of the lines are typically located in less densely populated areas of the city, where inter-station distances are longer and, consequently, the impact of other trains' presence is reduced. At the very end of each line, where no downstream segments exist, the corresponding estimated parameters closely resemble their prior values, effectively zero. The parameter associated with the terminal station Honor\'e-Beaugrand in both models in \zcref[S]{fig:gamma_11} approaches zero because no trains can be located ahead when a train departs from that station, rendering the delay process independent of distance. Similarly, for stations located near the end of the line but with a few downstream segments remaining, the induced delay effects from trains ahead remain relatively small, consistent with the behavior observed near the opposite end of the line.

A subtle distinction between the proposed models can be observed in \zcref[S]{fig:gamma_11}. In the skew-normal specification, the estimated $\gamma_{\ell,j}$ parameters exhibit more abrupt variations, particularly across distance, but also across location. A clear illustration of this behavior is observed at the Pie--IX station, where the estimated $\gamma_{\ell,j}$ values take large negative magnitudes for distances of $\ell = 1, 2,$ and $4$. This pattern can be attributed to the presence of outliers in the data, to which the skew-normal model is more susceptible and therefore more prone to overfitting relative to the skew-$t$ model. Owing to its heavier tails, the skew-$t$ model assigns less weight to outlier observations during parameter estimation, leading to smoother and more uniform variations across both location and distance when compared with the skew-normal model.
 
In \zcref[S]{fig:ridge11}, ridge plots of the posterior samples of the train spacing parameters are presented to compare train formation spacing effects across stations for Direction~1 of the Green metro line. The figure displays the density estimates obtained under both proposed models, namely the skew-normal and skew-$t$ specifications, for two station groups: (i) stations located in the downtown area and (ii) stations located outside the downtown area. The comparison is performed for spacing distances ranging from 1 to 5.

As expected, shorter spacing to the preceding train (i.e., smaller distance) is associated with larger delays. This effect is more pronounced for downtown stations than for non-downtown stations under both model specifications. Moreover, the posterior samples for downtown stations exhibit lower variability relative to those for non-downtown stations. The magnitude of the spacing effect decreases as the distance increases, and the difference in mean effects between the two station groups diminishes accordingly. The two proposed models yield similar mean effects across both station groups. However, the skew-$t$ model exhibits relatively lower posterior variance in $\gamma_{\ell,j}$ compared with the skew-normal model. Corresponding results for the remaining case studies, including the opposite direction of the Green line and both directions of the Orange line, are reported in \ref{appx:gamma}.

\begin{figure}
    \centering
    \includegraphics[width=1\linewidth]{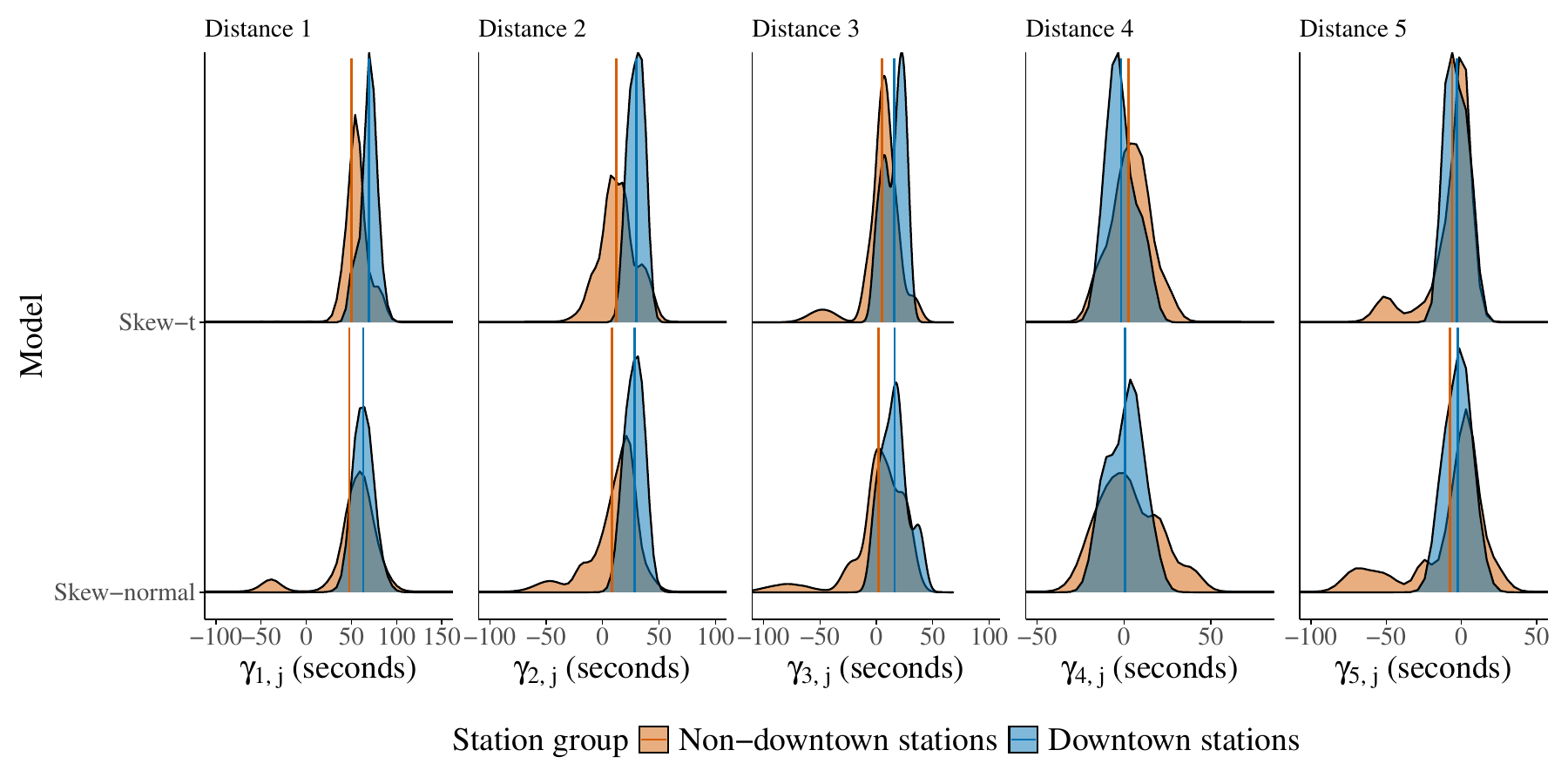}
    \caption{Ridge plots of the train formation parameters $\gamma_{\ell,j}$ for Direction~1 of the Green metro line under both proposed models and for two station groups, namely downtown and non-downtown stations. For each distance, the corresponding density represents the posterior samples of $\gamma_{\ell,j}$, where $j$ belongs to the set of station indices within the respective group. The vertical lines indicate the group-specific mean effect for each model, with colors distinguishing the station groups.}
    \label{fig:ridge11}
\end{figure}

\subsection{Distributional parameters}
In addition to the mean specification parameters discussed in the previous subsections ($\theta_m$ and $\gamma_{\ell,j}$), we examine the remaining distributional parameters of the models defined in eqn.~\ref{eq:sn_innov} and eqn.~\ref{eq:st_innov}. These include the scale specification parameters $\omega_0$ and $\omega_1$, as well as the skewness specification parameters $\alpha_0$ and $\alpha_1$, which are common to both the skew-normal and skew-$t$ models. In addition, the degrees of freedom  $\nu$ is considered for the skew-$t$ model. Posterior summaries of these parameters are reported in \zcref[S]{tab:distparam11} for both proposed models applied to the Green metro line in Direction~1.

\begin{table}[!htbp]
    \centering
    \tablesize
\begin{tabular}[t]{lrr@{}rrr@{}r}
\toprule
\multicolumn{1}{c}{ } & \multicolumn{3}{c}{Skew-normal} & \multicolumn{3}{c}{Skew-t} \\
\cmidrule(r{3pt}l{3pt}){2-4} \cmidrule(l{3pt}r{3pt}){5-7}
Parameter & \makecell[r]{Posterior mean} & (5\%, & 95\%) &
\makecell[r]{Posterior mean} & (5\%, & 95\%) \\
\midrule
$\omega_0$ & 2.300 & (\;2.168, & 2.436) & 0.460 & (\;0.424, & 0.497) \\
$\omega_1$ & 0.163 & (\;0.139, & 0.185) & 0.081 & (\;0.074, & 0.089) \\
\addlinespace
$\alpha_0$ & 2.158 & (\;2.064, & 2.254) & 2.329 & (\;2.157, & 2.508) \\
$\alpha_1$ & $-$0.031 & ($-$0.041, & $-$0.020) & $-$0.065 & ($-$0.076, & $-$0.054) \\
\addlinespace
$\nu$      &        &         &        & 2.666 & (\;2.561, & 2.773) \\
\bottomrule
\end{tabular}
    \caption{Posterior summaries of distributional parameters in our proposed skew-normal and skew-$t$ models for the Green metro line in Direction~1.}
    \label{tab:distparam11}
\end{table}

As described earlier in \zcref[S]{section:model_hier}, the squared scale parameter in both the skew-normal and skew-$t$ models is specified as a linear function of the traveled distance $k-j$, with an intercept $\omega_0$ and a slope $\omega_1$. A comparison of the estimation results reported in \zcref[S]{tab:distparam11} indicates that the skew-$t$ model yields smaller values for both $\omega_0$ and $\omega_1$. This implies that the scale of the error terms is lower across different travel distances and increases more slowly with distance in the skew-$t$ model relative to the skew-normal specification. In contrast, the skew-normal model assigns a larger scale parameter, which is directly related to variance, in order to accommodate the presence of large travel time realizations in the data. However, due to its heavy-tailed property, the skew-$t$ model is less influenced by outlier observations, ultimately leading to substantially smaller estimated scale parameters and narrower prediction intervals.

The estimated scale and skewness parameters are displayed in \zcref[S]{fig:sc_sk11}. As shown, the scale parameter follows a quadratic functional form in traveled distance, although its behavior is close to linear over the relevant range, with the skew-$t$ model consistently exhibiting smaller scale values. According to the estimates reported in \zcref[S]{tab:distparam11}, both models estimate a positive intercept in the skewness specification ($\alpha_0$) along with a small negative slope ($\alpha_1$). As illustrated in the right panel of \zcref[S]{fig:sc_sk11}, the estimated skewness parameter in the skew-$t$ model decreases more rapidly as traveled distance increases. The negative slope $\alpha_1$ in the skewness specification indicates that, for short travel distances immediately following a disruption, trains are more likely to experience substantial delays, leading to travel times that are considerably longer than expected while remaining bounded below. For longer journeys, however, trains are able to gradually recover from schedule deviations, and the resulting travel time distribution becomes closer to symmetric, exhibiting reduced skewness.

\begin{figure}
    \centering
    \includegraphics[width=\linewidth]{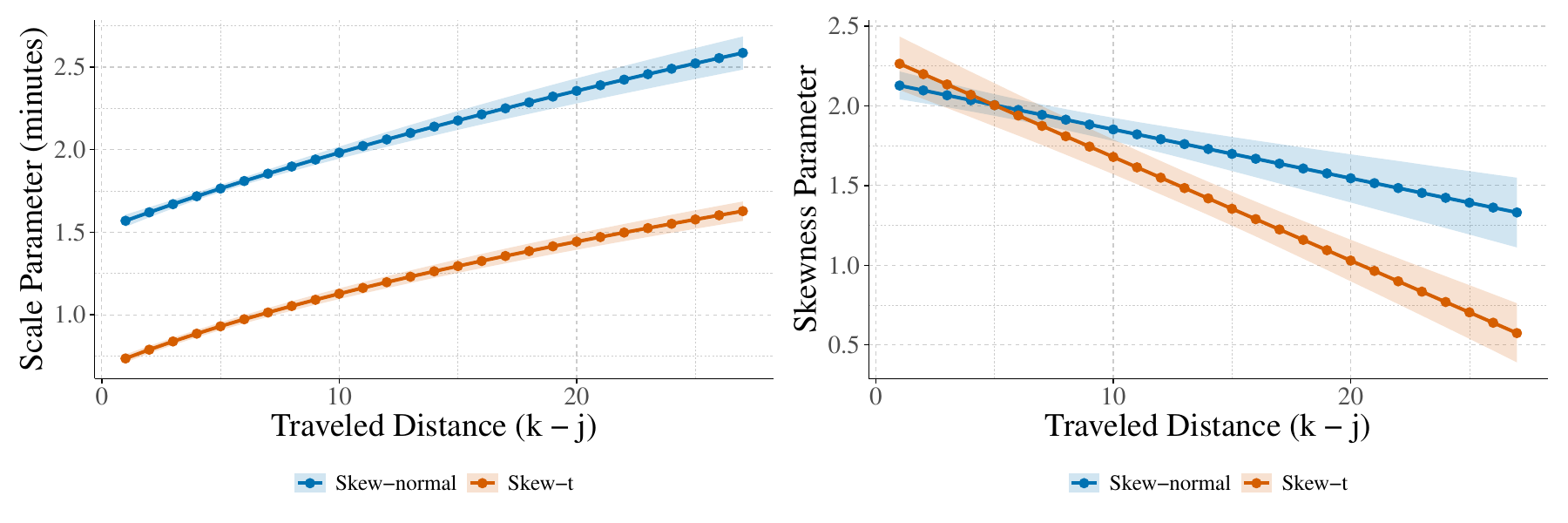}
    \caption{The left panel illustrates the estimated scale parameter $\sqrt{\omega_0 + \omega_1 (k - j)}$, while the right panel shows the estimated skewness parameter $\alpha_0 + \alpha_1 (k - j)$, together with pointwise 90\% credible intervals, for the skew-normal and skew-$t$ models applied to the Green metro line in Direction~1 across different traveled distances.}
    \label{fig:sc_sk11}
\end{figure}

The estimated degrees of freedom in the skew-$t$ model, $\nu = 2.66$, indicates that post-disruption travel times exhibit pronounced heavy-tailed behavior, reflecting the presence of substantial tail risk in the distribution.

\subsection{Error dependence parameters}
The final component of the proposed framework is the error-dependence structure introduced in \zcref[S]{subsec:err_dep}. As specified in eqn.~\ref{eq:rho}, the dependence between the travel time errors of two consecutive post-disruption trains is modeled using an exponentially decaying function of journey overlap, governed by the parameters $\rho$ and $\lambda$. Estimation results for Direction~1 of the Green metro line are reported in \zcref[S]{tab:errdep11}, while the corresponding results for the remaining case studies are provided in  \ref{appx:err_dep}.

\begin{table}[!htbp]
    \centering
    \tablesize
\begin{tabular}[t]{lrr@{}rrr@{}r}
\toprule
\multicolumn{1}{c}{ } & \multicolumn{3}{c}{Skew-normal} & \multicolumn{3}{c}{Skew-t} \\
\cmidrule(r{3pt}l{3pt}){2-4} \cmidrule(l{3pt}r{3pt}){5-7}
Parameter & \makecell[r]{Posterior mean} & (5\%, & 95\%) &
\makecell[r]{Posterior mean} & (5\%, & 95\%) \\
\midrule
$\rho$ & 0.966 & (0.940, &\;0.985) & 0.947 & (0.908, &\;0.977) \\
$\lambda$ & 1.552 & (1.477, &\;1.643) & 1.567 & (1.464, &\;1.701) \\
\bottomrule
\end{tabular}
    \caption{Posterior summaries of error dependence parameters, $\rho$ and $\lambda$ in our proposed skew-normal and skew-t models for the Green metro line in Direction~1.}
    \label{tab:errdep11}
\end{table}

Inspection of \zcref[S]{tab:errdep11} indicates that the estimated error-dependence parameters are very similar across the two proposed specifications, namely the skew-normal and skew-$t$ models. In particular, the estimated dependence magnitude $\rho$ is close to one, indicating strong dependence between trains that share a substantial portion of their journeys. A large overlap with respect to a destination station $k$ may arise either when two trains are located close to each other at the time the disruption is declared resolved, or when $k$ is sufficiently distant from both trains, corresponding to longer journeys. In both situations, it is expected that the travel time error of the preceding train largely propagates to the following train.

\subsection{Out-of-sample performance}

This section evaluates the out-of-sample performance of the fitted skew-normal and skew-$t$ models for Direction~1 of the Green metro line. Out-of-sample predictive accuracy is assessed using a range of metrics, including the mean squared error (MSE) and mean absolute error (MAE), as well as the length and empirical coverage of the highest density intervals (HDIs) of the predictive distributions, relative to the baseline model defined in eqn.~\ref{eq:baseline_model}. In addition, the unconditional posterior predictive performance of the fitted models is examined through the construction of probability-probability (P-P) and quantile-quantile (Q-Q) plots. Finally, the full posterior predictive distributions are compared using the continuous ranked probability score (CRPS).

\subsubsection{Realized vs. predicted travel times}

\zcref[S]{fig:pvr11} presents the realized post-disruption travel times against the predicted values, along with the corresponding model residuals, for both proposed specifications. As illustrated in the left column of \zcref[S]{fig:pvr11}, the predicted samples cluster closely around the main diagonal for both models, indicating strong predictive performance. Although the figure does not allow for a clear visual separation between the predictive accuracy of the skew-normal and skew-$t$ models, the skew-$t$ specification exhibits a slightly narrower band around the diagonal for smaller travel times, which also correspond to the most frequent observations in the dataset. Nevertheless, the improvement appears marginal when compared with the skew-normal model. Examining the right panel of \zcref[S]{fig:pvr11}, both models display small residuals across the entire sample. Moreover, the residual patterns provide visual evidence of the adopted distributional specifications, including the allowance for varying scale and skewness, as reflected in the structure of the residuals shown in \zcref[S]{fig:pvr11}. The results from the other case studies are presented in \ref{appx:rvp}.

\begin{figure}[!ht]
    \centering
    \includegraphics[width=1\linewidth]{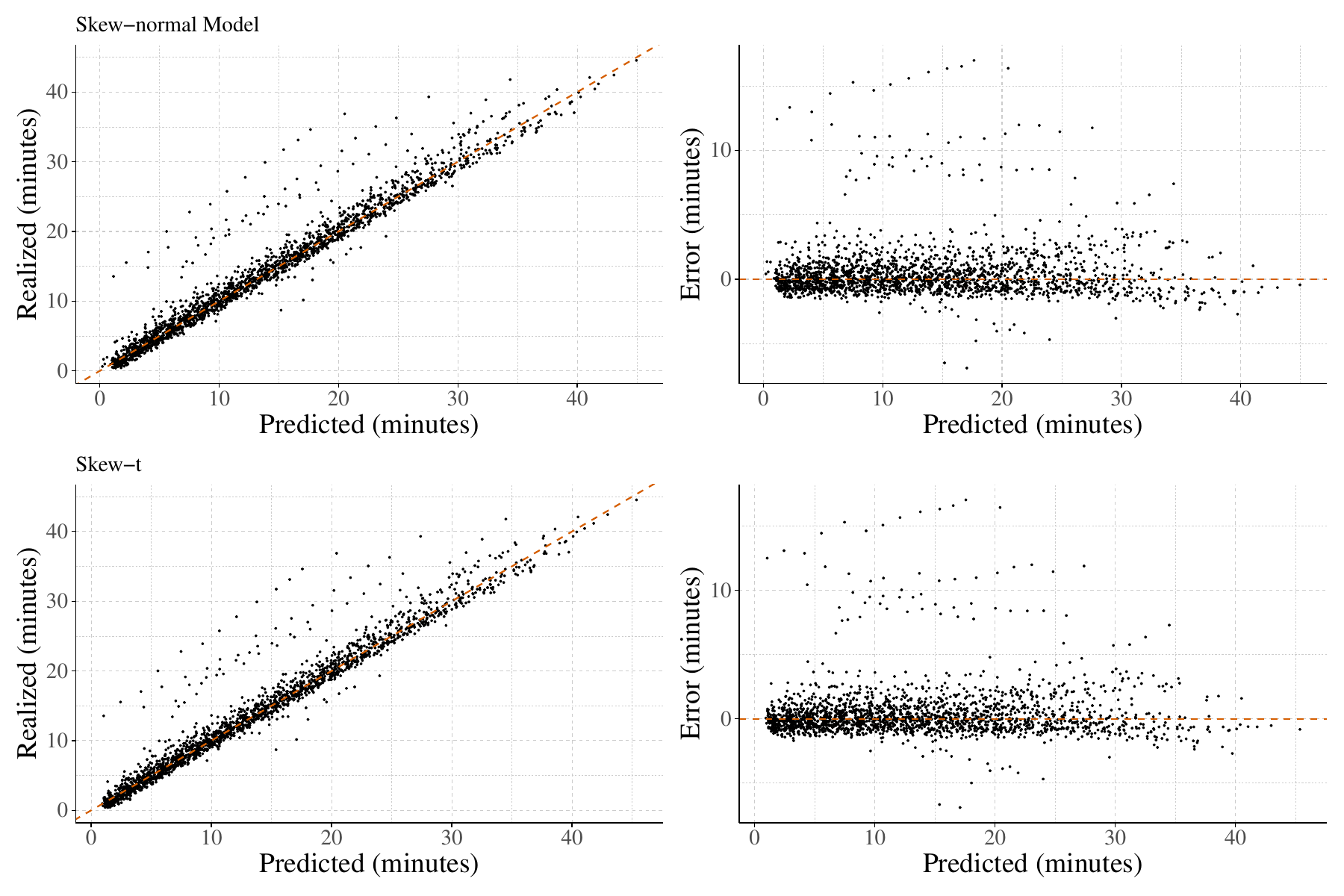}
    \caption{Realized versus predicted post-disruption travel times (left panel) and the corresponding model residuals (right panel) for the out-of-sample subset under the skew-normal and skew-$t$ specifications.}
    \label{fig:pvr11}
\end{figure}

Both models exhibit difficulty in predicting the extremely large travel time realizations in the sample set. We hypothesize that trains associated with these observations encountered additional disruptions during their post-disruption journeys, which further prolonged the travel time to downstream stations. Because such delays cannot be explained by either the train formation pattern along the metro line or the passenger accumulation at intermediate stations, the proposed model is unable to reproduce these samples accurately, resulting in large prediction errors.

An additional and noteworthy observation regarding these outliers is that their residuals appear to follow a dependence pattern with respect to the model mis-predictions: specifically, there is a sequence of points that lies approximately parallel to the main diagonal but shifted upward by a roughly constant amount. Recall that our objective is to forecast, at the moment a disruption is declared resolved, the travel time of a given train to each of its downstream stations, with the intent of informing passengers waiting at those stations about the expected arrival of that train. If, for any reason, that particular train experiences an unanticipated delay at some station after operations resume, then all subsequent travel time observations to downstream stations for that train would be shifted by approximately that additional delay. In this situation, the model may produce travel time predictions that are systematically shifted relative to the realized values. Since the model does not capture this additional delay component, the predicted travel times for subsequent stations become inaccurate. However, the fact that the slope of these outlier points remains similar to the main diagonal suggests that, conditional on the realization of the additional delay, the remaining downstream travel times would be predicted reasonably well. Nevertheless, because our modeling framework generates predictions at the level of an origin--destination station pair, without conditioning on an intervening realized delay, these travel time predictions remain erroneous for such cases.

\subsection{Prediction accuracy and uncertainty assessment}

In \zcref[S]{fig:met11}, four evaluation metrics are reported to assess the out-of-sample performance of the proposed models relative to the baseline specification introduced in eqn.~\ref{eq:baseline_model}. For each metric, the evaluation is conducted separately for each traveled distance and subsequently compared across models. This approach is motivated by the observation that, from a predictive standpoint, an error of a given magnitude has different implications depending on the traveled distance. For instance, a two-minute error for journeys covering a single station represents substantially poorer predictive performance than the same error for journeys spanning twenty stations. Accordingly, all metrics and error measures are evaluated and compared across journeys that share the same traveled distance. The results for the remaining case studies are presented in \ref{appx:met}.

\begin{figure}[!ht]
    \centering
    \includegraphics[width=1\linewidth]{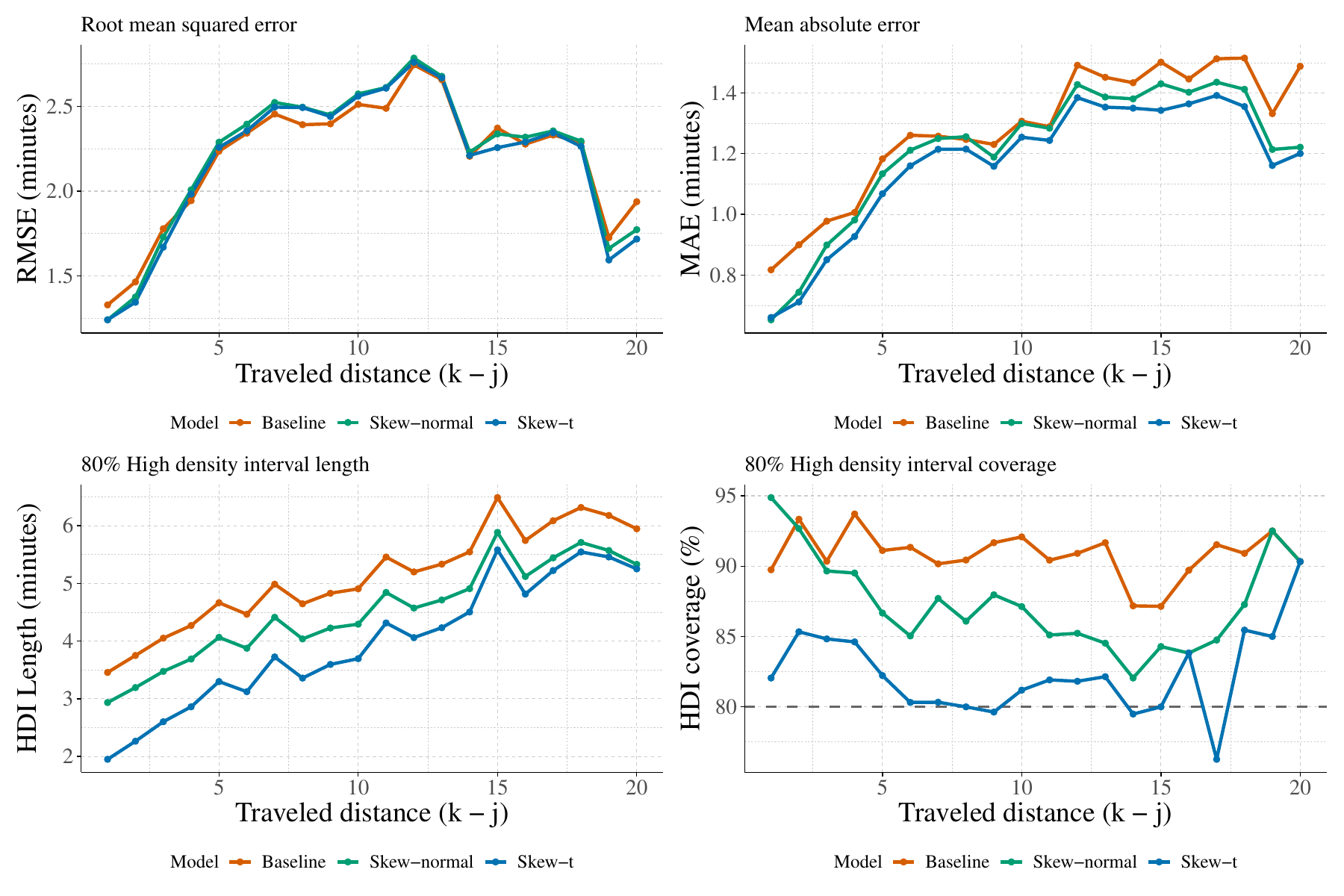}
    \caption{Comparison of error-based metrics and uncertainty assessments for the proposed models relative to the baseline specification. In each category, the evaluation is performed separately across traveled distances. The results indicate that both proposed models outperform the baseline in terms of MAE, HDI length, and empirical coverage, with the skew-$t$ model demonstrating superior performance compared to the skew-normal specification.}
    \label{fig:met11}
\end{figure}

The top-left panel of \zcref[S]{fig:met11} reports the root mean squared error (RMSE) of the predicted post-disruption travel times for all three models—the baseline, skew-normal, and skew-$t$ specifications. Under this metric, the three models exhibit broadly similar performance across traveled distances, with no clear out performance by any single model. In contrast, the top-right panel presents the mean absolute error (MAE), where both proposed models outperform the baseline across all traveled distances, with the skew-$t$ model consistently achieving the lowest MAE. The discrepancy between the RMSE and MAE results can be attributed to their differing sensitivities to extreme observations: RMSE places greater weight on outliers, whereas MAE is more robust to their presence.

As noted earlier, our objective is not only to obtain accurate point predictions but also to provide a reliable quantification of predictive uncertainty. To this end, we compare the length of the high-density intervals (HDIs) and the empirical coverage of these intervals across the competing models. The HDIs are constructed as the shortest intervals containing a specified probability mass of the posterior samples, using a data-driven, nonparametric approach based directly on posterior draws. For skewed distributions, uncertainty is inherently asymmetric, causing equal-tailed intervals to misrepresent coverage. In contrast, HDIs concentrate probability mass in regions of highest posterior density, providing a more informative and accurate representation of uncertainty. The bottom-left panel of \zcref[S]{fig:met11} shows that the $80\%$ HDI produced by the skew-$t$ model are consistently shorter than those obtained from both the skew-normal and the baseline specifications, while the corresponding coverage, shown in the bottom-right panel, remains close to the nominal $80\%$ level. When uncertainty is appropriately quantified, one expects shorter HDIs (at a fixed credibility level) paired with empirical coverage that closely matches the nominal interval probability. As illustrated in the bottom-right panel, the skew-$t$ model achieves coverage that is closer to $80\%$ than the other two models. It is also worth noting that, for both HDI length and coverage, the skew-normal model outperforms the baseline specification.

To assess the full posterior predictive distributions within our Bayesian framework, we construct P-P plots and Q-Q plots for the predictive outputs of both the proposed models and the baseline specification. Because posterior predictive samples correspond to different underlying distributions, a normalization step is required in order to map all samples onto a common scale. This normalization procedure, as well as the resulting reference distribution, differs across the considered models and is described below.

\begin{itemize}
    \item \textbf{Baseline model:}  
    Using eqns.~\ref{eq:baseline_model} and~\ref{eq:err_baseline_model}, the posterior predictive samples can be normalized using the following transformation, which maps all observations to a common scale:
    \begin{equation}
        z_{i,j,k} = \frac{y_{i,j,k} - \mu_{i,j,k}}{\sqrt{\omega_0 + \omega_1 (k - j)}} \sim \mathcal{N}(0,1).
        \label{eq:z_baseline}
    \end{equation}

    \item \textbf{Skew-normal model:}  
    For the skew-normal specification, a normalization procedure analogous to eqn.~\ref{eq:z_baseline} is applied. Using eqns.~\ref{eq:y}, \ref{eq:sn_mu} and \ref{eq:sn_innov}, the normalized predictive samples are obtained as
    \begin{equation}
        z_{i,j,k} =
        \begin{cases}
            \dfrac{y_{i,j,k} - \mu_{i,j,k}'}{\sqrt{\omega_0 + \omega_1 (k - j)}}, & \begin{aligned}[c]\genfrac{}{}{0pt}{}{\text{if train $i$ is the first\quad \,}}{\text{post-disruption train,}} \end{aligned}\\
            \dfrac{y_{i,j,k} - \mu_{i,j,k}' - \rho_{j,j',k}\,\varepsilon_{i-1,j',k}}{\sqrt{\omega_0 + \omega_1 (k - j)}}, & \text{otherwise},
        \end{cases}
        \label{eq:z_skew_n}
    \end{equation}
    
    which implies that $z_{i,j,k} \sim \mathcal{SN}\!\left\{0, 1, \alpha_0 + \alpha_1 (k - j)\right\}$. In this setting, the distribution of $z_{i,j,k}$ varies across samples due to differences in traveled distance. However, conditional on a fixed traveled distance $k-j$, all samples share the same distribution. This property allows P-P plots and Q-Q plots to be constructed separately for each traveled distance and subsequently aggregated into a single plot.

    \item \textbf{Skew-$t$ model:}  
    The normalization procedure for the skew-$t$ model follows the same structure as in eqn.~\ref{eq:z_skew_n}. The distinction lies in the resulting reference distribution, where the normalized samples satisfy
    \[
        z_{i,j,k} \sim \mathcal{ST}\!\left\{0, 1, \alpha_0 + \alpha_1 (k - j), \nu\right\}.
    \]
\end{itemize}

\begin{figure}[!ht]
    \centering
    \includegraphics[width=1\linewidth]{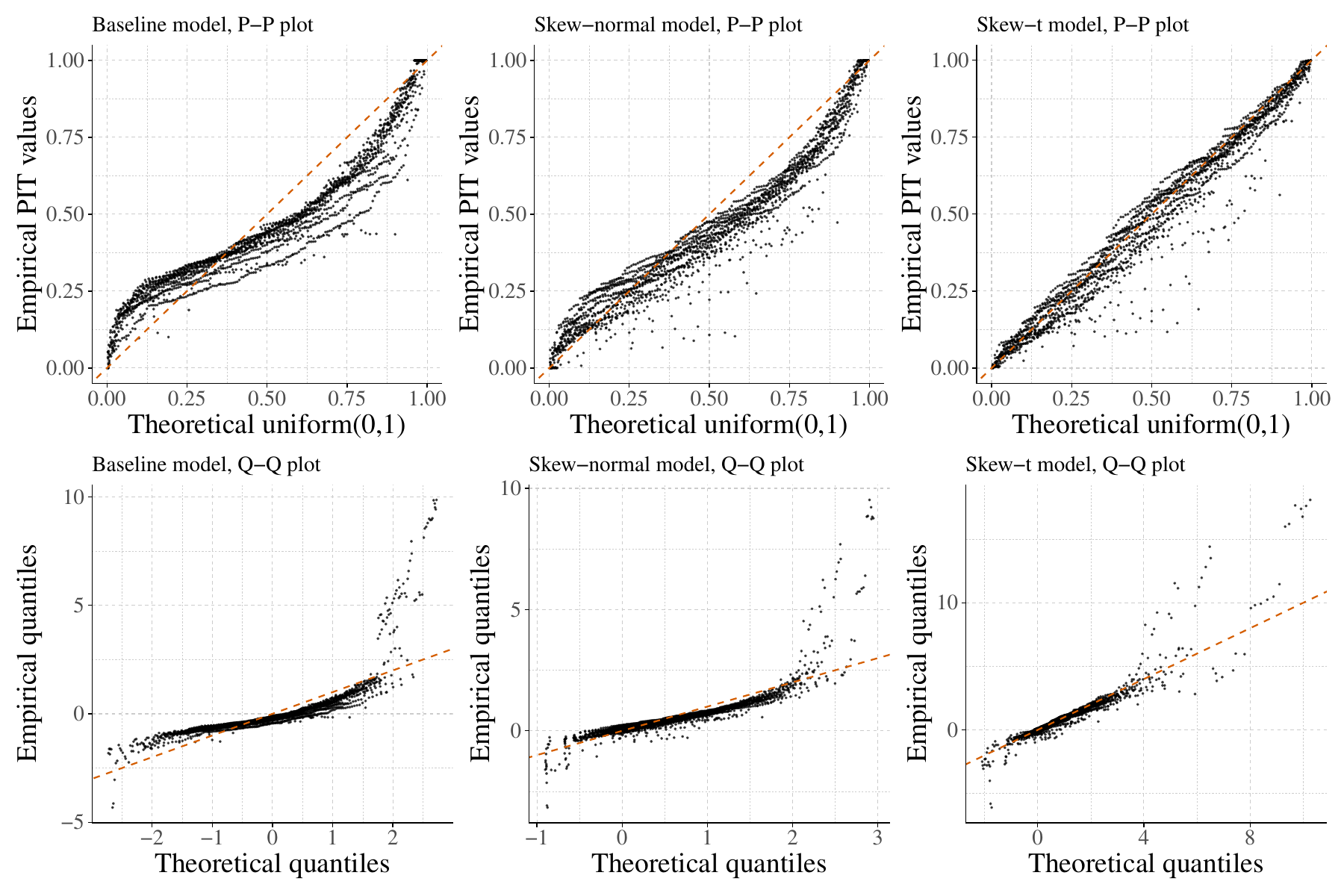}
    \caption{P-P plots (top row) and Q-Q plots (bottom row) for the baseline, skew-normal, and skew-$t$ model specifications are shown. The empirical and model-based cumulative probabilities and quantiles are computed using the transformations defined in eqn.~\ref{eq:z_baseline} and \ref{eq:z_skew_n} and are plotted against one another. For the skew-normal and skew-$t$ models, empirical cumulative probabilities and quantiles are evaluated separately for each traveled distance $(k-j)$. As illustrated in the rightmost panels, the skew-$t$ model exhibits comparatively better diagnostic performance, with the plotted points closely aligned with the main diagonal, indicating improved distributional fit relative to the other specifications.}
    \label{fig:ppqq11}
\end{figure}

\zcref[S]{fig:ppqq11} displays the P-P plots (top row) and Q-Q plots (bottom row) for the baseline, skew-normal, and skew-$t$ model specifications, constructed using the transformations defined in eqns.~\ref{eq:z_baseline} and~\ref{eq:z_skew_n}. The P-P plots compare the empirical cumulative probabilities of the observed data with the cumulative probabilities implied by each model. Under correct calibration, the P-P plot is expected to align closely with the main diagonal, while systematic deviations indicate under- or over-dispersion in the predictive distribution. As evidenced in \zcref[S]{fig:ppqq11}, the skew-$t$ model more accurately assigns probabilities that are consistent with the empirical distribution, outperforming both the baseline and skew-normal specifications in terms of calibration.

In contrast, Q-Q plots are used to assess the agreement between the shapes of the empirical and predictive distributions by comparing their respective quantiles, thereby providing insight into differences in location, scale, and skewness. As with P-P plots, departures from the diagonal indicate distributional mismatch between the model and the data. Inspection of the bottom row of \zcref[S]{fig:ppqq11} reveals that, for the baseline and skew-normal models, the predictive quantiles deviate systematically from the empirical quantiles, indicating a lack of fit across substantial portions of the distribution. By comparison, the skew-$t$ model captures the bulk of the empirical distribution more effectively, with noticeable discrepancies arising primarily in the most extreme observations.

Lastly, we employ the continuous ranked probability score (CRPS) to evaluate the full posterior predictive distributions generated by our models \citep{Matheson.Winkler:1976}. The CRPS is widely used as a quantitative measure for assessing probabilistic forecasts and a proper scoring rule. It is defined as a quadratic measure of discrepancy between the predictive cumulative distribution function (CDF), denoted by $F(x)$, and the empirical CDF of the realized observation $y$, represented by the indicator function $\mathbb{I}\{x \ge y\}$:
\begin{equation}
    \text{CRPS}(F, y) = \int_{-\infty}^{\infty} \left[ F(x) - \mathbb{I}\{x \ge y\} \right]^2 \, \mathrm{d}x.
\end{equation}
 In our analysis, the average CRPS across all observations is used as a summary performance metric. Lower CRPS values indicate superior predictive performance, as the score behaves analogously to an error metric while accounting for the entire predictive distribution. Importantly, the CRPS explicitly accounts for uncertainty quantification, which makes it particularly well suited for Bayesian and probabilistic modeling frameworks. 

In the context of post-disruption travel time modeling, CRPS is especially relevant because predictions are inherently distributional, tail behavior—such as the heavy tails captured by the skew-$t$ model—is of practical importance, and both predictive accuracy and uncertainty calibration are central objectives. Moreover, CRPS facilitates direct comparisons between competing probabilistic models, even when their predictive distributions differ in shape, variance, or tail behavior. \zcref[S]{fig:crps11} reports the mean CRPS for observations grouped by identical traveled distances across all three models. The results indicate that the proposed skew-$t$ model consistently outperforms both the baseline and the skew-normal specifications.

\begin{figure}[!ht]
    \centering
    \includegraphics[width=0.5\linewidth]{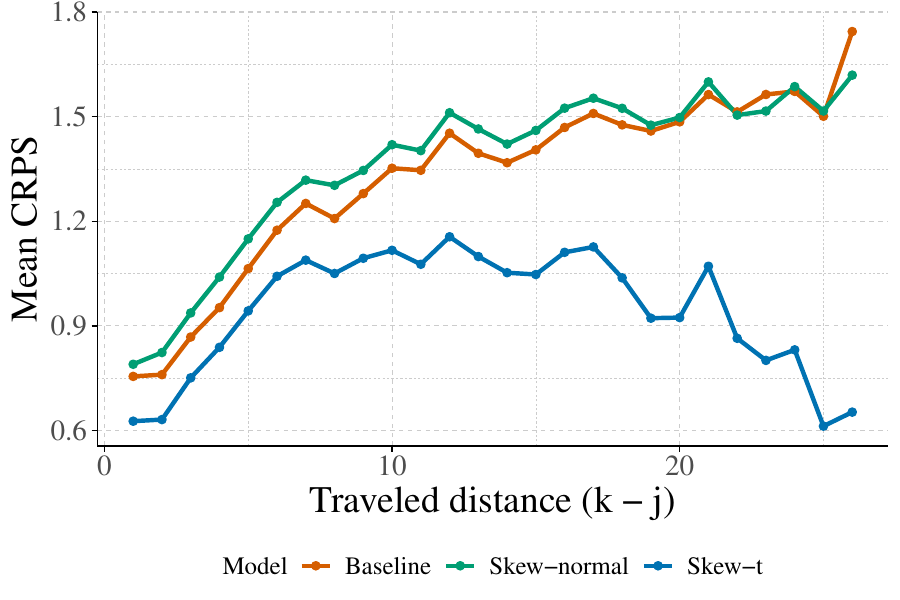}
    \caption{Mean CRPS for samples grouped by identical traveled distance.}
    \label{fig:crps11}
\end{figure}

\section{Conclusion} \label{sec:conclusion}

This paper examines the problem of predicting post-disruption train travel times in urban metro systems. We proposed a Bayesian hierarchical model that explicitly accounts for train interactions, headway imbalance due to disruption, and non-Gaussian features in travel time distributions observed during recovery periods.

Empirical results from our case study of the Montr\'eal metro system, across two lines and in both directions, indicate that post-disruption travel times exhibit pronounced heteroskedasticity, skewness, and heavy-tailed behavior. In addition, the results provide evidence of meaningful error dependence between consecutive trains, which is well captured by the proposed moving-average structure in the error specification. Both proposed models outperform the baseline specification in terms of point prediction accuracy and uncertainty quantification, with the skew-$t$ model providing the most robust performance for longer journeys and extreme delays.

From an operational perspective, the proposed framework enables more accurate and better-calibrated probabilistic travel time forecasts following disruptions. Such forecasts can improve passenger information systems and support operational decision-making during recovery phases, where uncertainty and dependence across trains play a critical role. Overall, this work provides a flexible and interpretable probabilistic framework for analyzing post-disruption railway operations, contributing both methodological advances and practical insights for resilient transit system management.

This study focuses on providing a one-time travel time prediction at the moment a disruption is reported as resolved. An important direction for future research is to extend the proposed framework to an online forecasting setting, in which predictions are continuously updated as new real-time operational information becomes available. Such an extension would allow the model to adapt dynamically to evolving recovery conditions, incorporate updated train positions and interactions, and further improve the accuracy and reliability of post-disruption travel time forecasts.

% \section*{Acknowledgements}
% This project was funded by CANSSI with support from NSERC through Collaborative Research Team (CRT) grant 25. 

\section*{Funding}
This project was funded by CANSSI (Canadian Statistical Sciences Institute) with support from NSERC (Natural Sciences and Engineering Research Council of Canada) through Collaborative Research Team (CRT) grant 25. 

\section*{Declaration of competing interest}
The authors declare that they have no known competing financial interests or personal relationships that could have appeared to influence the work reported in this paper.

\section*{Data availability}
The data used in this research article are not publicly available due to confidentiality agreements.

\section*{CRediT authorship contribution statement}
\textbf{Shayan Nazemi:} Conceptualization, Methodology, Software, Validation, Formal Analysis, Writing - Original Draft -- \textbf{Aurélie Labbe:} Conceptualization, Methodology, Writing - Review \& Editing, Supervision, Funding Acquisition -- \textbf{Stefan Steiner:} Conceptualization, Methodology -- \textbf{Pratheepa Jeganathan:} Conceptualization, Methodology, Writing - Review \& Editing -- \textbf{Martin Trépanier:} Conceptualization, Validation, Writing - Review \& Editing -- \textbf{Léo R. Belzile:} Conceptualization, Methodology, Software, Writing - Review \& Editing, Supervision, Funding Acquisition

\section*{Declaration of generative AI and AI-assisted technologies in the manuscript preparation process}
During the preparation of this manuscript, the authors used ChatGPT (https://chat.openai.com
) to assist with code debugging, language refinement (without introducing new ideas or content), grammatical corrections, and LaTeX syntax and formatting. All AI-assisted outputs were carefully reviewed and edited by the authors. The authors take full responsibility for the accuracy and integrity of the final published work.

\newpage
\bibliographystyle{elsarticle-num-names} 
\bibliography{bibliography}

\clearpage
\appendix
\section{Methodology}
\subsection{Variance Specification}
\label{appx:eda_var}
\begin{figure}[!ht]
    \centering
    \includegraphics[width=0.49\linewidth]{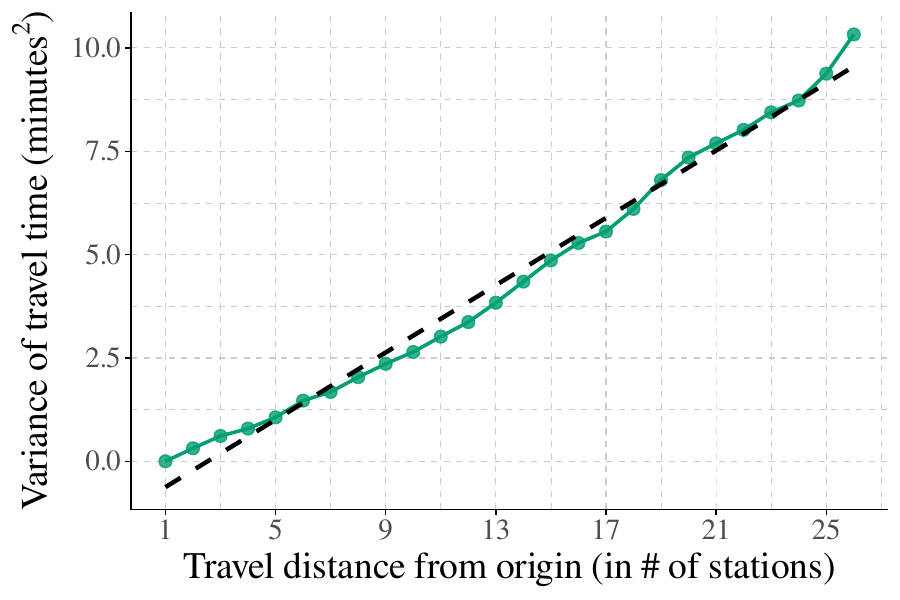}
    \caption{Green metro line -- Direction~2}
    \label{fig:eda_var_12}
\end{figure}

\begin{figure}[!ht]
    \centering
    \includegraphics[width=0.49\linewidth]{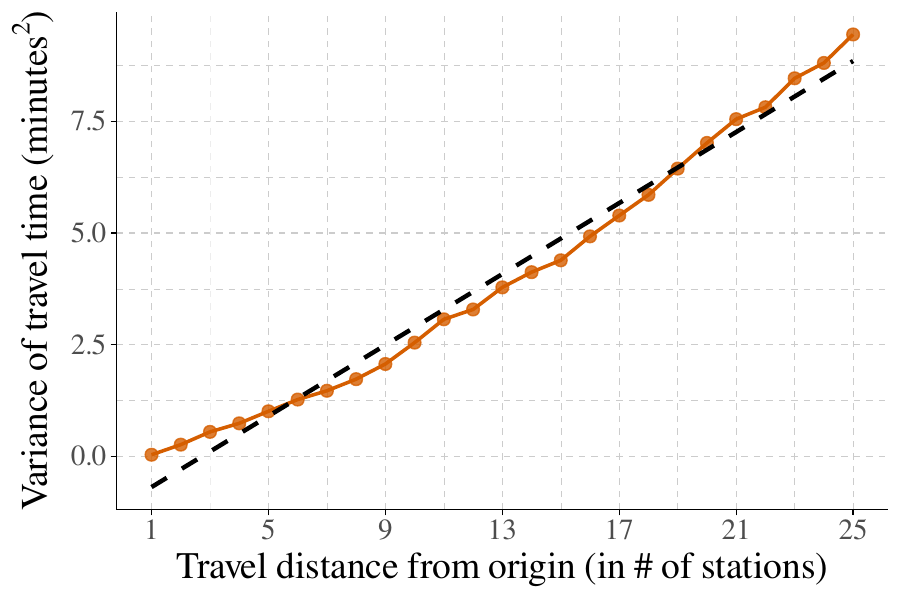}
    \includegraphics[width=0.49\linewidth]{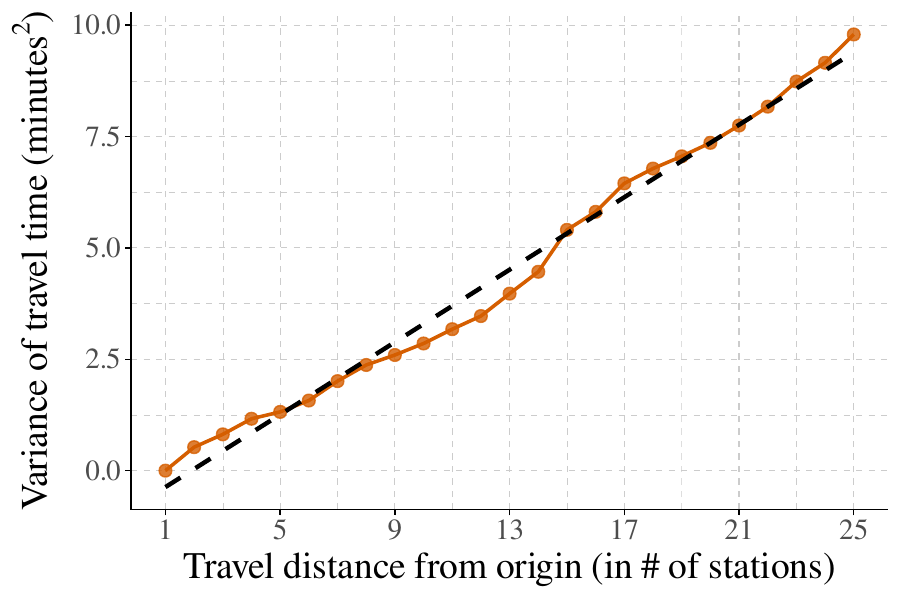}
    \caption{Orange metro line -- Direction~1 (left) and Direction~2 (right)}
    \label{fig:eda_var_2}
\end{figure}

The empirical variance of journey times for all train trajectories observed in 2018 on Direction~2 of the Green metro line is presented in \autoref{fig:eda_var_12}, while the corresponding results for both directions of the Orange line are shown in \autoref{fig:eda_var_2}. The horizontal axis denotes the traveled distance from the origin station, measured by the number of stations traversed, whereas the vertical axis reports the variance of the associated journey times.

\subsection{Baseline Residuals}
\label{appx:baseline_residuals}

\begin{figure}[!ht]
    \centering
    \includegraphics[width=\linewidth]{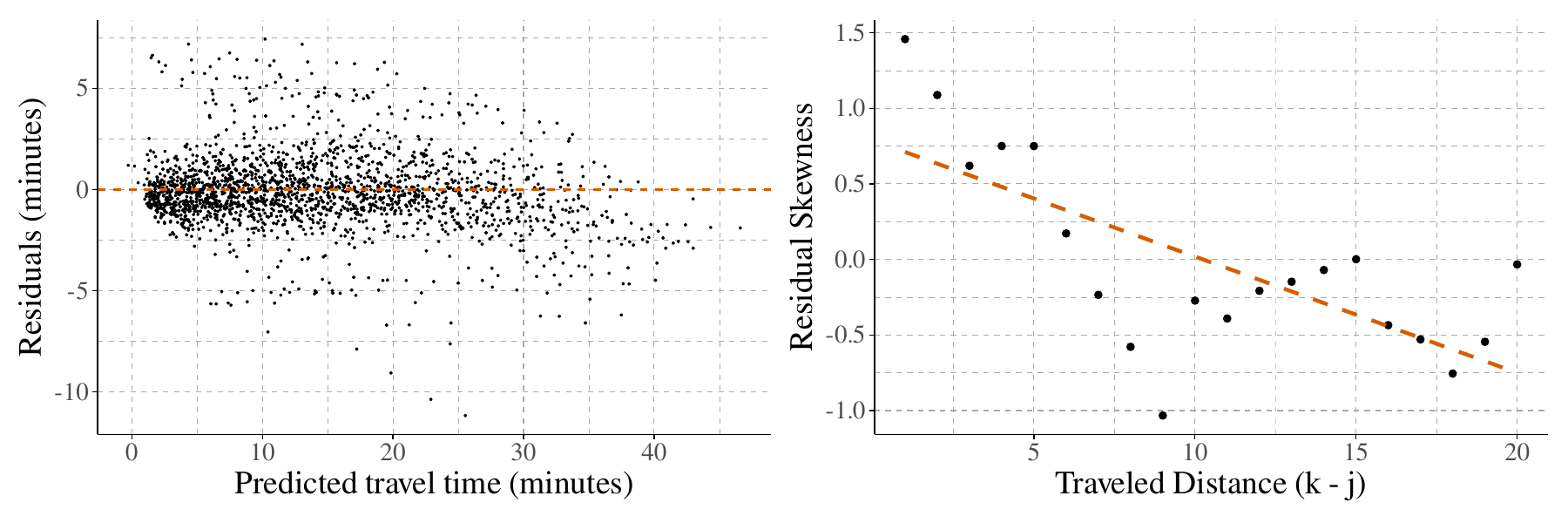}
    \caption{Green metro line -- Direction~2. The left panel presents baseline residuals versus predicted travel times, while the right panel depicts residual skewness as a function of traveled distance. The residuals-versus-predicted plot reveals noticeable misfit for longer travel times, accompanied by a decreasing skewness pattern as traveled distance increases.}
    \label{fig:baseline_residuals_12}
\end{figure}

\begin{figure}[!ht]
    \centering
    \includegraphics[width=\linewidth]{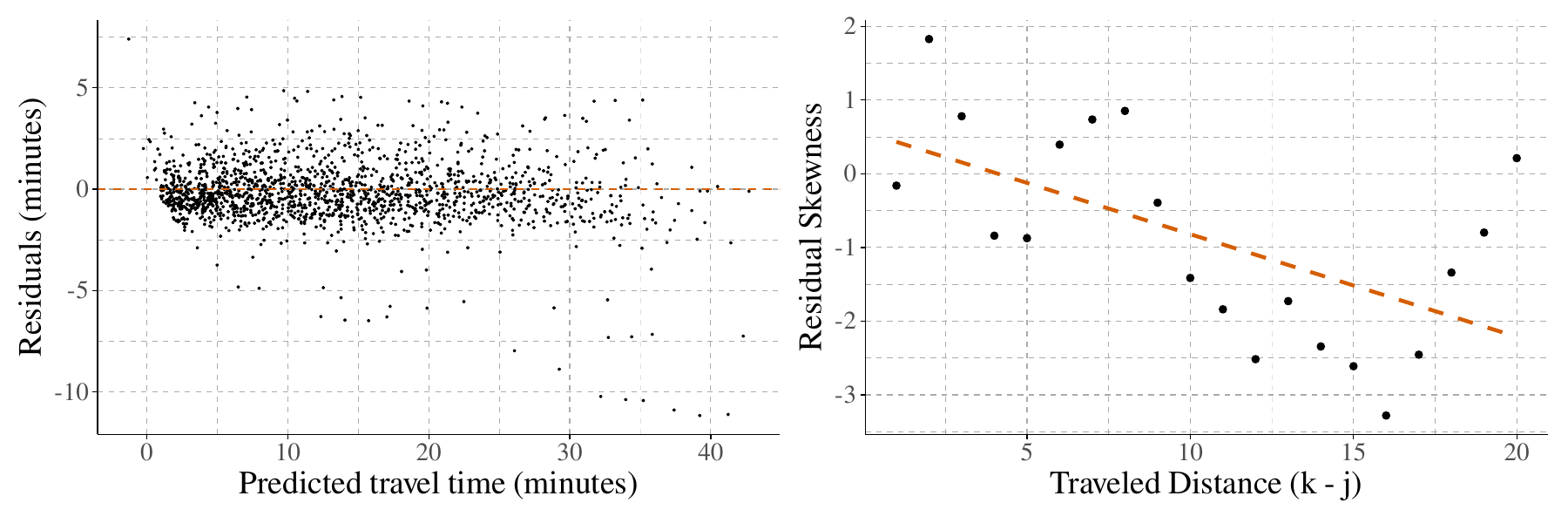}
    \includegraphics[width=\linewidth]{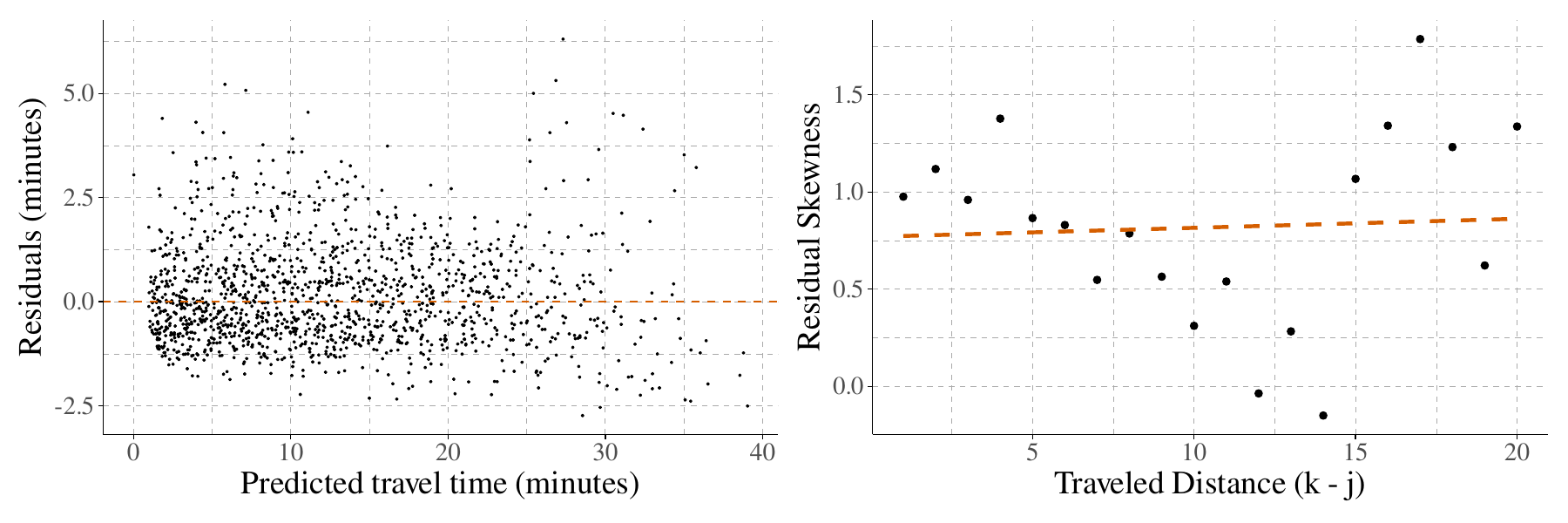}
    \caption{Orange metro line -- Direction~1 (top row) and Direction~2 (bottom row). The left column shows baseline residuals versus predicted travel times, while the right column presents residual skewness as a function of traveled distance. Comparing both directions reveals differences in residual variability and a shift in the decreasing skewness pattern at longer distances for the Orange line.}
    \label{fig:baseline_residuals_2}
\end{figure}

Here, we display the model residuals against the predicted values, along with the empirical skewness of the error terms as a function of traveled distance for Direction~2 of the Green line in \autoref{fig:baseline_residuals_12}, and the Orange line in \autoref{fig:baseline_residuals_2}.

\clearpage
\subsection{Travel Time Error Dependence}
\label{appx:err_dep}

Pearson and Spearman correlations between the baseline model residuals and those of the preceding trains, binned by the proportion of journey overlap. The results reveal positive correlations, indicating that delay and congestion propagate from one train to its successor. Train journeys with a larger overlap in traveled segments with the preceding train exhibit stronger positive correlations.

\begin{figure}[!ht]
  \centering
  \includegraphics[width=0.49\linewidth]{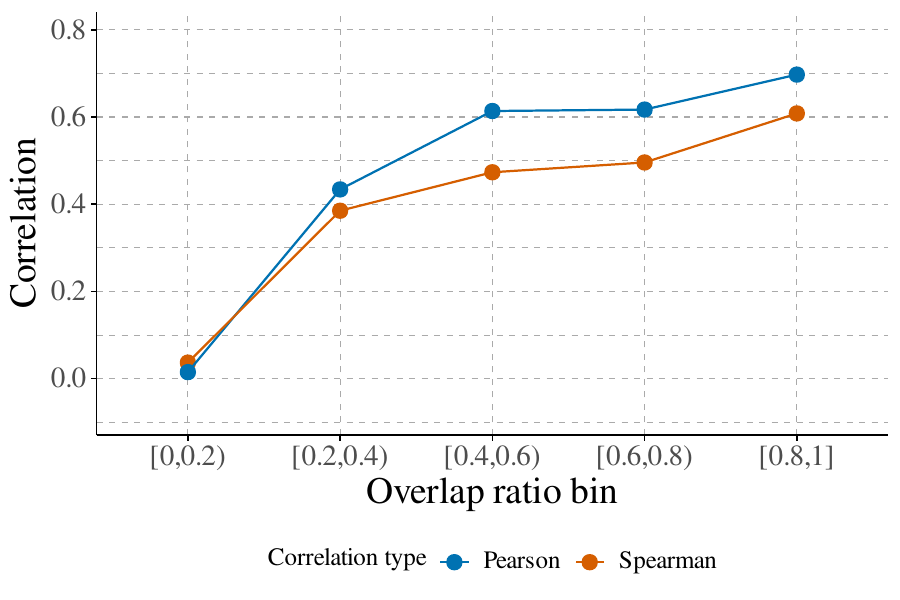}
    \caption{Green metro line -- Direction~2}
  \label{fig:error_dep_12}
\end{figure}

\begin{figure}[!ht]
  \centering
  \includegraphics[width=0.49\linewidth]{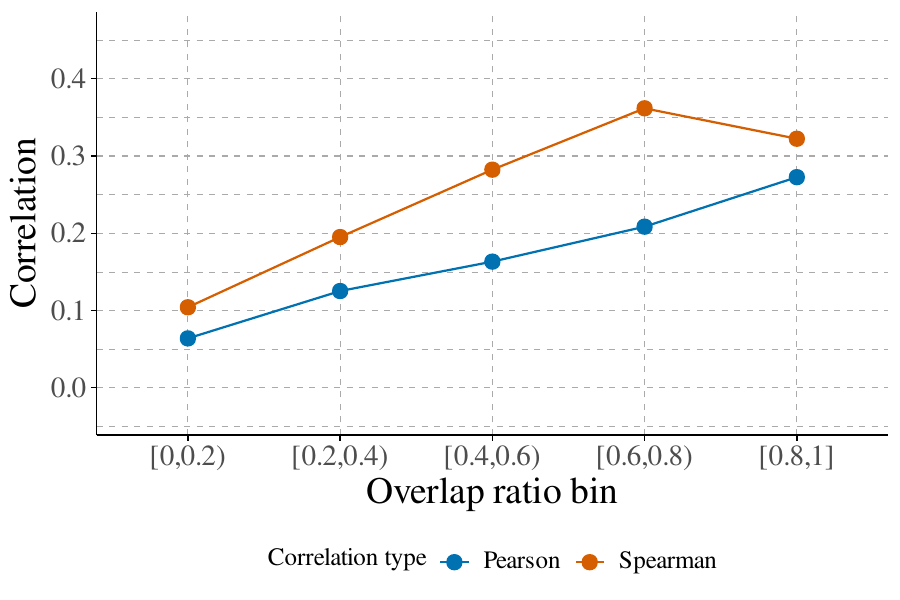}
  \includegraphics[width=0.49\linewidth]{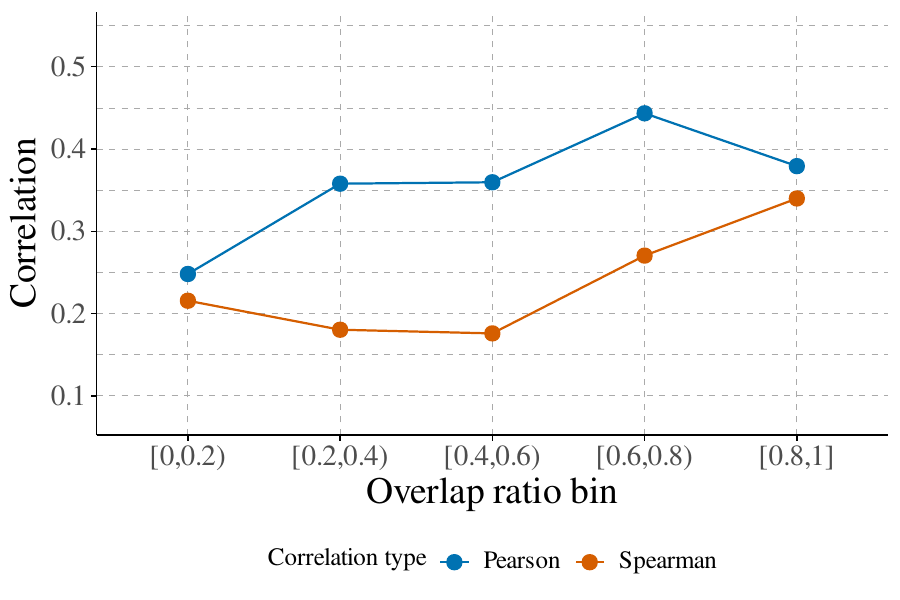}
  \caption{Orange metro line -- Direction~1 (left) and Direction~2 (right)}
  \label{fig:error_dep_2}
\end{figure}

\subsection{Model Estimation and Computation}
\label{appx:estimation}
\zcref[S]{fig:trace11_sn} and \zcref[S]{fig:trace11_st} display trace plots for the Green metro line in Direction~1 under the skew-normal and skew-$t$ specifications, covering a subset of key model parameters. These include the mean parameter $t_0$, the scale parameters $(\omega_0,\omega_1)$, the skewness parameters $(\alpha_0,\alpha_1)$, the dependence parameters $(\rho,\lambda)$, and, for the skew-$t$ model, the degrees-of-freedom parameter $\nu$. Owing to the large number of mean specification parameters, trace plots for $\theta_m$ and $\gamma_{\ell,j}$ are shown only for selected cases, namely $\theta_{\text{Guy--Concordia}}$, $\theta_{\text{Peel}}$, and $\theta_{\text{McGill}}$, as well as $\gamma_{1,\text{Peel}}$, $\gamma_{2,\text{Peel}}$, $\gamma_{1,\text{McGill}}$, and $\gamma_{2,\text{McGill}}$. Overall, the trace plots indicate good mixing and convergence across all parameters, with no evidence of divergence or convergence issues. Trace plots for the remaining parameters and for other case studies were also examined and exhibited similar convergence behavior; they are omitted here for brevity and to avoid redundancy.

\begin{figure}[!ht]
    \centering
    \includegraphics[width=0.95\linewidth]{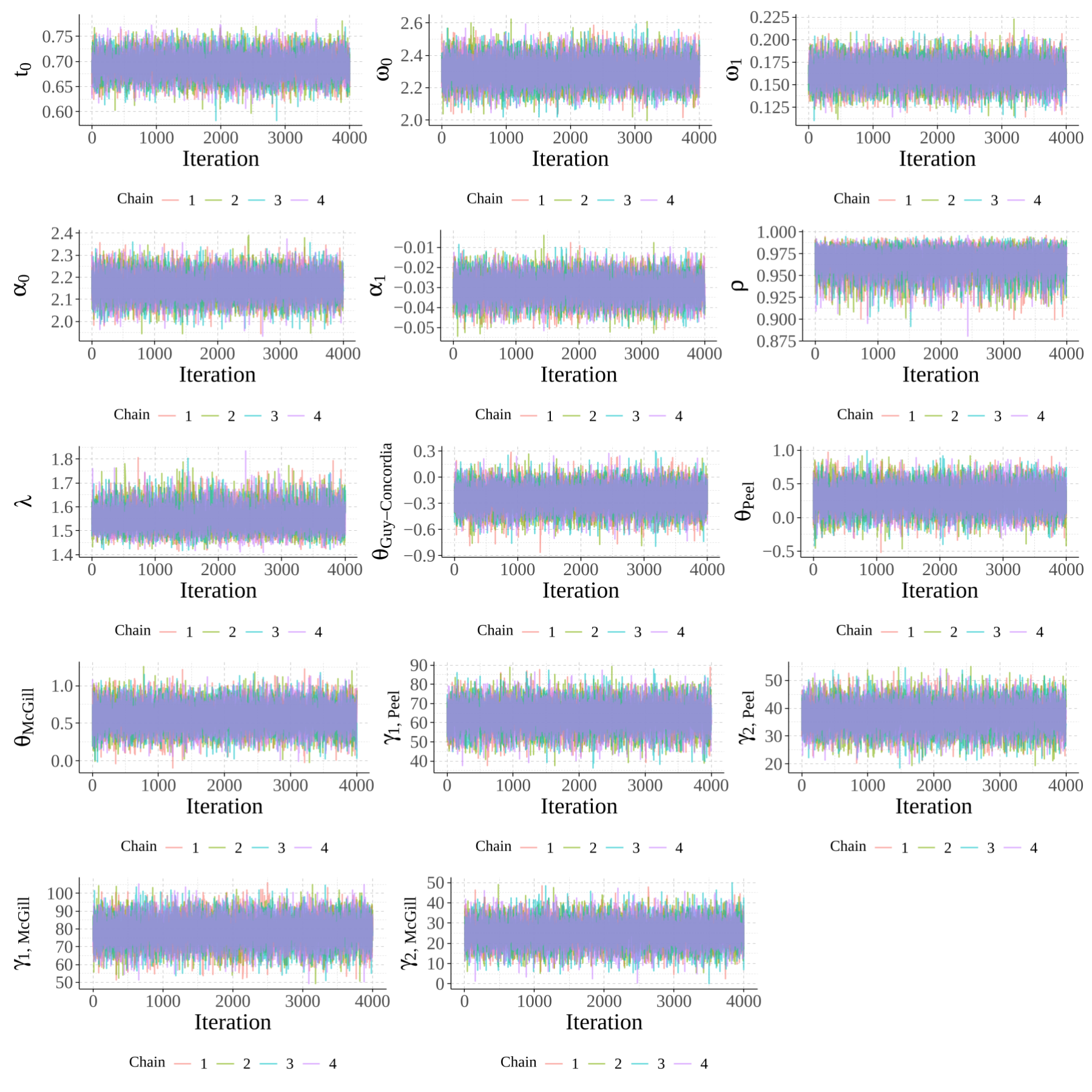}
    \caption{Trace plots for a selected subset of model parameters under the skew-normal specification.}
    \label{fig:trace11_sn}
\end{figure}

\begin{figure}[!ht]
    \centering
    \includegraphics[width=0.95\linewidth]{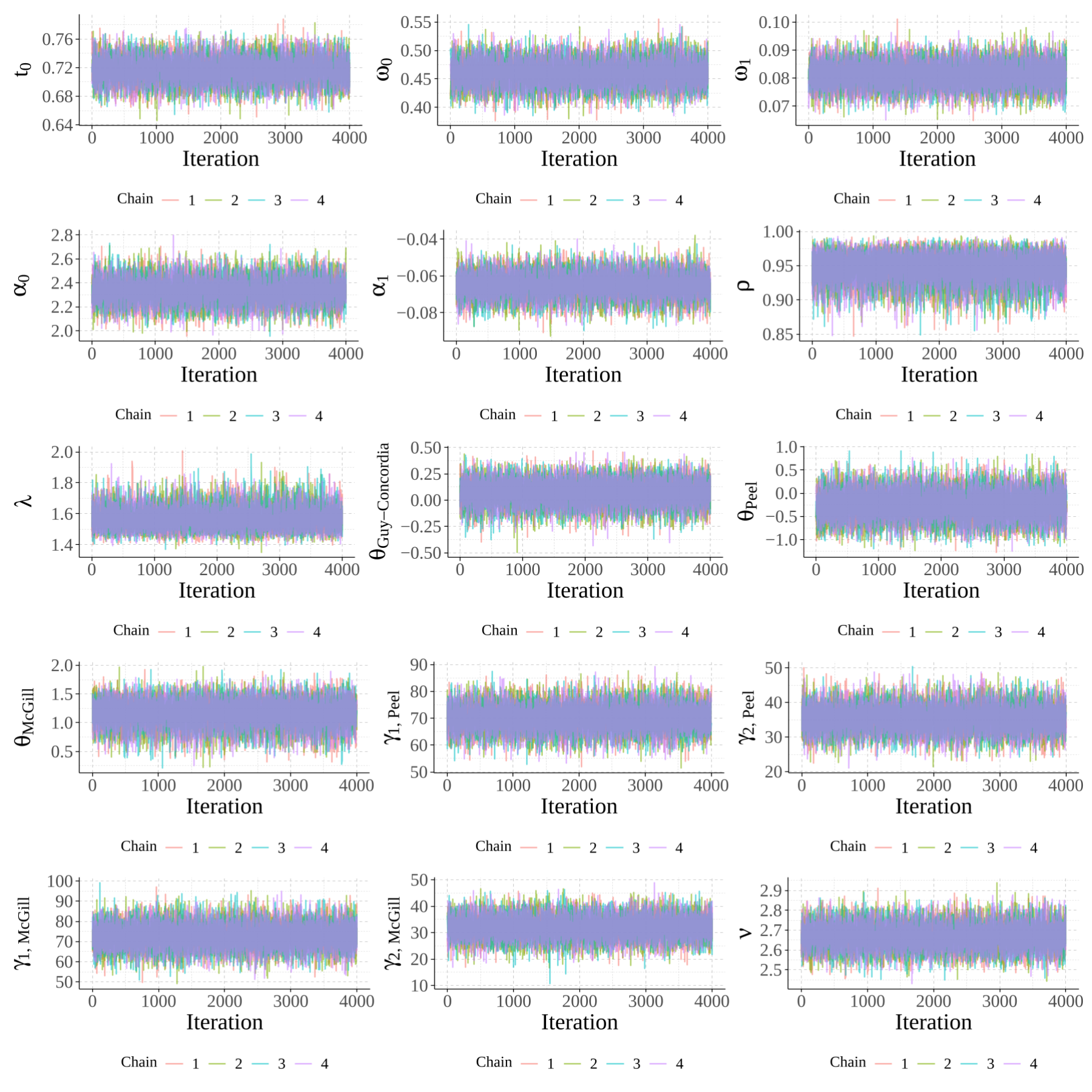}
    \caption{Trace plots for a selected subset of model parameters under the skew-$t$ specification.}
    \label{fig:trace11_st}
\end{figure}

\clearpage
\section{Results}
\subsection{\texorpdfstring{The effect of longer-than-usual headway, $\theta_m$}{The effect of longer-than-usual headway}}
Analogous to \autoref{tab:theta_11}, \autoref{tab:theta_12} reports the posterior summaries of the $\theta_m$ parameters for the skew-normal and skew-$t$ models applied to the Green line of the Montreal metro system in Direction~2. Corresponding results for the Orange metro line are presented in \autoref{tab:theta_21} for Direction~1 and in \autoref{tab:theta_22} for Direction~2. The estimates are interpreted as the change in travel time (in seconds) associated with a one-minute increase in headway. Posterior means marked with an asterisk ($\ast$) indicate strong evidence of either a positive or negative effect, as determined by the \(90\%\) credible interval. The column ``\% of median dwell time'' reports each $\theta_m$ estimate as a percentage of the median dwell time at the corresponding station. Rows shown in bold correspond to stations located in the downtown area.

\label{appx:theta}
\begin{table}[!ht]
\centering
\resizebox{\ifdim\width>\linewidth\linewidth\else\width\fi}{!}{%
\begin{tabular}[t]{rlr@{\quad(}r@{, }r@{)\quad}rr@{\quad(}r@{, }r@{)\quad}r}
\toprule
\multicolumn{2}{c}{ } & \multicolumn{4}{c}{Skew-normal} & \multicolumn{4}{c}{Skew-$t$} \\
\cmidrule(l{3pt}r{3pt}){3-6} \cmidrule(l{3pt}r{3pt}){7-10}
m & Parameter & \makecell[r]{Posterior mean\\(seconds)} & 5\% & 95\% & \makecell[r]{\% of median\\dwell time} & \makecell[r]{Posterior mean\\(seconds)} & 5\% & 95\% & \makecell[r]{\% of median\\dwell time}\\
\midrule
2 & $\theta_{\text{Radisson}}$ & $-1.1$$^{\ast}$ & $-2.0$ & $-0.3$ & $-2.93\%$ & $1.6$$^{\ast}$ & $1.0$ & $2.2$ & $4.20\%$\\
3 & $\theta_{\text{Langelier}}$ & $0.1$$^{\hphantom{\ast}}$ & $-0.6$ & $0.8$ & $0.25\%$ & $-0.7$$^{\ast}$ & $-1.1$ & $-0.2$ & $-1.71\%$\\
4 & $\theta_{\text{Cadillac}}$ & $-0.4$$^{\hphantom{\ast}}$ & $-1.0$ & $0.2$ & $-1.09\%$ & $0.4$$^{\hphantom{\ast}}$ & $0.0$ & $0.9$ & $1.06\%$\\
5 & $\theta_{\text{Assomption}}$ & $0.7$$^{\ast}$ & $0.1$ & $1.3$ & $1.83\%$ & $0.2$$^{\hphantom{\ast}}$ & $-0.1$ & $0.6$ & $0.67\%$\\
6 & $\theta_{\text{Viau}}$ & $-2.3$$^{\ast}$ & $-2.8$ & $-1.8$ & $-6.14\%$ & $-0.4$$^{\ast}$ & $-0.8$ & $-0.1$ & $-1.13\%$\\
\addlinespace
7 & $\theta_{\text{Pie-IX}}$ & $0.5$$^{\hphantom{\ast}}$ & $0.0$ & $1.0$ & $1.20\%$ & $-0.3$$^{\hphantom{\ast}}$ & $-0.7$ & $0.0$ & $-0.75\%$\\
8 & $\theta_{\text{Joliette}}$ & $-0.2$$^{\hphantom{\ast}}$ & $-0.7$ & $0.3$ & $-0.50\%$ & $0.9$$^{\ast}$ & $0.6$ & $1.3$ & $2.46\%$\\
9 & $\theta_{\text{Préfontaine}}$ & $0.8$$^{\ast}$ & $0.4$ & $1.3$ & $2.19\%$ & $0.5$$^{\ast}$ & $0.2$ & $0.8$ & $1.39\%$\\
10 & $\theta_{\text{Frontenac}}$ & $-0.6$$^{\ast}$ & $-1.2$ & $-0.1$ & $-1.66\%$ & $-0.3$$^{\hphantom{\ast}}$ & $-0.7$ & $0.0$ & $-0.79\%$\\
11 & $\theta_{\text{Papineau}}$ & $0.5$$^{\hphantom{\ast}}$ & $0.0$ & $1.0$ & $1.28\%$ & $0.6$$^{\ast}$ & $0.2$ & $0.9$ & $1.51\%$\\
\addlinespace
12 & $\theta_{\text{Beaudry}}$ & $-0.1$$^{\hphantom{\ast}}$ & $-0.6$ & $0.3$ & $-0.36\%$ & $-0.1$$^{\hphantom{\ast}}$ & $-0.4$ & $0.3$ & $-0.16\%$\\
13 & $\boldsymbol{\theta_{\textbf{Berri-UQAM}}}$ & $-0.1$$^{\hphantom{\ast}}$ & $-0.6$ & $0.4$ & $-0.19\%$ & $-0.2$$^{\hphantom{\ast}}$ & $-0.6$ & $0.2$ & $-0.39\%$\\
14 & $\boldsymbol{\theta_{\textbf{Saint-Laurent}}}$ & $0.2$$^{\hphantom{\ast}}$ & $-0.3$ & $0.7$ & $0.52\%$ & $0.3$$^{\hphantom{\ast}}$ & $-0.1$ & $0.7$ & $0.85\%$\\
15 & $\boldsymbol{\theta_{\textbf{Place-des-Arts}}}$ & $1.0$$^{\ast}$ & $0.6$ & $1.4$ & $2.36\%$ & $0.5$$^{\ast}$ & $0.1$ & $0.9$ & $1.16\%$\\
16 & $\boldsymbol{\theta_{\textbf{McGill}}}$ & $1.3$$^{\ast}$ & $0.8$ & $1.7$ & $2.94\%$ & $1.1$$^{\ast}$ & $0.7$ & $1.4$ & $2.52\%$\\
\addlinespace
17 & $\boldsymbol{\theta_{\textbf{Peel}}}$ & $-0.5$$^{\ast}$ & $-0.9$ & $0.0$ & $-1.13\%$ & $-0.5$$^{\ast}$ & $-0.8$ & $-0.1$ & $-1.08\%$\\
18 & $\boldsymbol{\theta_{\textbf{Guy-Concordia}}}$ & $0.4$$^{\ast}$ & $0.0$ & $0.9$ & $1.00\%$ & $0.5$$^{\ast}$ & $0.2$ & $0.8$ & $1.15\%$\\
19 & $\boldsymbol{\theta_{\textbf{Atwater}}}$ & $-0.5$$^{\ast}$ & $-1.0$ & $-0.1$ & $-1.24\%$ & $-0.7$$^{\ast}$ & $-1.1$ & $-0.3$ & $-1.57\%$\\
20 & $\boldsymbol{\theta_{\textbf{Lionel-Groulx}}}$ & $0.3$$^{\hphantom{\ast}}$ & $-0.2$ & $0.7$ & $0.63\%$ & $0.0$$^{\hphantom{\ast}}$ & $-0.4$ & $0.4$ & $-0.06\%$\\
21 & $\theta_{\text{Charlevoix}}$ & $0.8$$^{\ast}$ & $0.4$ & $1.2$ & $2.18\%$ & $1.2$$^{\ast}$ & $0.8$ & $1.6$ & $3.11\%$\\
\addlinespace
22 & $\theta_{\text{LaSalle}}$ & $-0.1$$^{\hphantom{\ast}}$ & $-0.5$ & $0.4$ & $-0.18\%$ & $-0.2$$^{\hphantom{\ast}}$ & $-0.6$ & $0.2$ & $-0.51\%$\\
23 & $\theta_{\text{De l’Église}}$ & $0.8$$^{\ast}$ & $0.4$ & $1.1$ & $1.90\%$ & $0.7$$^{\ast}$ & $0.3$ & $1.1$ & $1.72\%$\\
24 & $\theta_{\text{Verdun}}$ & $0.1$$^{\hphantom{\ast}}$ & $-0.1$ & $0.3$ & $0.25\%$ & $0.1$$^{\hphantom{\ast}}$ & $-0.1$ & $0.2$ & $0.22\%$\\
25 & $\theta_{\text{Jolicoeur}}$ & $-0.2$$^{\hphantom{\ast}}$ & $-0.6$ & $0.3$ & $-0.47\%$ & $-0.1$$^{\hphantom{\ast}}$ & $-0.4$ & $0.2$ & $-0.19\%$\\
26 & $\theta_{\text{Monk}}$ & $0.2$$^{\hphantom{\ast}}$ & $-0.3$ & $0.7$ & $0.42\%$ & $-0.1$$^{\hphantom{\ast}}$ & $-0.4$ & $0.2$ & $-0.23\%$\\
\bottomrule
\end{tabular}
}
    \caption{Green metro line -- Direction~2}
    \label{tab:theta_12}
\end{table}

\begin{table}[t]
\centering
\resizebox{\ifdim\width>\linewidth\linewidth\else\width\fi}{!}{%
\begin{tabular}[t]{rlr@{\quad(}r@{, }r@{)\quad}rr@{\quad(}r@{, }r@{)\quad}r}
\toprule
\multicolumn{2}{c}{ } & \multicolumn{4}{c}{Skew-normal} & \multicolumn{4}{c}{Skew-$t$} \\
\cmidrule(l{3pt}r{3pt}){3-6} \cmidrule(l{3pt}r{3pt}){7-10}
m & Parameter & \makecell[r]{Posterior mean\\(seconds)} & 5\% & 95\% & \makecell[r]{\% of median\\dwell time} & \makecell[r]{Posterior mean\\(seconds)} & 5\% & 95\% & \makecell[r]{\% of median\\dwell time}\\
\midrule
2 & $\theta_{\text{Crémazie}}$ & $-6.9$$^{\ast}$ & $-10.0$ & $-3.8$ & $-16.47\%$ & $-7.4$$^{\ast}$ & $-10.2$ & $-4.7$ & $-17.54\%$\\
3 & $\theta_{\text{Jarry}}$ & $6.0$$^{\ast}$ & $5.3$ & $6.6$ & $14.90\%$ & $4.2$$^{\ast}$ & $3.5$ & $6.4$ & $10.61\%$\\
4 & $\theta_{\text{Jean-Talon}}$ & $-2.1$$^{\ast}$ & $-3.1$ & $-1.2$ & $-4.74\%$ & $-6.0$$^{\ast}$ & $-7.1$ & $-5.0$ & $-13.31\%$\\
5 & $\theta_{\text{Beaubien}}$ & $3.8$$^{\ast}$ & $3.3$ & $4.3$ & $9.27\%$ & $2.6$$^{\ast}$ & $2.2$ & $3.1$ & $6.45\%$\\
6 & $\theta_{\text{Rosemont}}$ & $-5.3$$^{\ast}$ & $-6.1$ & $-4.4$ & $-12.91\%$ & $0.7$$^{\hphantom{\ast}}$ & $0.0$ & $1.5$ & $1.79\%$\\
\addlinespace
7 & $\theta_{\text{Laurier}}$ & $1.0$$^{\ast}$ & $0.2$ & $1.8$ & $2.45\%$ & $1.7$$^{\ast}$ & $1.0$ & $2.3$ & $4.11\%$\\
8 & $\theta_{\text{Mont-Royal}}$ & $0.4$$^{\hphantom{\ast}}$ & $-0.3$ & $1.1$ & $1.04\%$ & $0.0$$^{\hphantom{\ast}}$ & $-0.5$ & $0.5$ & $0.07\%$\\
9 & $\theta_{\text{Sherbrooke}}$ & $-1.1$$^{\ast}$ & $-1.8$ & $-0.4$ & $-2.54\%$ & $0.6$$^{\ast}$ & $0.1$ & $1.2$ & $1.49\%$\\
10 & $\boldsymbol{\theta_{\textbf{Berri-UQAM}}}$ & $-0.5$$^{\hphantom{\ast}}$ & $-1.3$ & $0.3$ & $-0.97\%$ & $0.1$$^{\hphantom{\ast}}$ & $-0.6$ & $0.7$ & $0.10\%$\\
11 & $\boldsymbol{\theta_{\textbf{Champ-de-Mars}}}$ & $1.1$$^{\ast}$ & $0.3$ & $1.9$ & $2.47\%$ & $0.5$$^{\hphantom{\ast}}$ & $-0.1$ & $1.1$ & $1.09\%$\\
\addlinespace
12 & $\boldsymbol{\theta_{\textbf{Place-d'Armes}}}$ & $-2.1$$^{\ast}$ & $-3.0$ & $-1.1$ & $-4.75\%$ & $-1.4$$^{\ast}$ & $-2.2$ & $-0.7$ & $-3.27\%$\\
13 & $\boldsymbol{\theta_{\textbf{Square-Victoria-OACI}}}$ & $1.4$$^{\ast}$ & $0.4$ & $2.3$ & $3.28\%$ & $1.5$$^{\ast}$ & $0.7$ & $2.3$ & $3.56\%$\\
14 & $\boldsymbol{\theta_{\textbf{Bonaventure}}}$ & $-1.1$$^{\ast}$ & $-2.1$ & $-0.1$ & $-2.79\%$ & $1.3$$^{\ast}$ & $0.5$ & $2.2$ & $3.43\%$\\
15 & $\boldsymbol{\theta_{\textbf{Lucien-L'Allier}}}$ & $0.7$$^{\hphantom{\ast}}$ & $-0.3$ & $1.7$ & $1.64\%$ & $-1.4$$^{\ast}$ & $-2.3$ & $-0.6$ & $-3.53\%$\\
16 & $\boldsymbol{\theta_{\textbf{Georges-Vanier}}}$ & $2.7$$^{\ast}$ & $1.8$ & $3.7$ & $7.02\%$ & $2.1$$^{\ast}$ & $1.4$ & $2.9$ & $5.47\%$\\
\addlinespace
17 & $\boldsymbol{\theta_{\textbf{Lionel-Groulx}}}$ & $0.8$$^{\hphantom{\ast}}$ & $-0.1$ & $1.6$ & $1.63\%$ & $0.8$$^{\ast}$ & $0.1$ & $1.4$ & $1.62\%$\\
18 & $\theta_{\text{Place-Saint-Henri}}$ & $2.3$$^{\ast}$ & $1.7$ & $3.0$ & $5.68\%$ & $2.6$$^{\ast}$ & $2.0$ & $3.1$ & $6.27\%$\\
19 & $\theta_{\text{Vendôme}}$ & $0.3$$^{\hphantom{\ast}}$ & $-0.2$ & $0.9$ & $0.77\%$ & $0.1$$^{\hphantom{\ast}}$ & $-0.3$ & $0.4$ & $0.15\%$\\
20 & $\theta_{\text{Villa-Maria}}$ & $0.2$$^{\hphantom{\ast}}$ & $-0.4$ & $0.7$ & $0.43\%$ & $0.2$$^{\hphantom{\ast}}$ & $-0.2$ & $0.5$ & $0.39\%$\\
21 & $\theta_{\text{Snowdon}}$ & $1.1$$^{\ast}$ & $0.5$ & $1.7$ & $2.47\%$ & $0.7$$^{\ast}$ & $0.4$ & $1.1$ & $1.62\%$\\
\addlinespace
22 & $\theta_{\text{Côte-Sainte-Catherine}}$ & $-0.6$$^{\hphantom{\ast}}$ & $-1.2$ & $0.0$ & $-1.51\%$ & $-0.8$$^{\ast}$ & $-1.2$ & $-0.4$ & $-1.89\%$\\
23 & $\theta_{\text{Plamondon}}$ & $1.0$$^{\ast}$ & $0.2$ & $1.7$ & $2.22\%$ & $0.9$$^{\ast}$ & $0.4$ & $1.4$ & $2.13\%$\\
24 & $\theta_{\text{Namur}}$ & $0.8$$^{\ast}$ & $0.0$ & $1.5$ & $1.96\%$ & $0.3$$^{\hphantom{\ast}}$ & $-0.2$ & $0.8$ & $0.74\%$\\
25 & $\theta_{\text{De La Savane}}$ & $0.3$$^{\hphantom{\ast}}$ & $-0.5$ & $1.1$ & $0.79\%$ & $0.1$$^{\hphantom{\ast}}$ & $-0.4$ & $0.7$ & $0.34\%$\\
\bottomrule
\end{tabular}

}
    \caption{Orange metro line -- Direction~1}
    \label{tab:theta_21}
\end{table}

\clearpage
\begin{table}[t]
\centering
\resizebox{\ifdim\width>\linewidth\linewidth\else\width\fi}{!}{%
\begin{tabular}[t]{rlr@{\quad(}r@{, }r@{)\quad}rr@{\quad(}r@{, }r@{)\quad}r}
\toprule
\multicolumn{2}{c}{ } & \multicolumn{4}{c}{Skew-normal} & \multicolumn{4}{c}{Skew-$t$} \\
\cmidrule(l{3pt}r{3pt}){3-6} \cmidrule(l{3pt}r{3pt}){7-10}
m & Parameter & \makecell[r]{Posterior mean\\(seconds)} & 5\% & 95\% & \makecell[r]{\% of median\\dwell time} & \makecell[r]{Posterior mean\\(seconds)} & 5\% & 95\% & \makecell[r]{\% of median\\dwell time}\\
\midrule
2 & $\theta_{\text{De La Savane}}$ & $-1.8$$^{\hphantom{\ast}}$ & $-6.5$ & $3.6$ & $-4.50\%$ & $-3.0$$^{\hphantom{\ast}}$ & $-6.8$ & $2.6$ & $-7.56\%$\\
3 & $\theta_{\text{Namur}}$ & $0.3$$^{\hphantom{\ast}}$ & $-1.4$ & $2.0$ & $0.81\%$ & $1.1$$^{\hphantom{\ast}}$ & $-0.3$ & $2.4$ & $2.64\%$\\
4 & $\theta_{\text{Plamondon}}$ & $-2.0$$^{\ast}$ & $-3.6$ & $-0.3$ & $-4.69\%$ & $-0.1$$^{\hphantom{\ast}}$ & $-1.2$ & $1.2$ & $-0.13\%$\\
5 & $\theta_{\text{Côte-Sainte-Catherine}}$ & $4.1$$^{\ast}$ & $3.4$ & $4.9$ & $11.85\%$ & $3.3$$^{\ast}$ & $2.6$ & $4.0$ & $9.41\%$\\
6 & $\theta_{\text{Snowdon}}$ & $1.7$$^{\ast}$ & $0.8$ & $2.5$ & $3.77\%$ & $2.0$$^{\ast}$ & $1.2$ & $2.8$ & $4.48\%$\\
\addlinespace
7 & $\theta_{\text{Villa-Maria}}$ & $-0.2$$^{\hphantom{\ast}}$ & $-0.9$ & $0.6$ & $-0.36\%$ & $0.0$$^{\hphantom{\ast}}$ & $-0.8$ & $0.7$ & $-0.05\%$\\
8 & $\theta_{\text{Vendôme}}$ & $0.3$$^{\hphantom{\ast}}$ & $-0.3$ & $0.9$ & $0.60\%$ & $0.0$$^{\hphantom{\ast}}$ & $-0.6$ & $0.6$ & $-0.01\%$\\
9 & $\theta_{\text{Place-Saint-Henri}}$ & $0.3$$^{\hphantom{\ast}}$ & $-0.4$ & $0.9$ & $0.62\%$ & $-0.6$$^{\hphantom{\ast}}$ & $-1.2$ & $0.1$ & $-1.40\%$\\
10 & $\boldsymbol{\theta_{\textbf{Lionel-Groulx}}}$ & $0.7$$^{\ast}$ & $0.0$ & $1.3$ & $1.42\%$ & $2.0$$^{\ast}$ & $1.4$ & $2.7$ & $4.19\%$\\
11 & $\boldsymbol{\theta_{\textbf{Georges-Vanier}}}$ & $-1.3$$^{\ast}$ & $-1.9$ & $-0.7$ & $-3.20\%$ & $-2.0$$^{\ast}$ & $-2.5$ & $-1.5$ & $-4.97\%$\\
\addlinespace
12 & $\boldsymbol{\theta_{\textbf{Lucien-L'Allier}}}$ & $2.2$$^{\ast}$ & $1.7$ & $2.7$ & $5.35\%$ & $3.0$$^{\ast}$ & $2.7$ & $3.4$ & $7.41\%$\\
13 & $\boldsymbol{\theta_{\textbf{Bonaventure}}}$ & $-0.8$$^{\ast}$ & $-1.2$ & $-0.4$ & $-1.82\%$ & $-0.3$$^{\hphantom{\ast}}$ & $-0.7$ & $0.1$ & $-0.66\%$\\
14 & $\boldsymbol{\theta_{\textbf{Square-Victoria-OACI}}}$ & $-0.3$$^{\hphantom{\ast}}$ & $-0.8$ & $0.3$ & $-0.67\%$ & $-0.4$$^{\hphantom{\ast}}$ & $-0.9$ & $0.1$ & $-0.98\%$\\
15 & $\boldsymbol{\theta_{\textbf{Place-d'Armes}}}$ & $1.3$$^{\ast}$ & $0.7$ & $1.9$ & $3.04\%$ & $0.7$$^{\ast}$ & $0.3$ & $1.2$ & $1.72\%$\\
16 & $\boldsymbol{\theta_{\textbf{Champ-de-Mars}}}$ & $-1.4$$^{\ast}$ & $-2.1$ & $-0.8$ & $-3.33\%$ & $-0.9$$^{\ast}$ & $-1.4$ & $-0.4$ & $-2.05\%$\\
\addlinespace
17 & $\boldsymbol{\theta_{\textbf{Berri-UQAM}}}$ & $0.1$$^{\hphantom{\ast}}$ & $-0.3$ & $0.5$ & $0.17\%$ & $0.3$$^{\hphantom{\ast}}$ & $0.0$ & $0.6$ & $0.53\%$\\
18 & $\theta_{\text{Sherbrooke}}$ & $1.7$$^{\ast}$ & $1.2$ & $2.3$ & $3.98\%$ & $1.4$$^{\ast}$ & $0.9$ & $1.8$ & $3.19\%$\\
19 & $\theta_{\text{Mont-Royal}}$ & $-0.6$$^{\ast}$ & $-1.1$ & $0.0$ & $-1.30\%$ & $-0.4$$^{\hphantom{\ast}}$ & $-0.9$ & $0.1$ & $-0.92\%$\\
20 & $\theta_{\text{Laurier}}$ & $0.8$$^{\ast}$ & $0.1$ & $1.4$ & $1.75\%$ & $0.5$$^{\hphantom{\ast}}$ & $0.0$ & $1.1$ & $1.20\%$\\
21 & $\theta_{\text{Rosemont}}$ & $0.6$$^{\hphantom{\ast}}$ & $-0.1$ & $1.3$ & $1.39\%$ & $0.6$$^{\ast}$ & $0.1$ & $1.2$ & $1.56\%$\\
\addlinespace
22 & $\theta_{\text{Beaubien}}$ & $0.5$$^{\hphantom{\ast}}$ & $0.0$ & $1.1$ & $1.30\%$ & $0.1$$^{\hphantom{\ast}}$ & $-0.4$ & $0.7$ & $0.36\%$\\
23 & $\theta_{\text{Jean-Talon}}$ & $-0.3$$^{\hphantom{\ast}}$ & $-0.9$ & $0.3$ & $-0.65\%$ & $0.0$$^{\hphantom{\ast}}$ & $-0.7$ & $0.8$ & $-0.04\%$\\
24 & $\theta_{\text{Jarry}}$ & $0.2$$^{\hphantom{\ast}}$ & $-0.4$ & $0.8$ & $0.55\%$ & $0.8$$^{\ast}$ & $0.1$ & $1.4$ & $2.06\%$\\
25 & $\theta_{\text{Crémazie}}$ & $1.6$$^{\ast}$ & $0.8$ & $2.4$ & $3.75\%$ & $0.7$$^{\hphantom{\ast}}$ & $-0.1$ & $1.7$ & $1.71\%$\\
\bottomrule
\end{tabular}
}
    \caption{Orange metro line -- Direction~2}
    \label{tab:theta_22}
\end{table}

\clearpage
\subsection{\texorpdfstring{The effect of train formation parameters, $\gamma_{\ell, j}$}{The effect of train formation parameters}}
\label{appx:gamma}

\begin{figure}[!ht]
    \centering
    \includegraphics[width=0.9\linewidth]{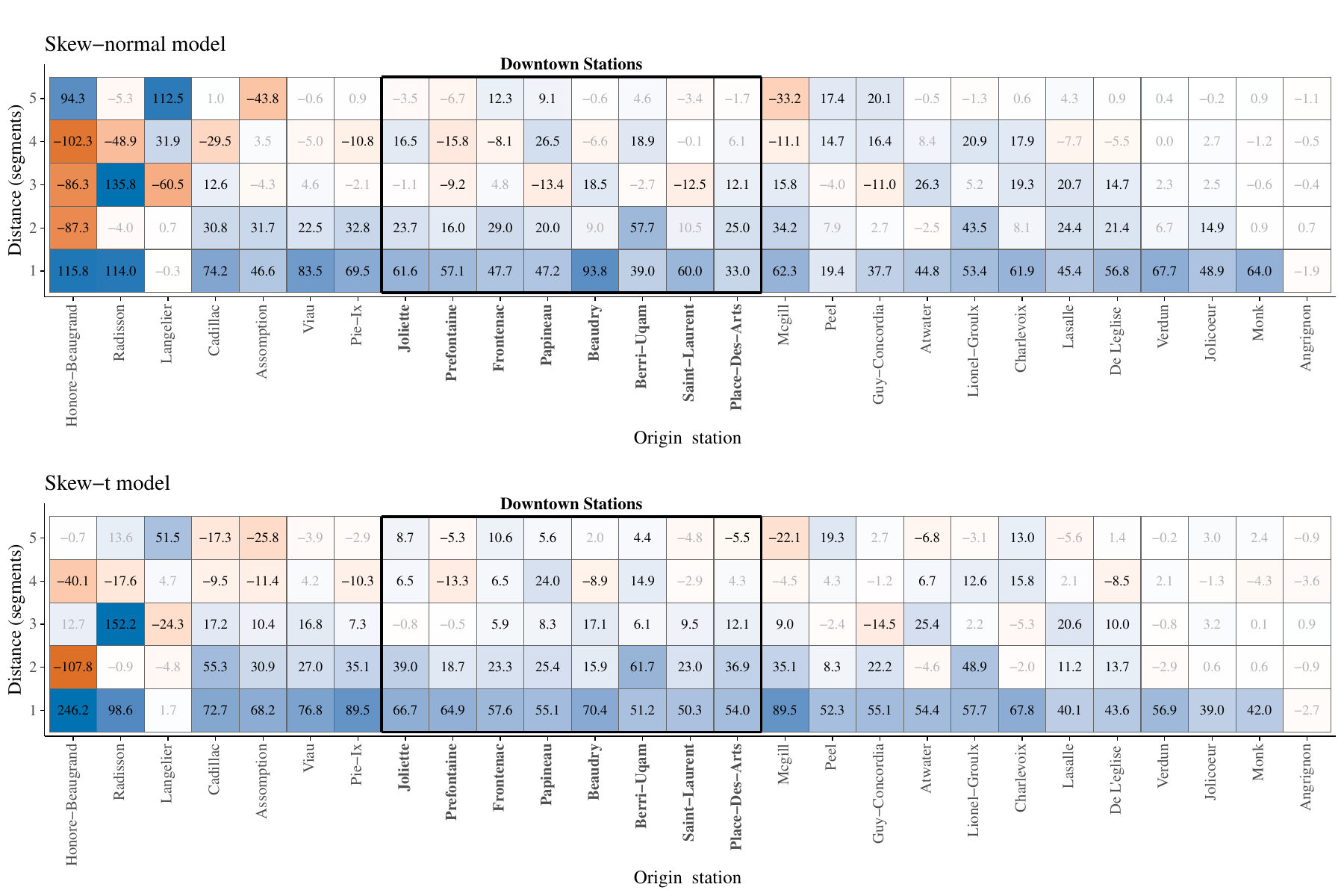}
    \caption{Green metro line -- Direction~2}
    \label{fig:gamma_12}
\end{figure}

\begin{figure}[!ht]
    \centering
    \includegraphics[width=0.9\linewidth]{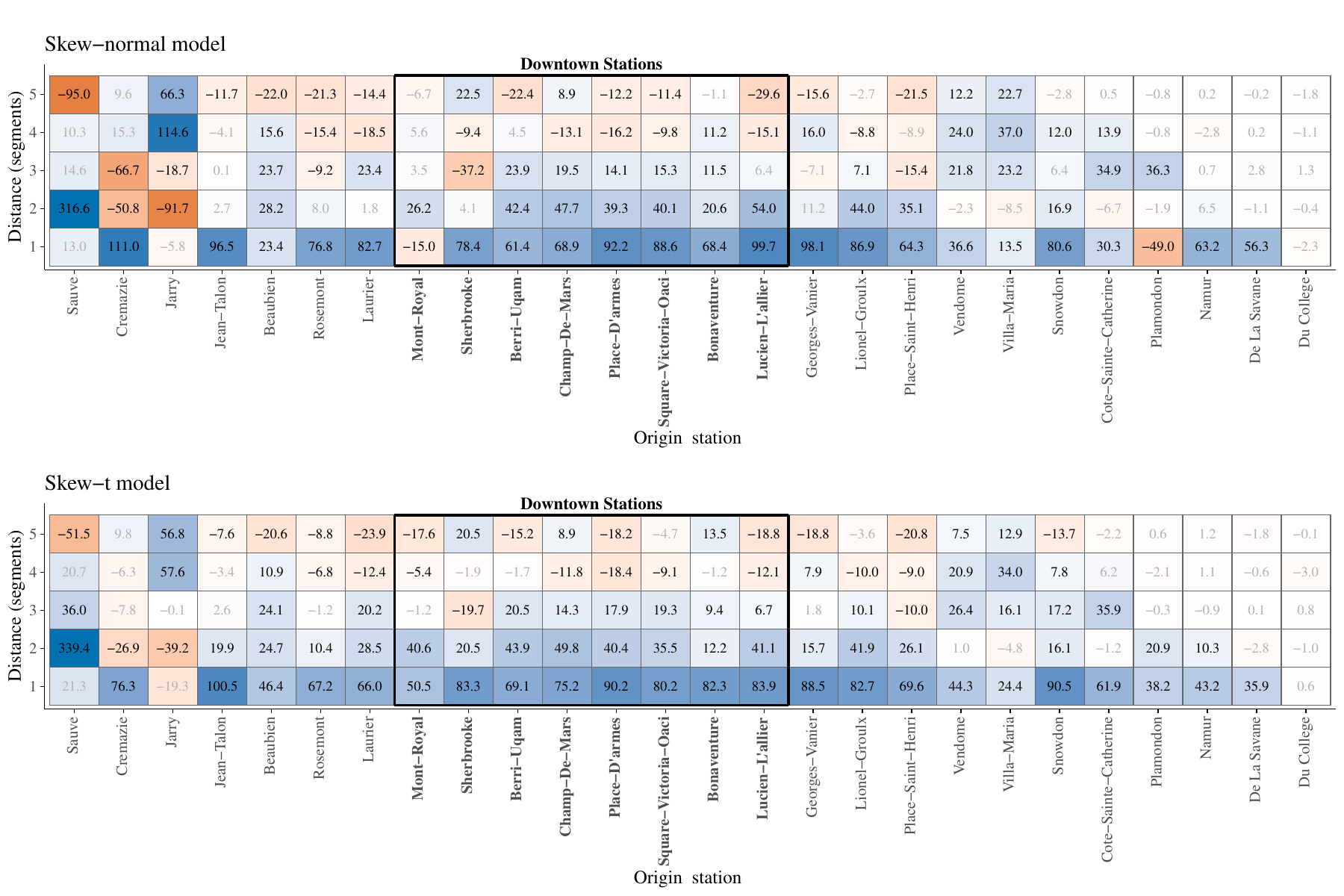}
    \caption{Orange metro line -- Direction~1}
    \label{fig:gamma_21}
\end{figure}

\begin{figure}[!ht]
    \centering
    \includegraphics[width=0.9\linewidth]{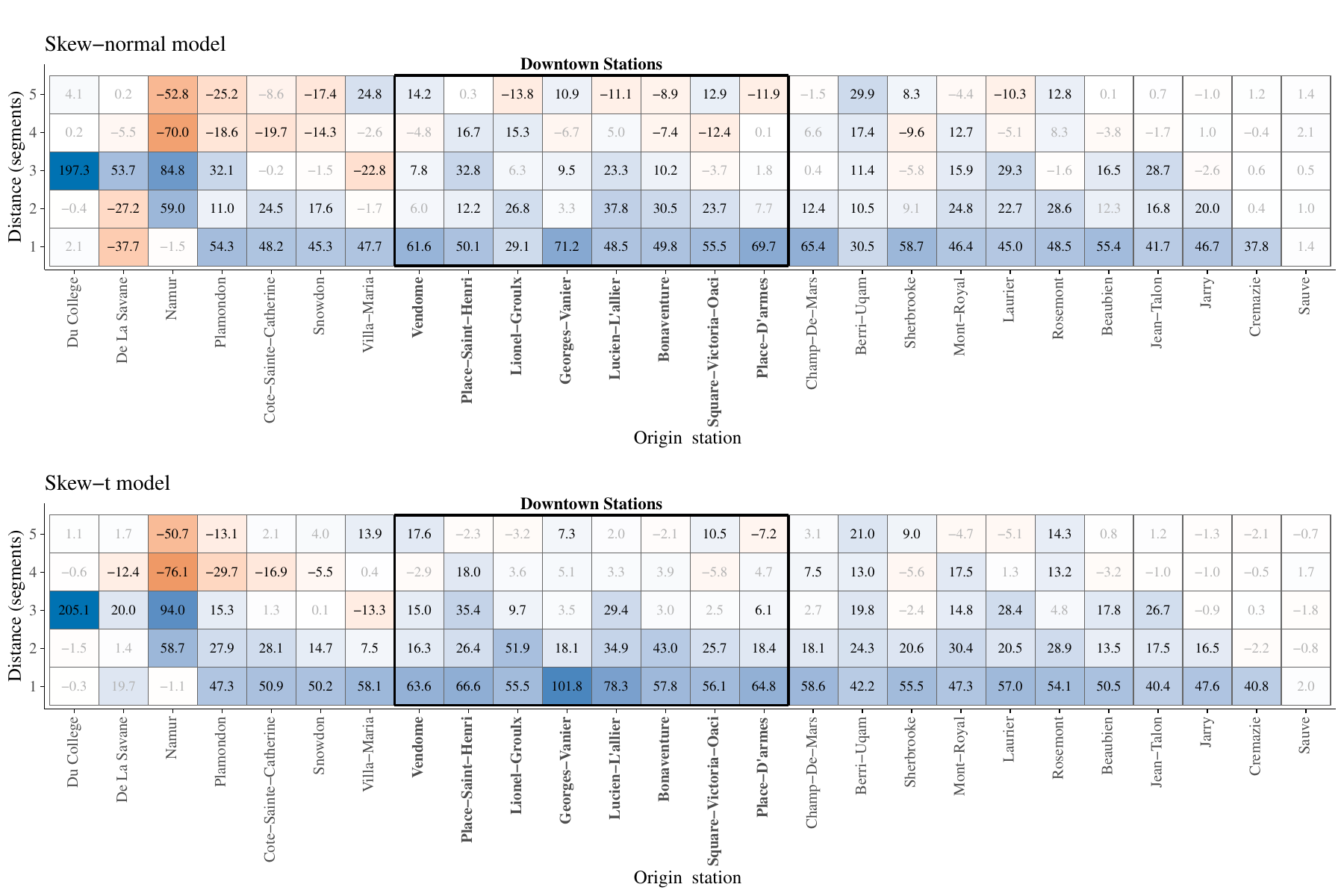}
    \caption{Orange metro line -- Direction~2}
    \label{fig:gamma_22}
\end{figure}

Posterior means of the train formation parameters $\gamma_{\ell,j}$ under post-disruption operations are presented for both the skew-normal and skew-$t$ models in \autoref{fig:gamma_12} for Direction~2 of the Green line, and in \autoref{fig:gamma_21} and \autoref{fig:gamma_22} for Directions~1 and~2 of the Orange line, respectively. As in \autoref{fig:gamma_11}, the results reveal substantial variation in the induced delay effects across both distance and location. Consistent with earlier findings, the skew-normal estimates display more abrupt changes across distances and stations, which may again be attributed to the presence of outliers in the data—features that are more effectively accommodated by the heavier-tailed skew-$t$ distribution.

\begin{figure}[!htbp]
    \centering
    \includegraphics[width=1\linewidth]{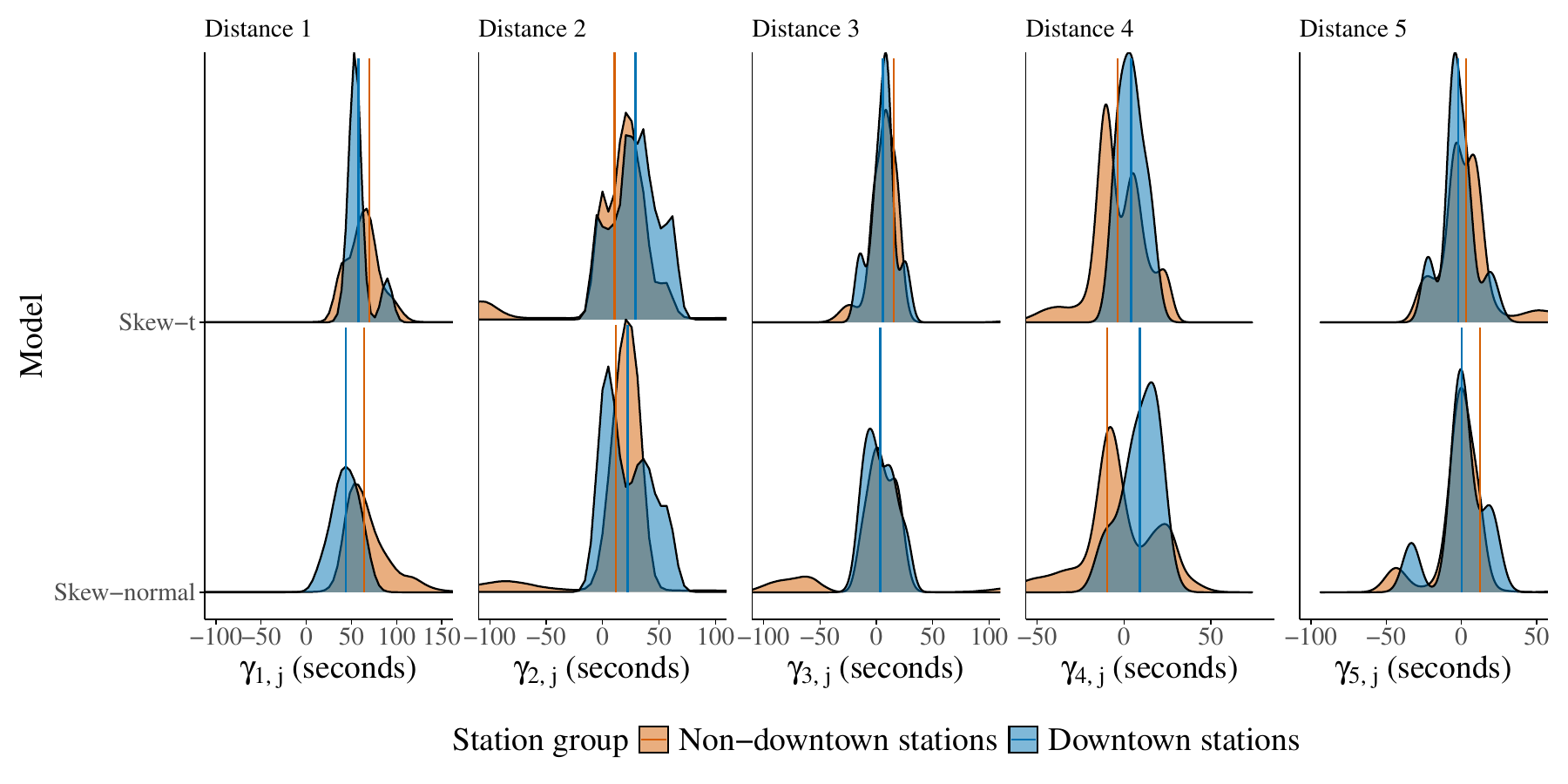}
    \caption{Green metro line -- Direction~2}
    \label{fig:ridge12}
\end{figure}

\begin{figure}[!htbp]
    \centering
    \includegraphics[width=1\linewidth]{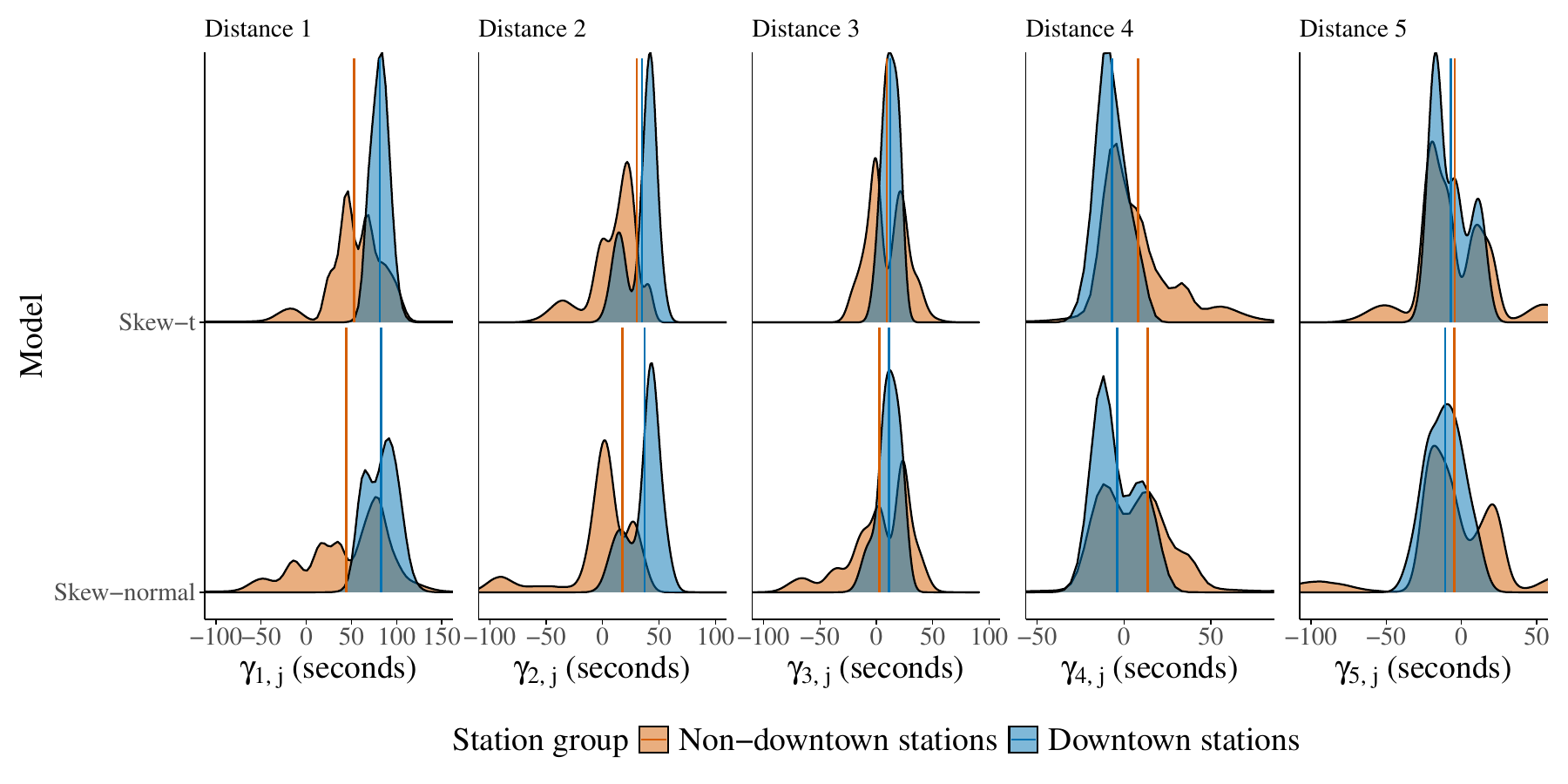}
    \caption{Orange metro line -- Direction~1}
    \label{fig:ridge21}
\end{figure}

\begin{figure}[!htbp]
    \centering
    \includegraphics[width=1\linewidth]{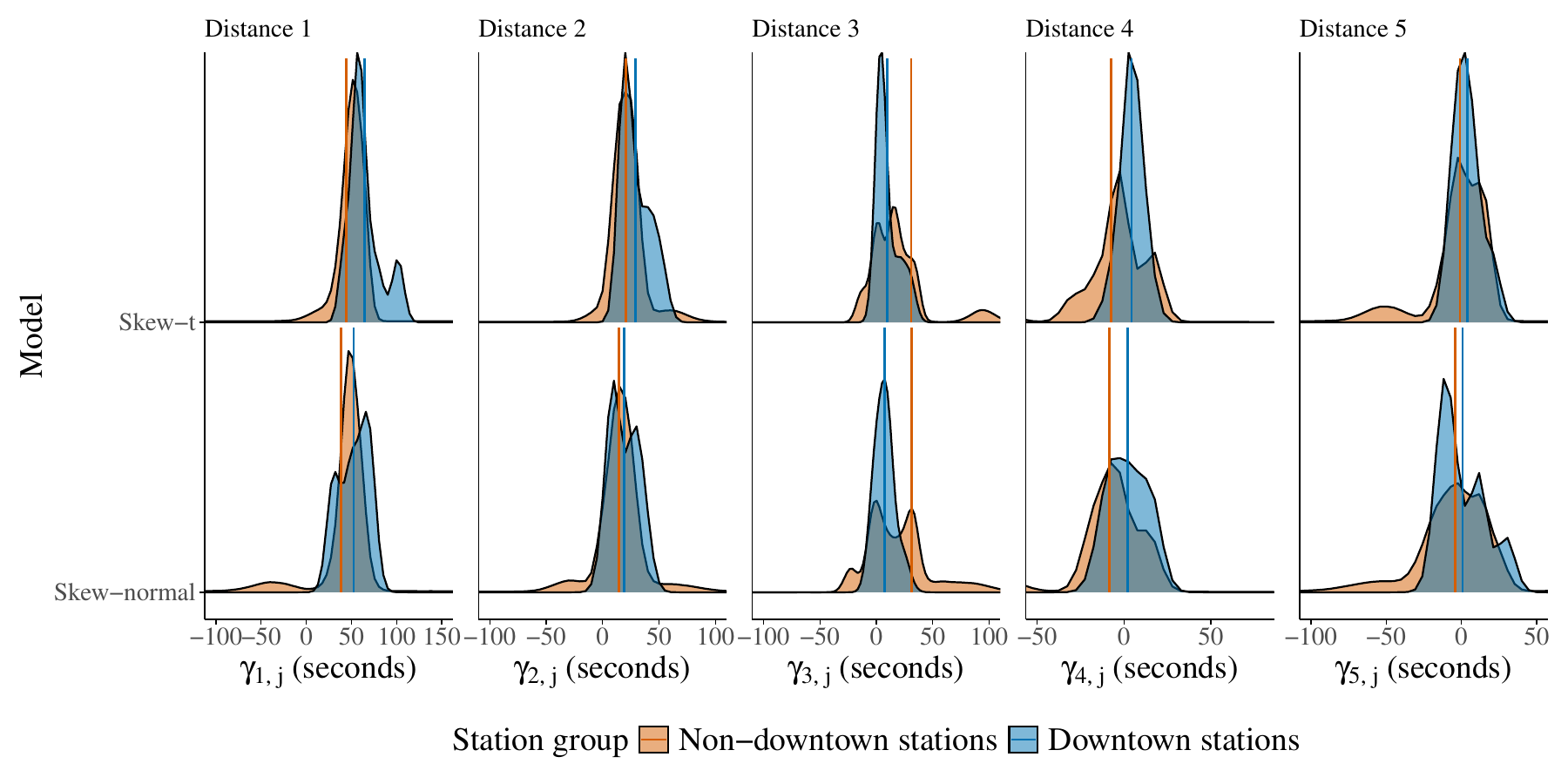}
        \caption{Orange metro line -- Direction~2}
    \label{fig:ridge22}
\end{figure}

\clearpage
The ridge plots of the train formation parameters $\gamma_{\ell,j}$ for Direction~2 of the Green metro line and both directions of the Orange line are shown in Figures \ref{fig:ridge12}, \ref{fig:ridge21}, and \ref{fig:ridge22}. The densities of the posterior samples under both proposed models and for two station groups, namely downtown and non-downtown stations are shown. For each distance, the corresponding density represents the posterior samples of $\gamma_{\ell,j}$, where $j$ belongs to the set of station indices within the respective group. The vertical lines indicate the group-specific mean effect for each model, with colors distinguishing the station groups.

\clearpage

\subsection{Distributional Parameters}
\label{appx:distparam}
This section reports the estimated distributional parameters for Direction~2 of the Green metro line (\autoref{tab:distparam12}), Direction~1 of the Orange metro line (\autoref{tab:distparam21}), and Direction~2 of the Orange metro line (\autoref{tab:distparam22}). A comparison of the estimated degrees-of-freedom parameters in the skew-$t$ models across all cases indicates pronounced heavy-tailed behavior in post-disruption travel times, as the estimated values of $\nu$ remain small throughout. 
 
\begin{table}[!htbp]
    \centering
    \tablesize
\begin{tabular}[t]{lrr@{}rrr@{}r}
\toprule
\multicolumn{1}{c}{ } & \multicolumn{3}{c}{Skew-normal} & \multicolumn{3}{c}{Skew-t} \\
\cmidrule(r{3pt}l{3pt}){2-4} \cmidrule(l{3pt}r{3pt}){5-7}
Parameter & \makecell[r]{Posterior mean} & (5\%, & 95\%) &
\makecell[r]{Posterior mean} & (5\%, & 95\%) \\
\midrule
$\omega_0$ & 2.875 & (\;2.744, & 3.008) & 0.354 & (\;0.326, & 0.384) \\
$\omega_1$ & 0.176 & (\;0.157, & 0.196) & 0.084 & (\;0.078, & 0.090) \\
\addlinespace
$\alpha_0$ & 3.990 & (\;3.826, & 4.162) & 2.183 & (\;2.027, & 2.342) \\
$\alpha_1$ & $-$0.229 & ($-$0.244, & $-$0.214) & $-$0.067 & ($-$0.077, & $-$0.057) \\
\addlinespace
$\nu$      &        &           &        & 2.039 & (\;2.003, & 2.094) \\
\bottomrule
\end{tabular}
    \caption{Posterior summaries of distributional parameters in our proposed skew-normal and skew-$t$ models for the Green metro line in Direction~2.}
    \label{tab:distparam12}
\end{table}
\begin{table}[!htbp]
\centering
    \tablesize
\begin{tabular}[t]{lrr@{}rrr@{}r}
\toprule
\multicolumn{1}{c}{ } & \multicolumn{3}{c}{Skew-normal} & \multicolumn{3}{c}{Skew-t} \\
\cmidrule(r{3pt}l{3pt}){2-4} \cmidrule(l{3pt}r{3pt}){5-7}
Parameter & \makecell[r]{Posterior mean} & (5\%, & 95\%) &
\makecell[r]{Posterior mean} & (5\%, & 95\%) \\
\midrule
$\omega_0$ & 2.260 & (\;2.149, & 2.373) & 0.332 & (\;0.299, & 0.366) \\
$\omega_1$ & 0.029 & (\;0.015, & 0.043) & 0.037 & (\;0.031, & 0.043) \\
\addlinespace
$\alpha_0$ & 2.293 & (\;2.174, & 2.416) & 1.442 & (\;1.275, & 1.614) \\
$\alpha_1$ & $-$0.086 & ($-$0.097, & $-$0.075) & $-$0.037 & ($-$0.050, & $-$0.024) \\
\addlinespace
$\nu$      &        &           &        & 2.485 & (\;2.375, & 2.601) \\
\bottomrule
\end{tabular}
    \caption{Posterior summaries of distributional parameters in our proposed skew-normal and skew-$t$ models for the Orange metro line in Direction~1.}
    \label{tab:distparam21}
\end{table}
\begin{table}[!htbp]
\centering
    \tablesize
\begin{tabular}[t]{lrr@{}rrr@{}r}
\toprule
\multicolumn{1}{c}{ } & \multicolumn{3}{c}{Skew-normal} & \multicolumn{3}{c}{Skew-t} \\
\cmidrule(r{3pt}l{3pt}){2-4} \cmidrule(l{3pt}r{3pt}){5-7}
Parameter & \makecell[r]{Posterior mean} & (5\%, & 95\%) &
\makecell[r]{Posterior mean} & (5\%, & 95\%) \\
\midrule
$\omega_0$ & 1.613 & (\;1.524, & 1.703) & 0.387 & (\;0.346, & 0.430) \\
$\omega_1$ & 0.032 & (\;0.020, & 0.046) & 0.032 & (\;0.025, & 0.039) \\
\addlinespace
$\alpha_0$ & 2.253 & (\;2.116, & 2.390) & 1.509 & (\;1.355, & 1.671) \\
$\alpha_1$ & -0.118 & ($-$0.134, & $-$0.100) & $-$0.054 & ($-$0.070, & $-$0.039) \\
\addlinespace
$\nu$      &        &           &        & 2.720 & (\;2.579, & 2.867) \\
\bottomrule
\end{tabular}
    \caption{Posterior summaries of distributional parameters in our proposed skew-normal and skew-$t$ models for the Orange metro line in Direction~2.}
    \label{tab:distparam22}
\end{table}

\begin{figure}[!ht]
    \centering
    \includegraphics[width=0.9\linewidth]{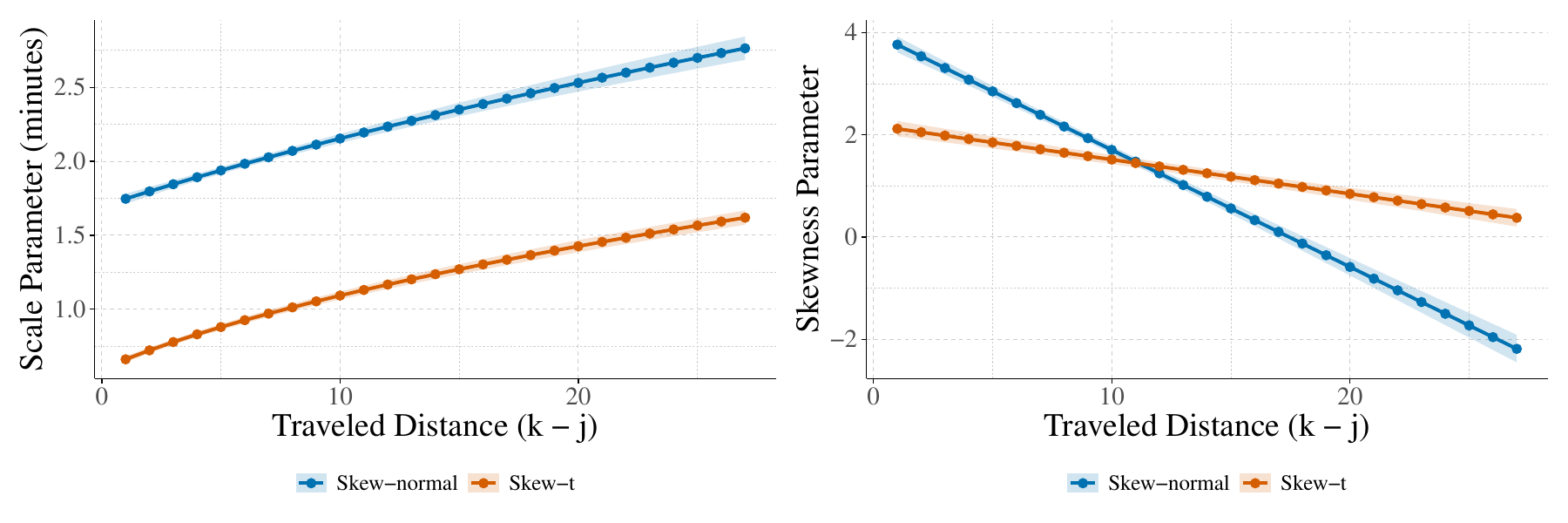}
    \caption{Green metro line -- Direction~2}
    \label{fig:sc_sk12}
\end{figure}

\begin{figure}[!ht]
    \centering
    \includegraphics[width=0.9\linewidth]{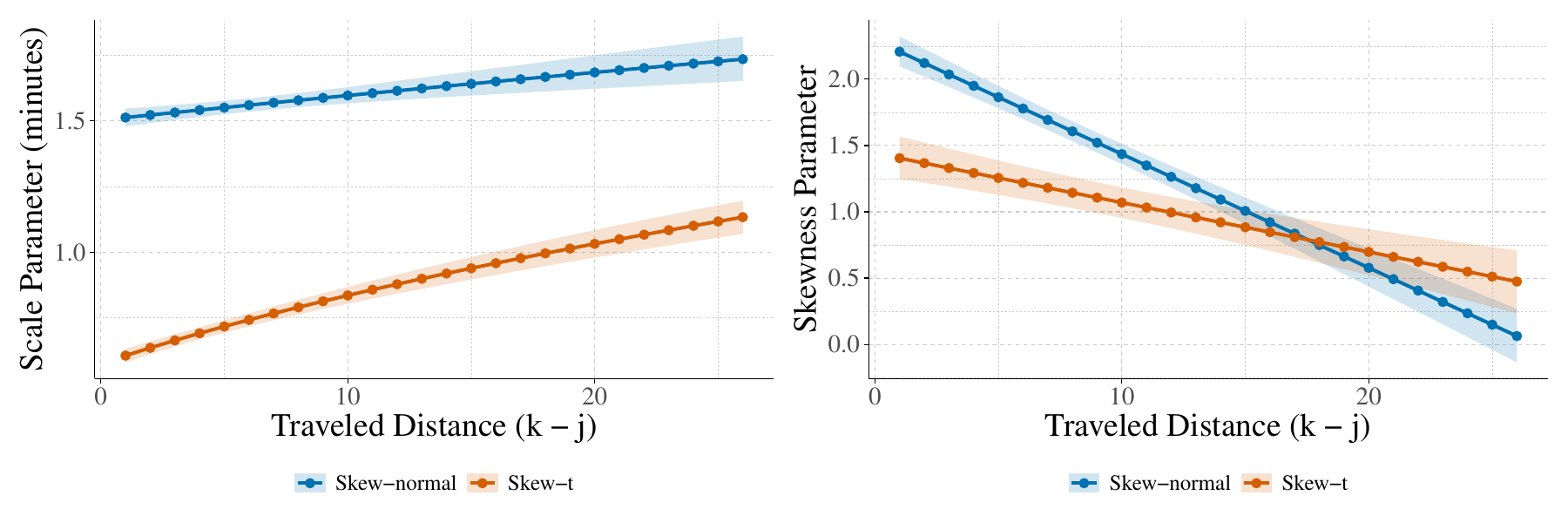}
    \caption{Orange metro line -- Direction~1}
    \label{fig:sc_sk21}
\end{figure}

\begin{figure}[!ht]
    \centering
    \includegraphics[width=0.9\linewidth]{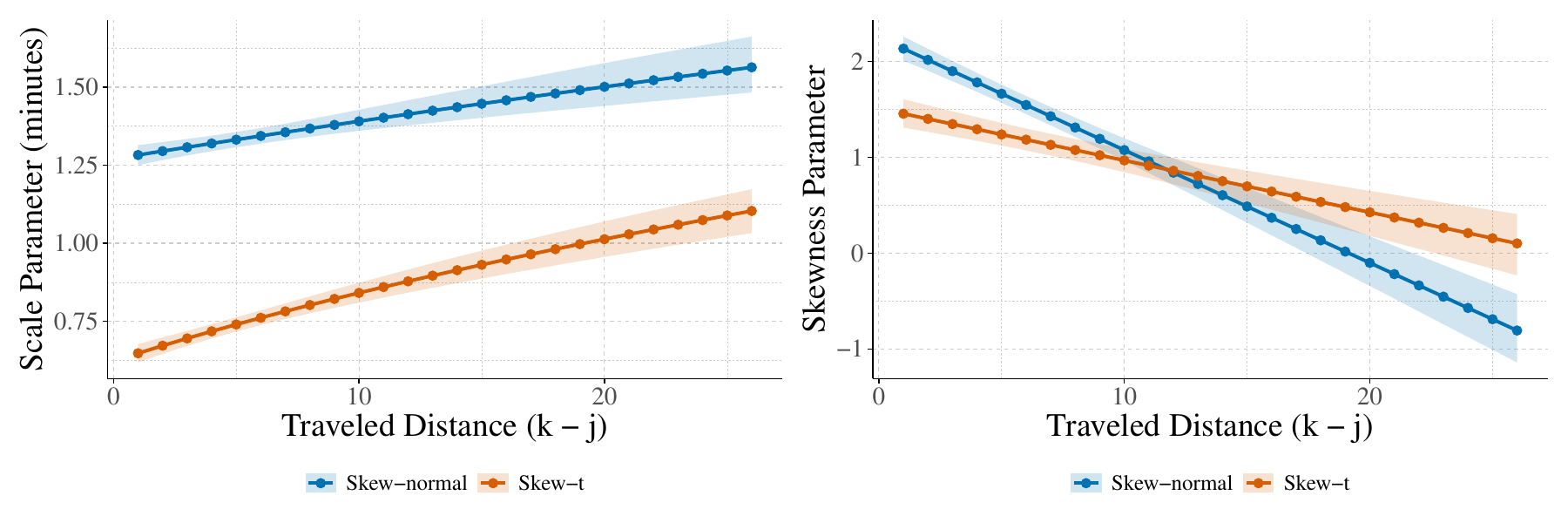}
    \caption{Orange metro line -- Direction~2}
    \label{fig:sc_sk22}
\end{figure}

In a manner analogous to \autoref{fig:sc_sk11}, Figures~\ref{fig:sc_sk12}, \ref{fig:sc_sk21}, and \ref{fig:sc_sk22} present the estimated scale (left panels) and skewness (right panels) parameters, with the corresponding 90\% credible intervals, for the skew-normal and skew-$t$ models across the remaining case studies considered in this work. Across all cases, the estimated scale parameter is consistently smaller under the skew-$t$ specification than under the skew-normal model. In addition, the skewness parameter exhibits a decreasing pattern as traveled distance increases, with the rate of decrease being steeper for the skew-normal models relative to the skew-$t$ models.

\clearpage
\subsection{Error dependence parameters}
Tables~\ref{tab:errdep12}, \ref{tab:errdep21}, and \ref{tab:errdep22} report the posterior summaries of the error-dependence parameters, $\rho$ and $\lambda$, for the proposed skew-normal and skew-$t$ models across the remaining case studies. When these results are compared with those in \autoref{tab:errdep11}, the estimated dependence parameters are broadly similar across the two proposed model specifications. Nevertheless, the degree of post-disruption error dependence varies noticeably across metro lines and directions.

A comparison between \autoref{tab:errdep11} and \autoref{tab:errdep12} indicates that Direction~2 of the Green line exhibits a faster decay in dependence with increasing journey overlap between consecutive trains, while the magnitude of dependence $\rho$ remains comparable. For the Orange line, Direction~1 is characterized by a larger decay rate $\lambda$ and a smaller dependence magnitude $\rho$ relative to Direction~2.

\begin{table}[!htbp]
    \centering
    \tablesize
\begin{tabular}[t]{lrr@{}rrr@{}r}
\toprule
\multicolumn{1}{c}{ } & \multicolumn{3}{c}{Skew-normal} & \multicolumn{3}{c}{Skew-t} \\
\cmidrule(r{3pt}l{3pt}){2-4} \cmidrule(l{3pt}r{3pt}){5-7}
Parameter & \makecell[r]{Posterior mean} & (5\%, & 95\%) &
\makecell[r]{Posterior mean} & (5\%, & 95\%) \\
\midrule
$\rho$ & 0.943 & (0.903, &\;0.975) & 0.952 & (0.917, &\;0.979) \\
$\lambda$ & 1.722 & (1.600, &\;1.875) & 1.745 & (1.634, &\;1.887) \\
\bottomrule
\end{tabular}
    \caption{Green metro line - Direction~2.}
    \label{tab:errdep12}
\end{table}
\begin{table}[!htbp]
    \centering
    \tablesize
\begin{tabular}[t]{lrr@{}rrr@{}r}
\toprule
\multicolumn{1}{c}{ } & \multicolumn{3}{c}{Skew-normal} & \multicolumn{3}{c}{Skew-t} \\
\cmidrule(r{3pt}l{3pt}){2-4} \cmidrule(l{3pt}r{3pt}){5-7}
Parameter & \makecell[r]{Posterior mean} & (5\%, & 95\%) &
\makecell[r]{Posterior mean} & (5\%, & 95\%) \\
\midrule
$\rho$ & 0.791 & (0.735, &\;0.857) & 0.808 & (0.742, &\;0.882) \\
$\lambda$ & 2.112 & (1.763, &\;2.491) & 2.039 & (1.666, &\;2.471) \\
\bottomrule
\end{tabular}
    \caption{Orange metro line - Direction~1}
    \label{tab:errdep21}
\end{table}
\begin{table}[!htbp]
    \centering
    \tablesize
\begin{tabular}[t]{lrr@{}rrr@{}r}
\toprule
\multicolumn{1}{c}{ } & \multicolumn{3}{c}{Skew-normal} & \multicolumn{3}{c}{Skew-t} \\
\cmidrule(r{3pt}l{3pt}){2-4} \cmidrule(l{3pt}r{3pt}){5-7}
Parameter & \makecell[r]{Posterior mean} & (5\%, & 95\%) &
\makecell[r]{Posterior mean} & (5\%, & 95\%) \\
\midrule
$\rho$ & 0.874 & (0.797, &\;0.943) & 0.860 & (0.775, &\;0.938) \\
$\lambda$ & 1.592 & (1.380, &\;1.900) & 1.567 & (1.339, &\;1.933) \\
\bottomrule
\end{tabular}
    \caption{Orange metro line - Direction~2}
    \label{tab:errdep22}
\end{table}

\clearpage
\subsection{Realized vs. predicted travel times}
\label{appx:rvp}

\begin{figure}[!ht]
    \centering
    \includegraphics[width=0.9\linewidth]{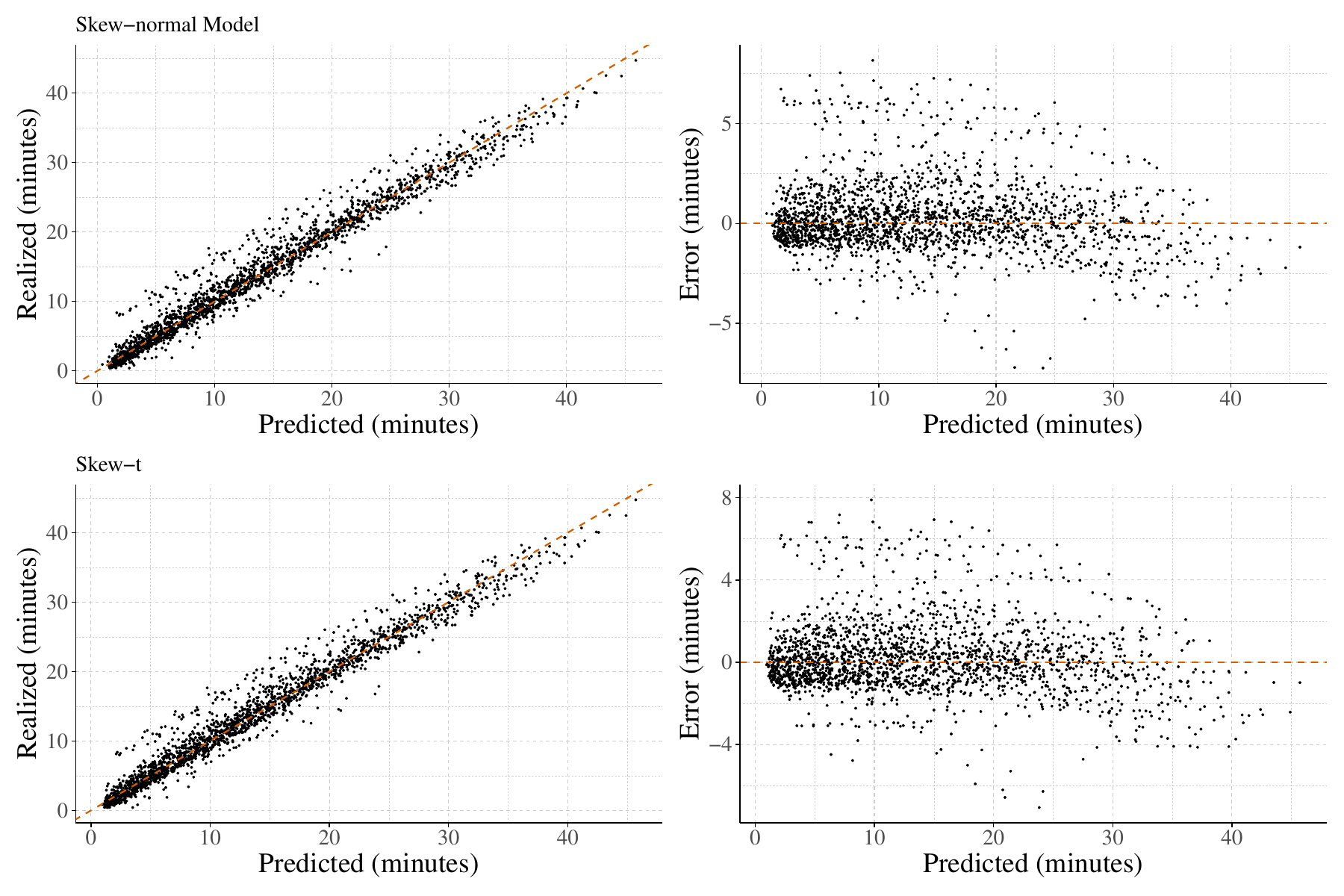}
    \caption{Green metro line -- Direction~2}
    \label{fig:pvr12}
\end{figure}

\begin{figure}[!ht]
    \centering
    \includegraphics[width=0.9\linewidth]{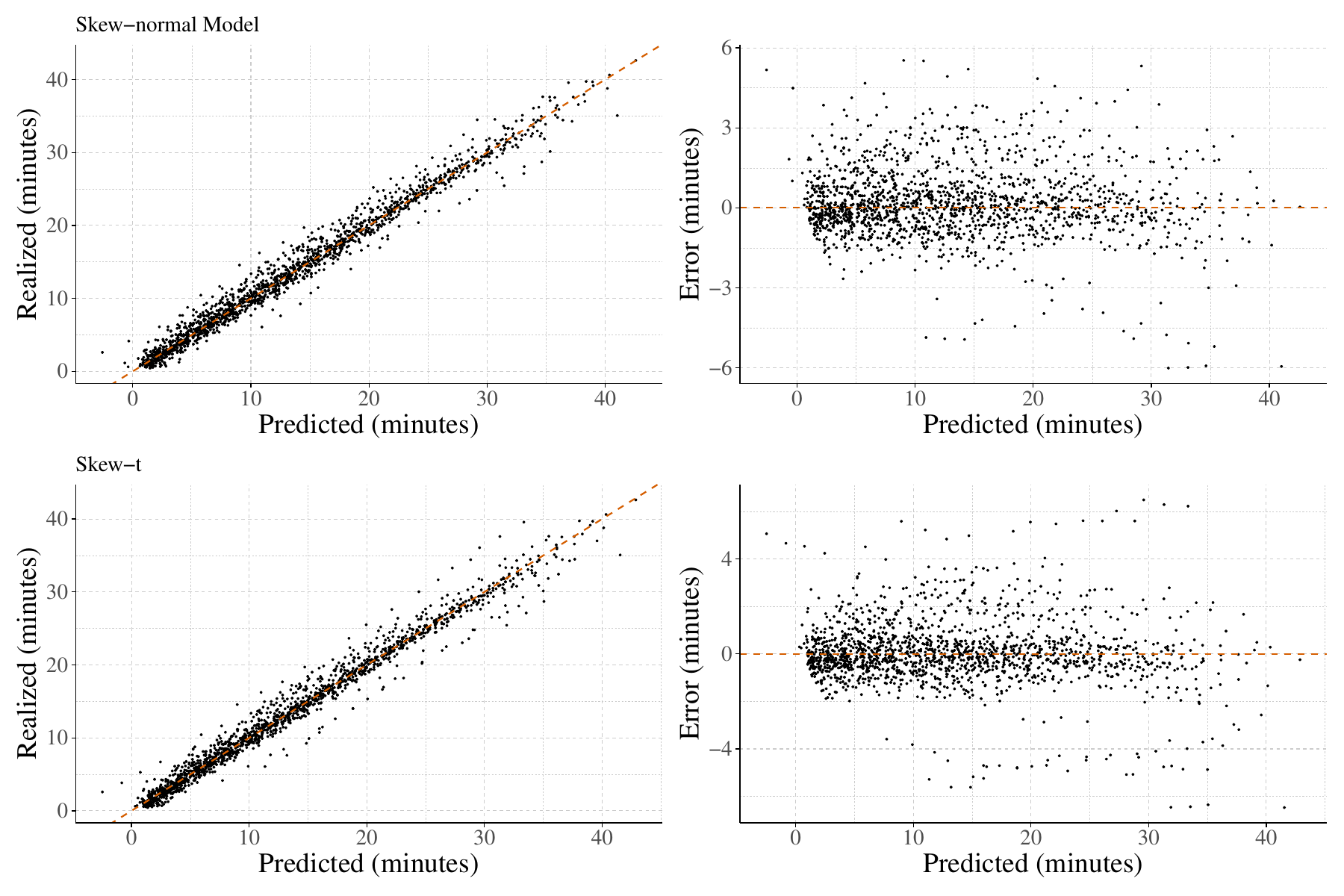}
    \caption{Orange metro line -- Direction~1}
    \label{fig:pvr21}
\end{figure}

\begin{figure}[!ht]
    \centering
    \includegraphics[width=0.9\linewidth]{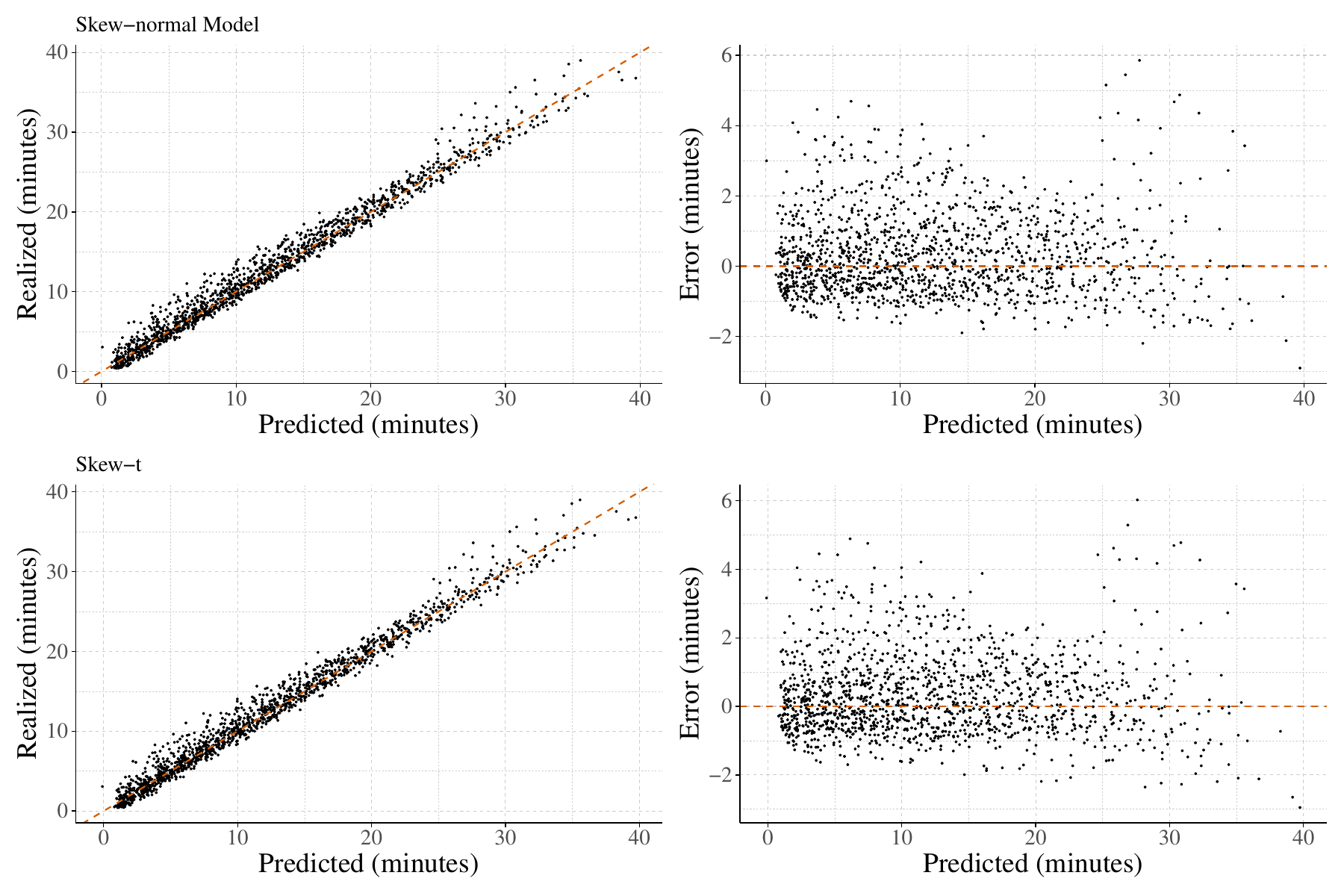}
    \caption{Orange metro line -- Direction~2}
    \label{fig:pvr22}
\end{figure}

Figures~\ref{fig:pvr12}, \ref{fig:pvr21}, and \ref{fig:pvr22} present the realized versus predicted post-disruption travel times (left panels) and the corresponding model residuals (right panels) for the out-of-sample subsets under the skew-normal and skew-$t$ specifications for the remaining case studies, namely Direction~2 of the Green line and both directions of the Orange line. Overall, the plots indicate strong predictive performance, as the predicted values cluster closely around the main diagonal. Nevertheless, the dispersion of residuals appears larger for the Green line (Figures \ref{fig:pvr11} and \ref{fig:pvr12})) than for the Orange line (Figures \ref{fig:pvr21} and \ref{fig:pvr22})), which may point to more stable post-disruption operations on the Orange line.

\clearpage
\subsection{Prediction accuracy and uncertainty assessment}
\label{appx:met}
\begin{figure}[!htbp]
    \centering
    \includegraphics[width=0.9\linewidth]{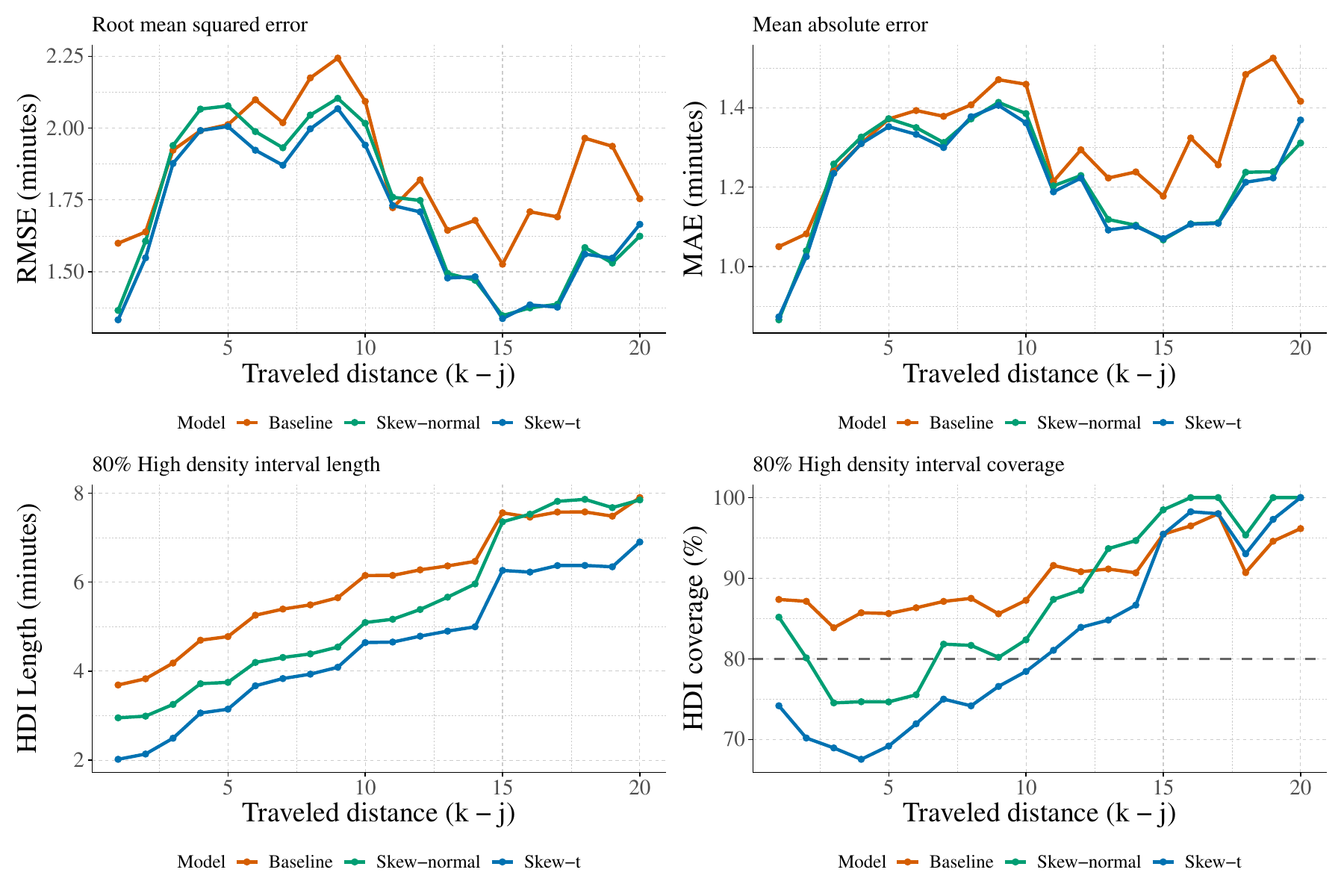}
    \caption{Green metro line -- Direction~2}
    \label{fig:met12}
\end{figure}

\begin{figure}[!htbp]
    \centering
    \includegraphics[width=0.9\linewidth]{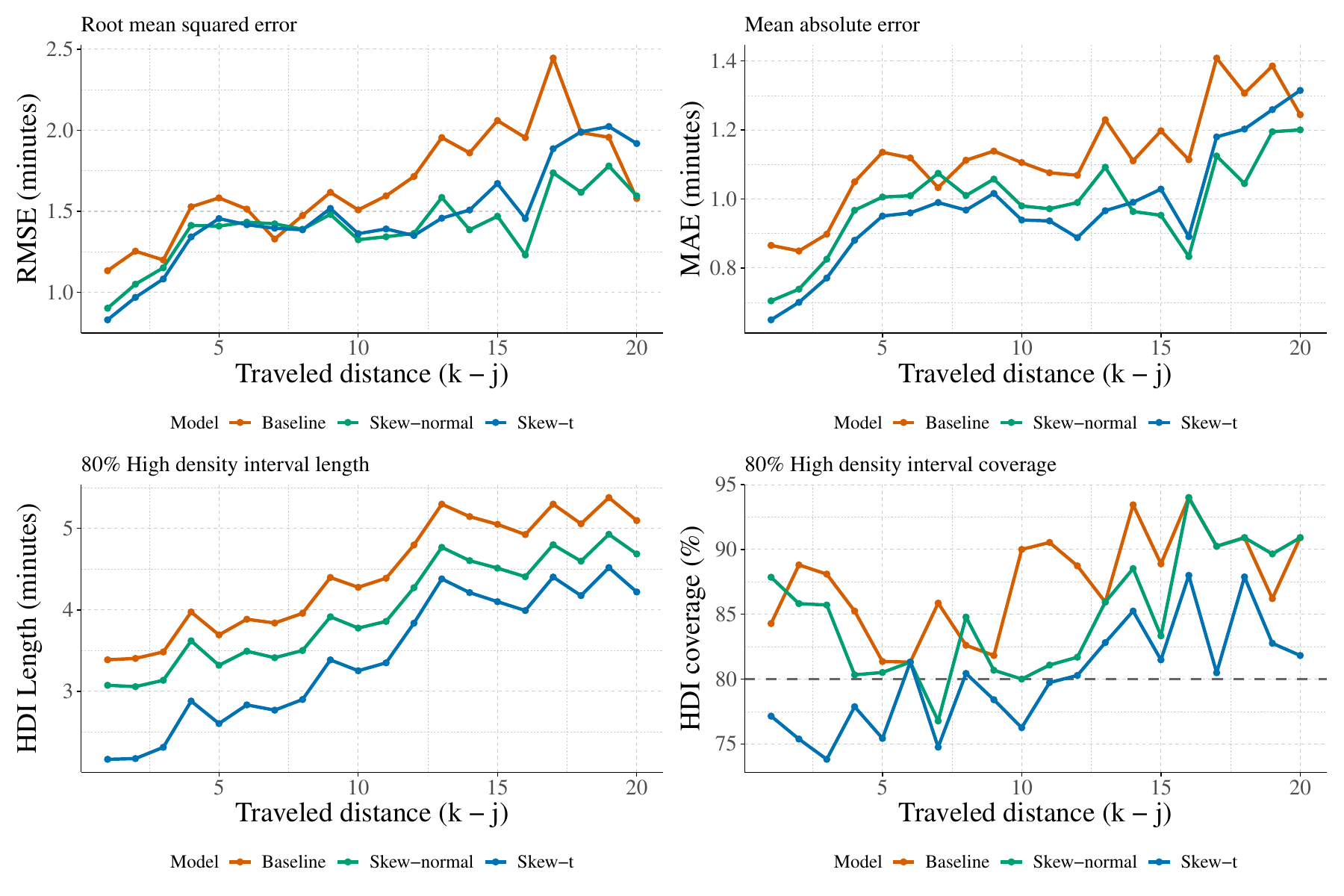}
    \caption{Orange metro line -- Direction~1}
    \label{fig:met21}
\end{figure}

\begin{figure}[!htbp]
    \centering
    \includegraphics[width=0.9\linewidth]{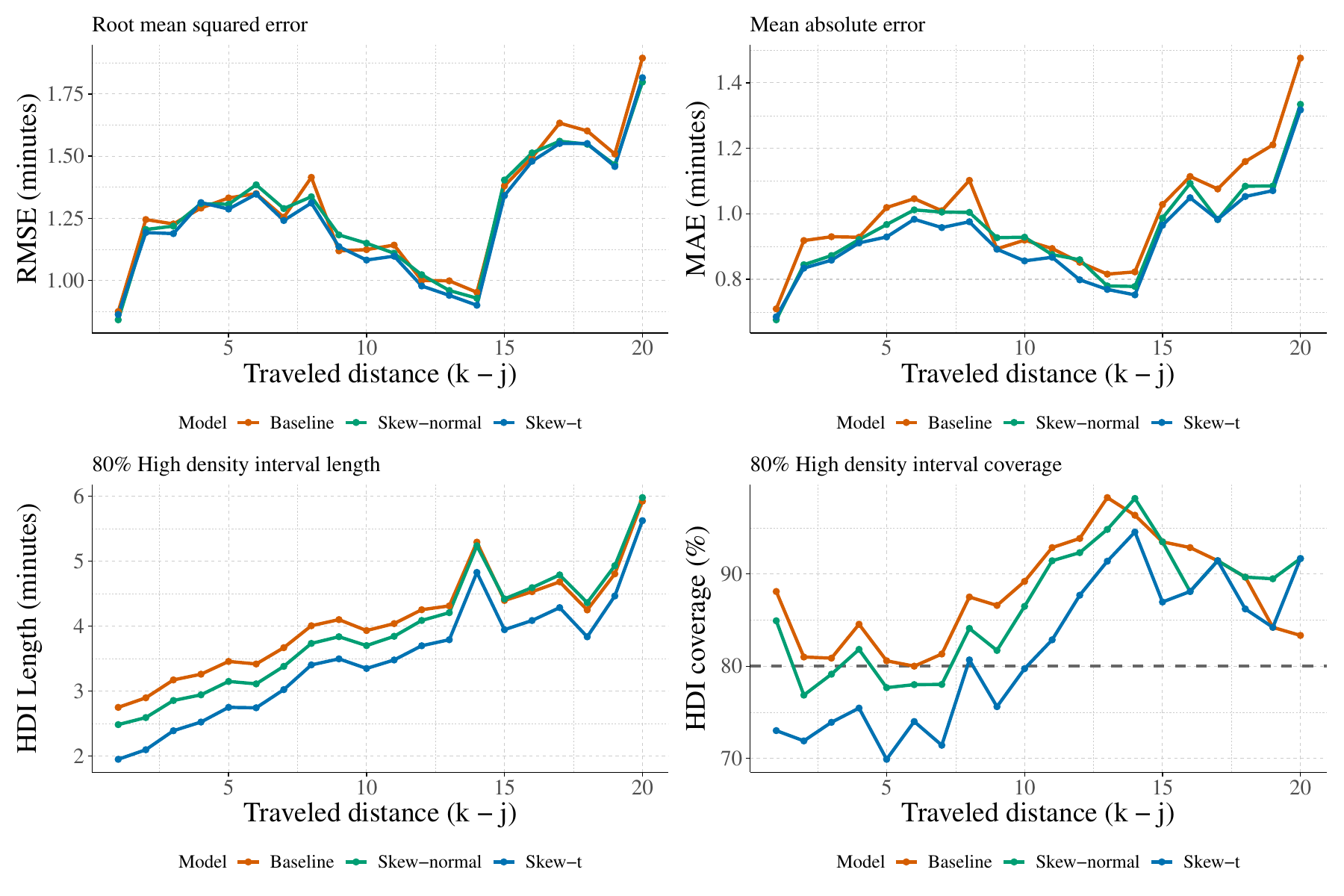}
    \caption{Orange metro line -- Direction~2}
    \label{fig:met22}
\end{figure}

Comparison of error-based metrics and uncertainty assessments for the proposed models relative to the baseline specification for the Direction~2 of the Green line in \autoref{fig:met12}, Direction~1 of Orange line in \autoref{fig:met21} and Direction~2 of Orange line in \autoref{fig:met22}. In each category, the evaluation is performed separately across traveled distances. The results indicate that both proposed models outperform the baseline in terms of MAE, HDI length, and empirical coverage, with the skew-$t$ model demonstrating superior performance compared to the skew-normal specification.

Although the skew-$t$ models yield smaller HDI lengths across all case studies, for the Orange line case studies the skew-normal models exhibit coverage closer to the 80\% level for shorter-distance journeys. Nevertheless, in all cases the skew-$t$ models achieve coverage closer to the nominal 80\% level overall, indicating that the statistical behavior of post-disruption travel times, for longer journeys, exhibits pronounced heavy-tailed characteristics.

\begin{figure}[!htbp]
    \centering
    \includegraphics[width=0.9\linewidth]{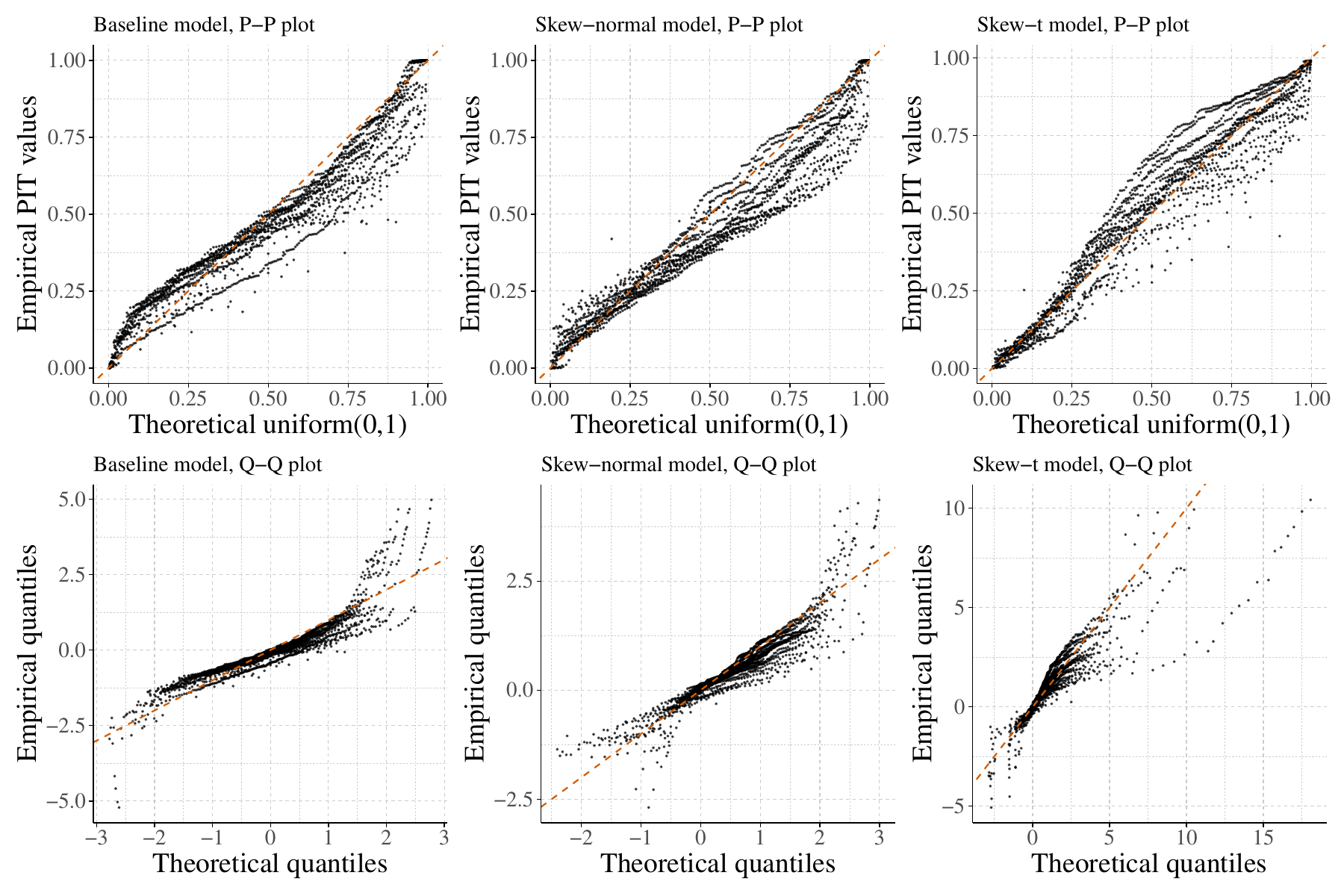}
    \caption{Green metro line -- Direction~2}
    \label{fig:ppqq12}
\end{figure}

\begin{figure}[!htbp]
    \centering
    \includegraphics[width=0.9\linewidth]{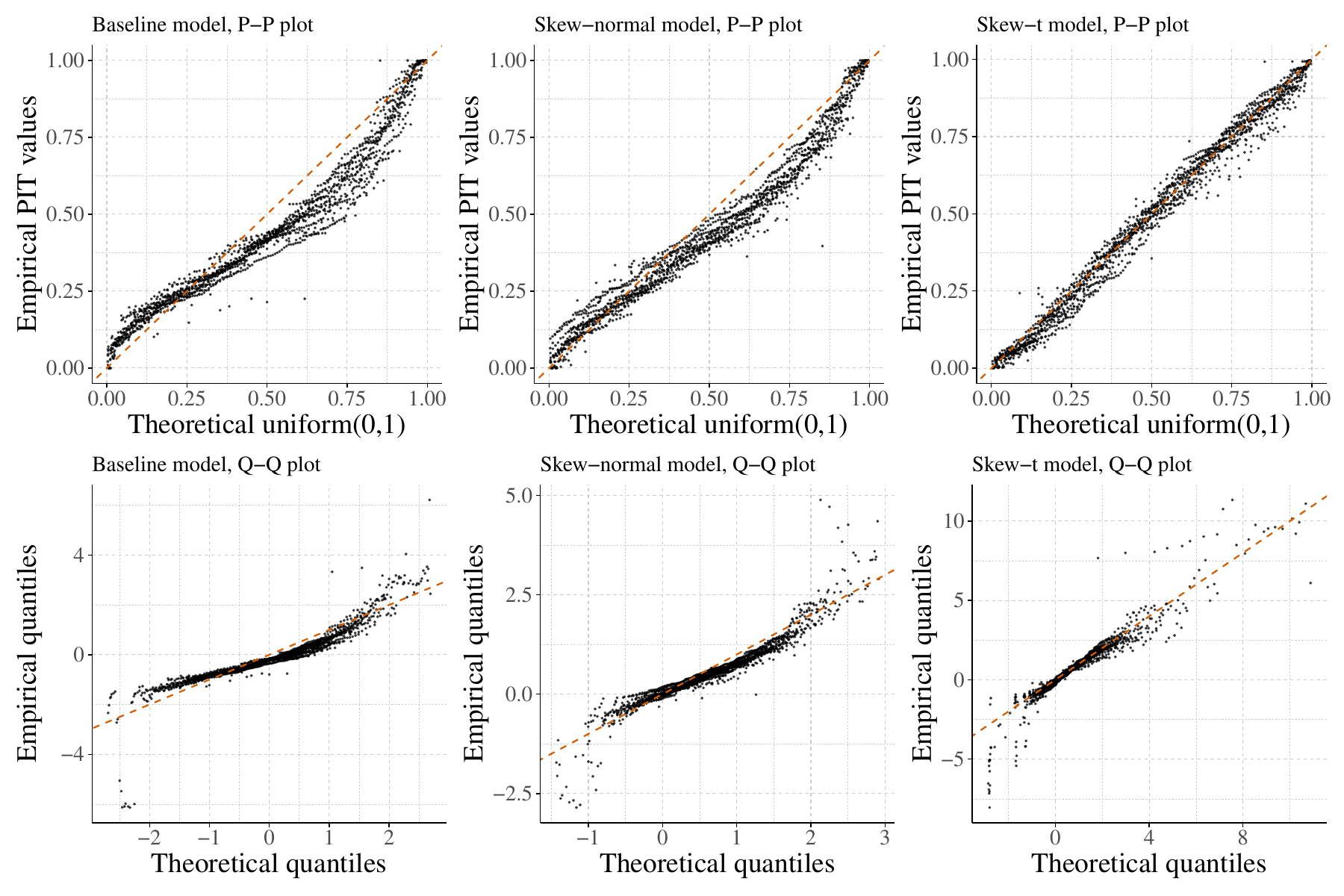}
    \caption{Orange metro line -- Direction~1}
    \label{fig:ppqq21}
\end{figure}

\begin{figure}[!htbp]
    \centering
    \includegraphics[width=0.9\linewidth]{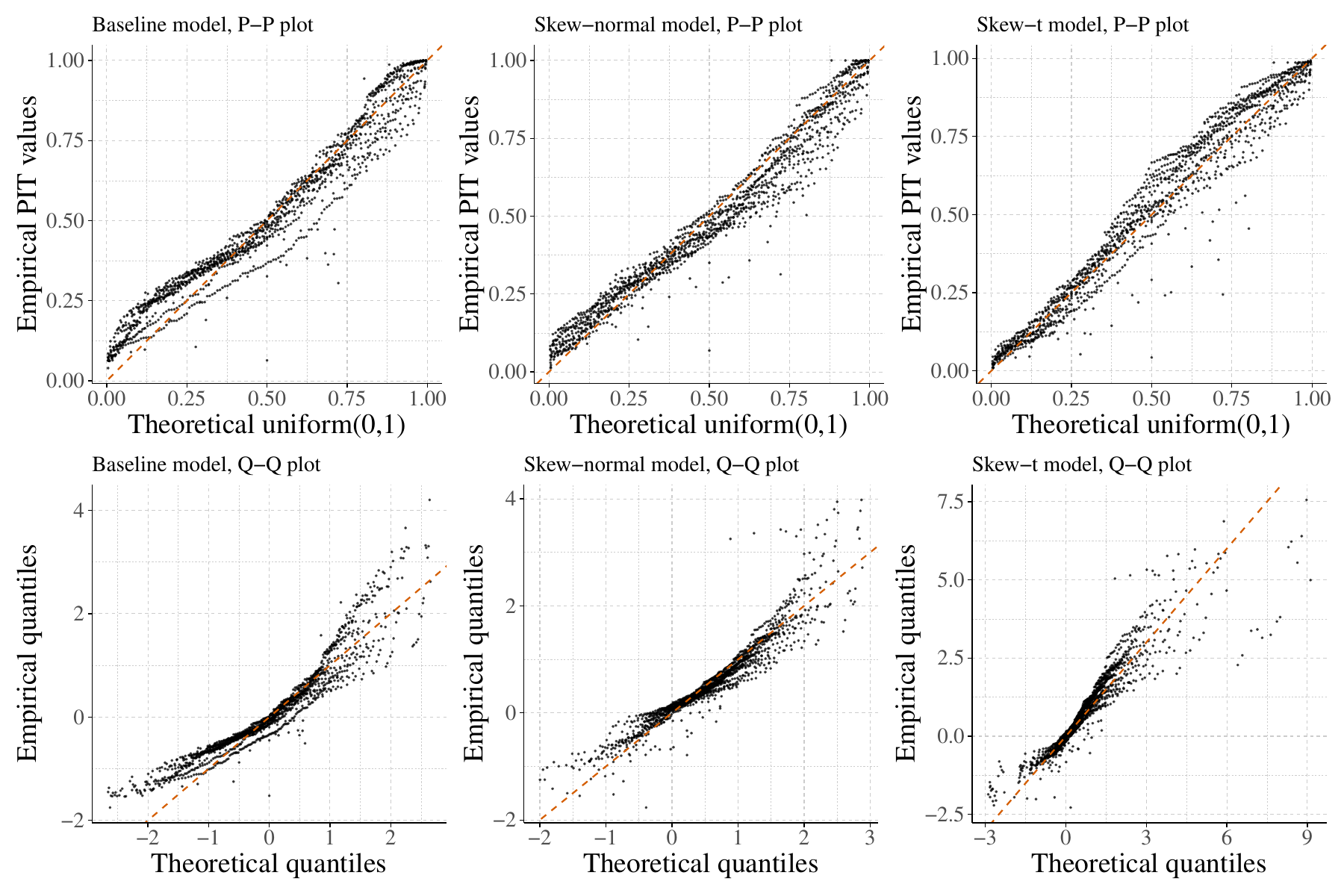}
    \caption{Orange metro line -- Direction~2}
    \label{fig:ppqq22}
\end{figure}

\clearpage
Figures~\ref{fig:ppqq12}, \ref{fig:ppqq21}, and \ref{fig:ppqq22} present the P–P plots (top rows) and Q–Q plots for the baseline, skew-normal, and skew-$t$ model specifications for the remaining three case studies, namely Direction~2 of the Green line and both directions of the Orange line. The empirical and model-based cumulative probabilities and quantiles are computed using the transformations defined in \autoref{eq:z_baseline} and \autoref{eq:z_skew_n} and are plotted against one another. For the skew-normal and skew-$t$ models, empirical cumulative probabilities and quantiles are evaluated separately for each traveled distance $(k-j)$. 

Across all case studies, the proposed models provide a better characterization of both the central mass and the tails of the post-disruption travel time distributions than the baseline specification, as evidenced by the closer alignment of points with the main diagonal in the P–P and Q–Q plots. However, for Direction~2 of the Green line and Direction~2 of the Orange line, the skew-normal model offers a more accurate fit to both the bulk and the tails than the skew-$t$ model, whereas for Direction~1 of the Orange line the skew-$t$ specification exhibits superior performance.

\clearpage

\begin{figure}[!htbp]
    \centering
    \includegraphics[width=0.45\linewidth]{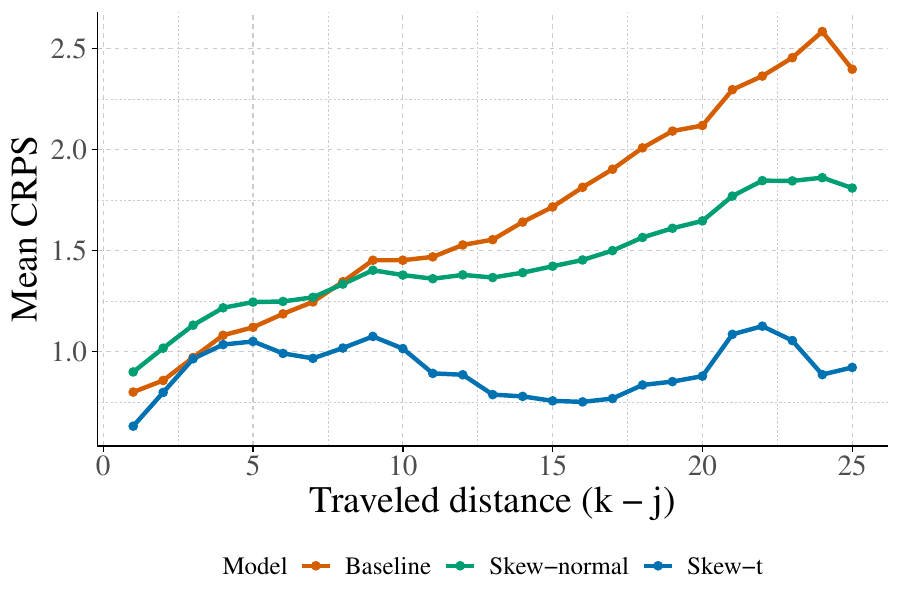}
    \caption{Green metro line -- Direction~2}
    \label{fig:crsp12}
\end{figure}

\begin{figure}[!htbp]
    \centering
    \includegraphics[width=0.45\linewidth]{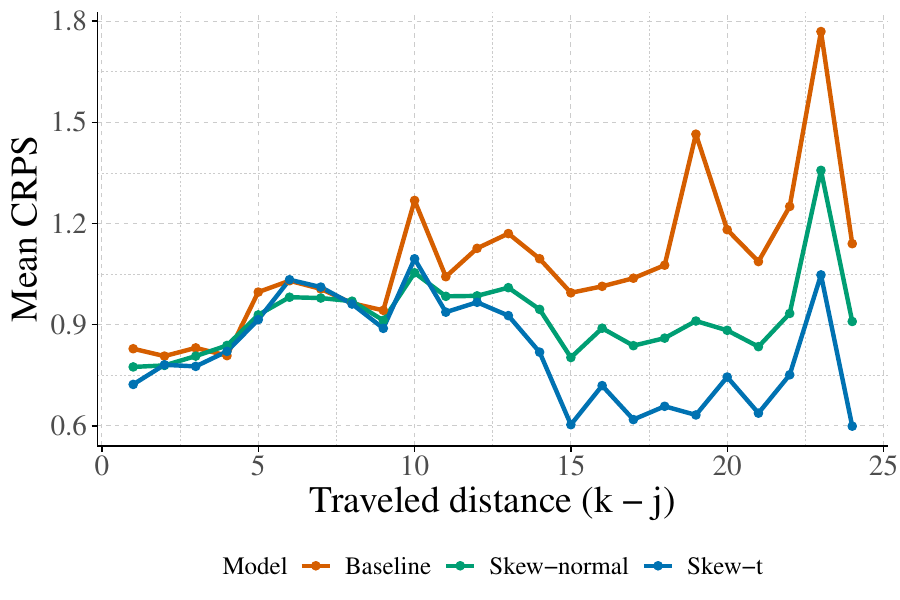}
    \includegraphics[width=0.45\linewidth]{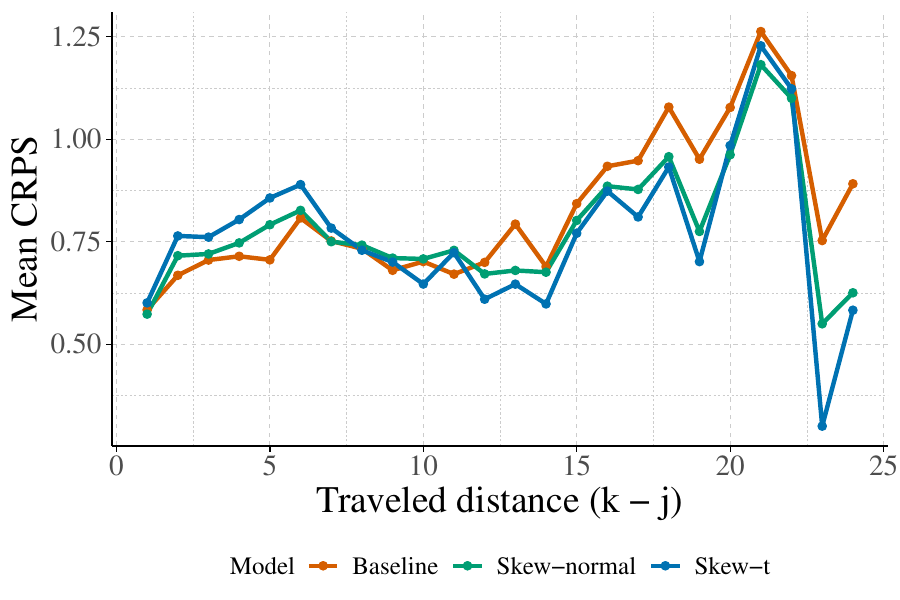}
    \caption{Orange metro line -- (left) Direction~1 and (right) Direction~2}
    \label{fig:crsp2}
\end{figure}

Figures~\ref{fig:crsp12} and~\ref{fig:crsp2} compare the CRPS values across different traveled distances for the baseline, skew-normal, and skew-$t$ specifications for the remaining three case studies, namely Direction~2 of the Green line and both directions of the Orange line. Overall, the proposed models achieve lower average CRPS values across distances in all cases. The skew-$t$ model generally exhibits superior performance relative to both the baseline and the skew-normal specifications. While the skew-$t$ model consistently outperforms the alternatives for Direction~2 of the Green line, its performance for shorter journeys on both directions of the Orange line (\autoref{fig:crsp2}) is comparable to the other models, before clearly outperforming them for longer journeys.

\end{document}